%% file: main.tex
\documentclass[11pt,a4paper]{article}
\pdfoutput=1
\usepackage{jinstpub}
\usepackage{lscape}
\usepackage{epsfig}
\usepackage{url}
\usepackage{graphicx}
\usepackage{multirow}
\usepackage{subfigure}
\usepackage{natbib}
\usepackage{graphicx}
\usepackage{comment}
\usepackage{rotating}
\usepackage{lineno}
\usepackage{textgreek}

\title{A Low Mass Optical Grid for the PROSPECT Reactor Antineutrino Detector}

\input{AuthorList.tex}

\abstract{
PROSPECT, the Precision Reactor Oscillation and SPECTrum experiment, is a short-baseline reactor antineutrino experiment designed to provide precision measurements of the $^{235}$U product $\overline{\nu}_e$ spectrum, utilizing an optically segmented 4-ton liquid scintillator detector.  
PROSPECT's segmentation system, the optical grid, plays a central role in reconstructing the position and energy of $\overline{\nu}_e$ interactions in the detector.  
This paper is the technical reference for this PROSPECT subsystem, describing its design, fabrication, quality assurance, transportation and assembly in detail. 
In addition, the dimensional, optical and mechanical characterizations of optical grid components and the assembled PROSPECT target are also presented.  
The technical information and characterizations detailed here will inform geometry-related inputs for PROSPECT physics analysis, and can guide a variety of future particle detection development efforts, such as those using optically reflecting materials or filament-based 3D printing.
}

\date{December 2017}

\begin{document}

\maketitle

\section{Introduction}
    
PROSPECT, the Precision Reactor Oscillation and SPECTrum experiment, is a short-baseline reactor antineutrino experiment designed to measure the flux and spectrum of the electron antientrino ($\overline{\nu}_e$) generated by a Highly Enriched $^{235}$U (HEU) reactor, the High Flux Isotope Reactor (HFIR) at Oak Ridge National Laboratory (ORNL)\cite{bib:prospect_physics, bib:prospect_nim}. PROSPECT's goal is to probe the possible $\overline{\nu}_e$ oscillation to 1 eV scale sterile neutrinos at short baselines, without reliance on absolute $\overline{\nu}_e$ flux models~\cite{bib:prospect_physics, bib:prospect_first}.  
Operating at HFIR, PROSPECT is able to precisely measure the $\overline{\nu}_e$ spectrum and flux solely from the fission products of $^{235}$U~\cite{bib:prospect_spectrum}. 
By comparing $\overline{\nu}_e$ flux and spectrum measurements with those from low enriched Uranium reactors, where $\overline{\nu}_e$ are generated from a mixture of fission isotopes, PROSPECT will also be able to investigate the isotopic origin of reactor flux anomaly~\cite{bib:mention2011,bib:huber,Dyb_Evol,Giunti_235239,bib:Giunti2017yid,surukuchi_flux} and the local spectral excess at 5~MeV to 7~MeV $\overline{\nu}_e$ energy~\cite{bib:DYB,bib:DC,bib:RENO,bib:haser,bib:huber2016xis}.

\begin{figure}[h!]
\centering
\includegraphics[width=0.45\textwidth]{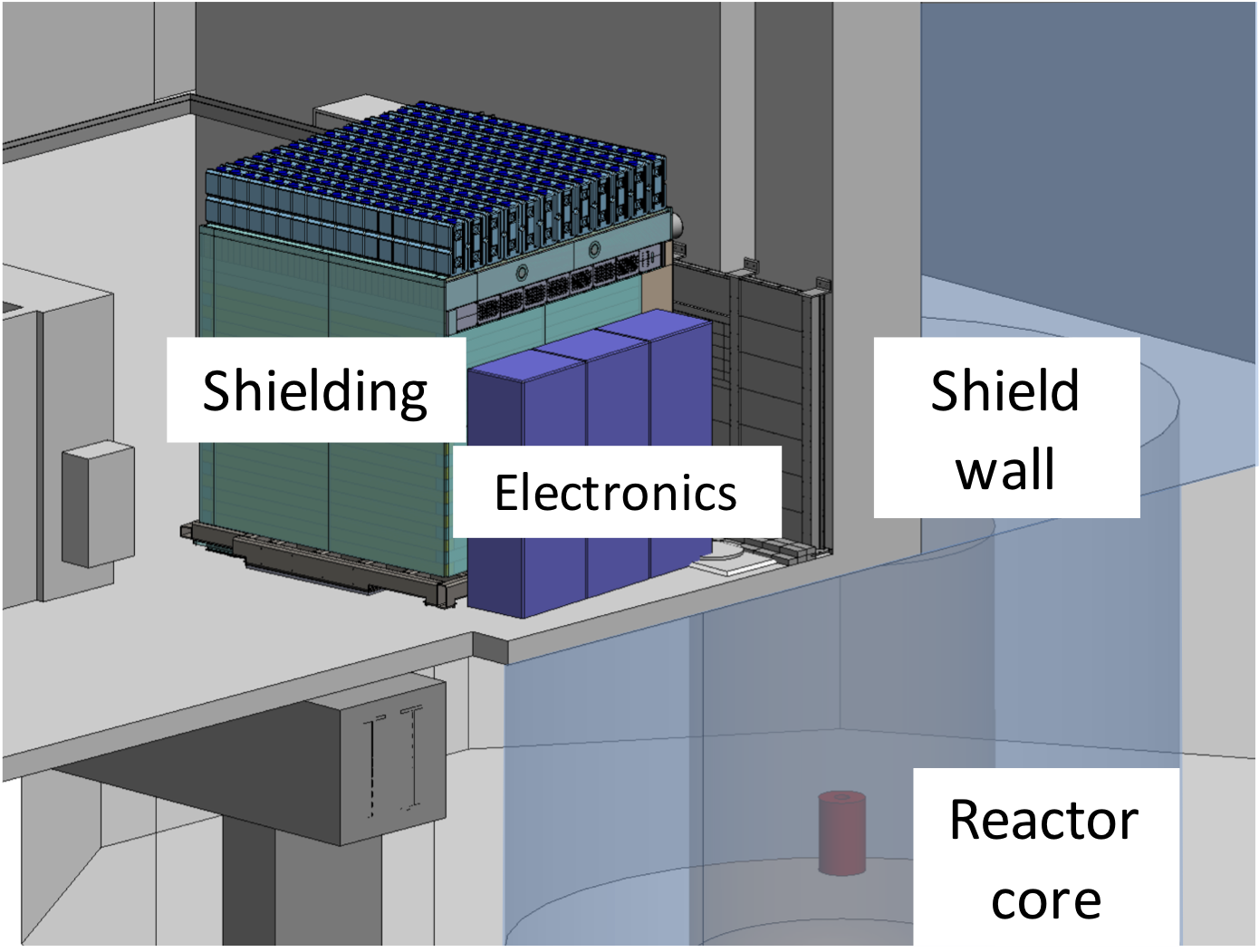}\quad
\includegraphics[trim = 0cm 2cm 0cm 2cm, clip,width=0.6\textwidth]{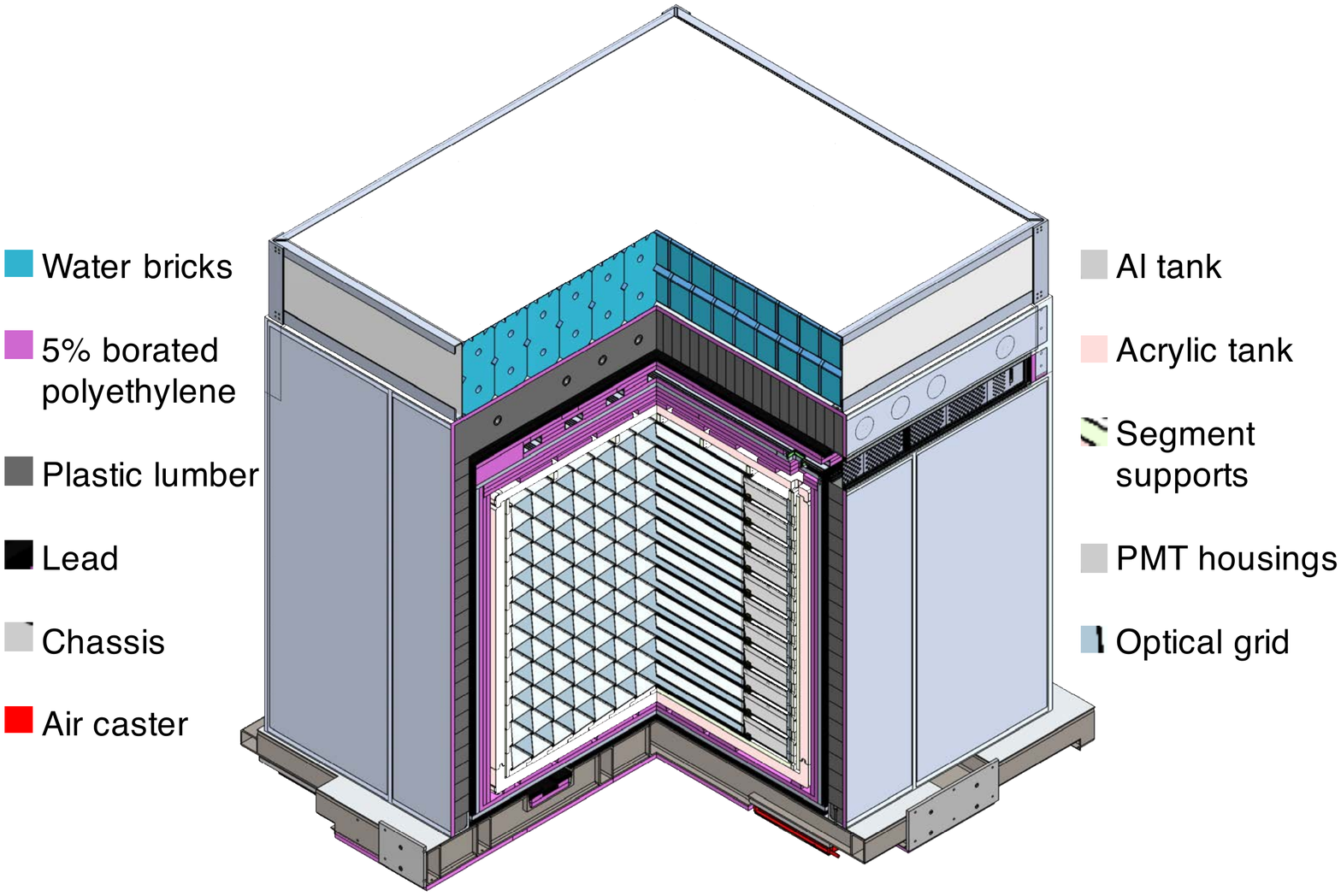}
\caption{(Top) The layout of PROSPECT experiment, where  detector is $\approx7.9$~m from reactor center to detector center. 
(Bottom) The detailed schematic of the PROSPECT AD}
\label{fig:prospect}
\end{figure}

To achieve these physics goals, PROSPECT uses a $\approx$4~ton $^{6}$Li-doped liquid scintillator ($^{6}$LiLS) antineutrino detector (AD) deployed at 7~m to 9~m baselines from the HFIR core.
The $^{6}$LiLS of PROSPECT AD is made from an EJ-309 base \cite{bib:lspaper}.
A labelled model of the PROSPECT AD and its location relative to the HFIR reactor core is shown in Figure~\ref{fig:prospect}; a detailed description of the full experimental layout can found in Reference~\cite{bib:prospect_nim}.  
Antineutrinos are detected via the inverse beta decay (IBD) interaction on protons in the scintillator, producing a correlated positron and neutron pair.  
The IBD positron deposits energy and produces scintillation light in the immediate vicinity of its creation point, and subsequently annihilates, producing two 511~keV gammas.
The neutron is thermalized in the scintillator and captures on $^{6}$Li ($\approx 80\%$) or hydrogen ($\approx 20\%$) on a 40~\textmu s to 50~\textmu s timescale within a few centimeters of the IBD interaction vertex.  

Within the detector, the signals of the IBD interaction are recorded as time- and position-correlated scintillation light pulses~\cite{bib:prospect_physics}.  
Due to the Pulse Shape Discrimination (PSD) capabilities of the scintillator~\cite{bib:prospect_50}, the first (positron-produced) pulse is composed of a large proportion of scintillation light with a short (ns scale) time constant.  
The second pulse, if produced by neutron-$^{6}$Li capture, will exhibit a substantially larger contribution of long-time-constant (>40~ns) scintillation light.  
Thus, by using common PSD techniques, reactor-related and cosmogenic backgrounds to IBD detection can be highly suppressed in the PROSPECT detector.  

\begin{figure}[h!]
\centering
\includegraphics[clip, width=0.4\textwidth]{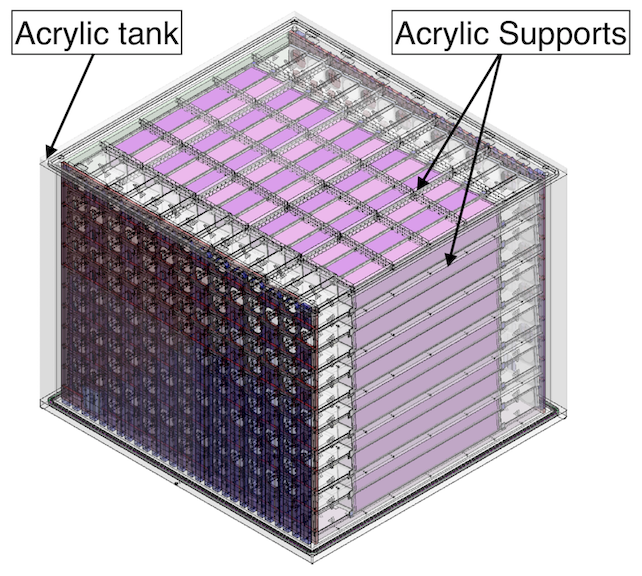} \\
\includegraphics[trim = 2cm 2cm 2cm 2cm,clip, width=0.47\textwidth]{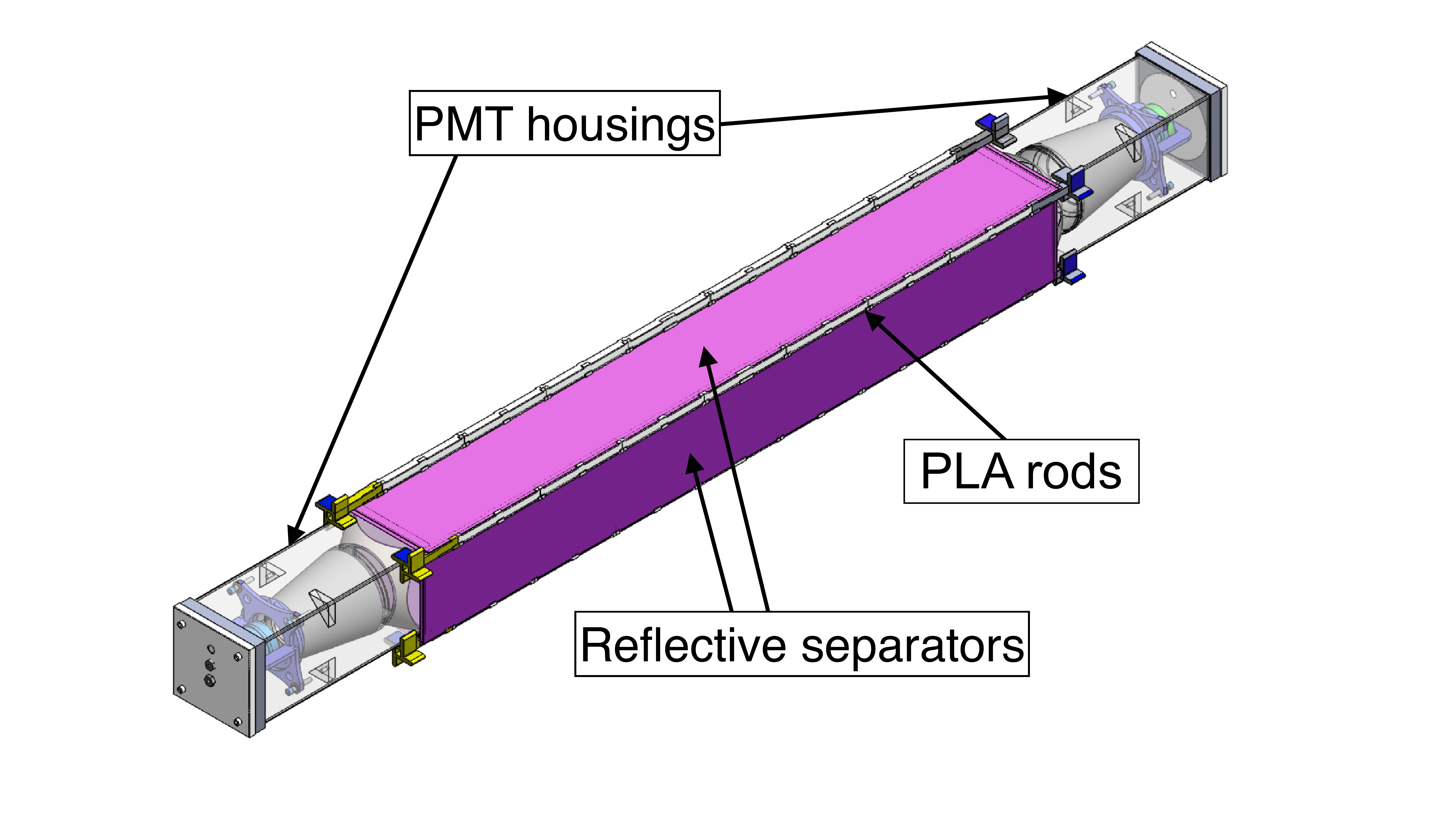} 
\includegraphics[trim = 2cm 2cm 2cm 2cm,clip, width=0.52\textwidth]{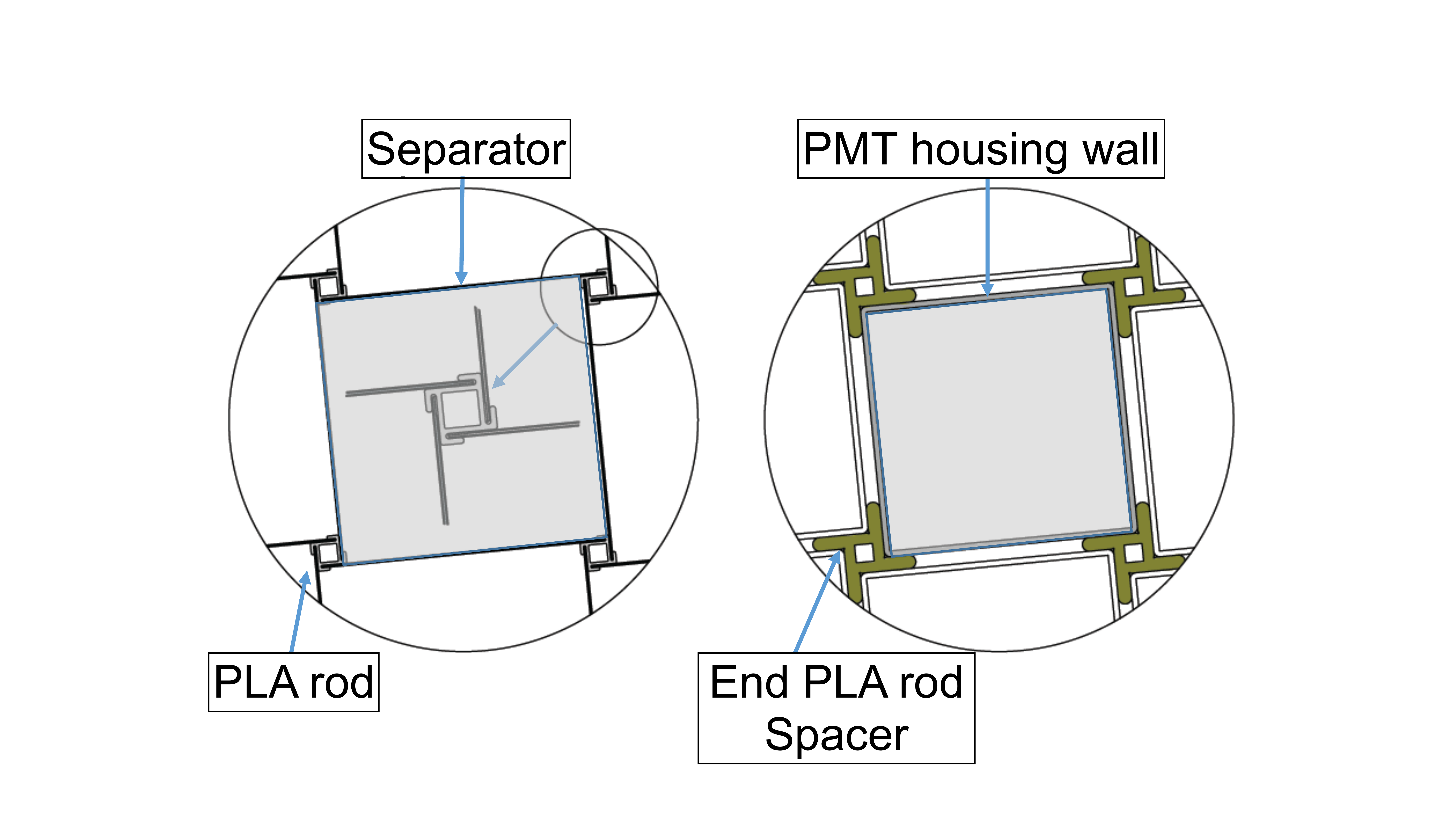}
\caption{Detailed PROSPECT AD schematic. (Top) The active detector enclosed by liquid-tight sealed acrylic tank. (Bottom left) The individual segment with a 12.7~cm (5~in) diameter PMT on each end and enclosed by 4 reflective separators. (Bottom right) The cross section view of the PLA rods and segment, where the separators are slotted on the PLA rods and the PLA rods are hollow to allow calibration sources to be inserted.}
\label{fig:detector}
\end{figure}

The 1.176\,m wide $\times$ 2.045\,m long  $\times$ 1.607\,m tall PROSPECT antineutrino target is optically divided into a 14 (long) $\times$ 11 (tall) grid of longitudinal segments with a long axis nearly perpendicular with respect to a line formed by the core-detector baseline. 
Isometric drawings of the segmented PROSPECT target volume and an individual segment are provided in Figure~\ref{fig:detector}. 
Each segment is 1.176~m in length and has a 14.5~cm $\times$ 14.5~cm square cross-sectional area. 
PROSPECT measures the baseline-dependent flux and spectrum of $\overline{\nu}_e$ by reconstructing the energy and 3D position of the IBD positron and neutron. 
The coordinates of event position reconstruction are defined with respect to the segments' axial direction, where x-direction is the vertical direction perpendicular to the segment axis, and y-direction is the horizontal direction perpendicular to the direction perpendicular to the segment axis, and z-direction is the horizontal direction parallel to the segments' axial direction.
The segment with largest energy deposit is identified as the interaction point (x- and y-direction) of the incident particle.
The readout of photomultiplier tubes (PMTs) housed in mineral-oil filled acrylic modules (PMT optical modules) on both sides of each segment, allowing for timing- and charge-based position reconstruction along the axis (z-direction) of each segment~\cite{bib:prospect_50}.  
The hit segment topologies and reconstructed z-positions are also used in the IBD selection process to reject reactor-related and cosmogenic backgrounds.

The PROSPECT optical grid subsystem is designed to contain the scintillation light generated within a segment and efficiently guide it to the PMTs.
This optical grid consists of low-mass, highly specularly reflective separators (henceforth referred to as separators) held in position by white 3D printed polylactic acid (PLA) rods (henceforth referred to as PLA rods).  
The primary optical grid components are further supported and constrained on two ends by the PMT housings. 
The optical grid is supported by acrylic support plates on the other four external sides. 

This paper gives a detailed description of the design, fabrication, quality assurance and control (QA/QC), transportation, assembly and characterization of the PROSPECT optical grid subsystem.  
    
\section{Optical Grid Design}
\label{sec:Design}
    
To achieve the physics goals of the experiment, the components of the PROSPECT optical grid must be designed to meet the following requirements:
\begin{itemize}
\item{Components should have high reflectance for optical photons in the 400~nm to 550~nm region to ensure efficient transport of scintillation light to the PMTs.  
According to optical simulations and PROSPECT prototyping efforts~\cite{bib:P20}, specularly reflecting components produce a superior light collection compared to diffuse (Lambertian) reflection.}
\item{Components should be opaque, so that the $^{6}$LiLS within each segment is optically isolated from its neighbors.}
\item{Since the energy of the IBD positron is directly correlated to the $\overline{\nu}_e$ that produced it, component thicknesses and volumes should be small to minimize the target's non-scintillating volume and a subsequent fraction of undetected IBD positron energy.}
\item{The optical grid must be mechanically stable, and be able to withstand vibrations during detector shipping and movement with minimal variation in realized segment dimensions.}
\item{Components must exhibit a high degree of dimensional uniformity to enable assembly of the detector and ensure uniformity of segment volumes.}
\item{Component surfaces exposed directly to liquid scintillator must be chemically compatible with it.}
\item{The optical grid's structure must accommodate the deployment of radioactive sources and optical calibration tubes freely in the detector target interior.}
\item{The optical grid must interface properly with nearby detector components.}
\end{itemize}
    
In the following sub-sections, we will detail the design of the separators and PLA rods, placing emphasis on how the aforementioned requirements are fulfilled.

\subsection{Separators}
\begin{figure}[h!]
\centering
\includegraphics[trim = 0.0cm 8.0cm 0.0cm 10.0cm, clip=true, width=0.9\textwidth]{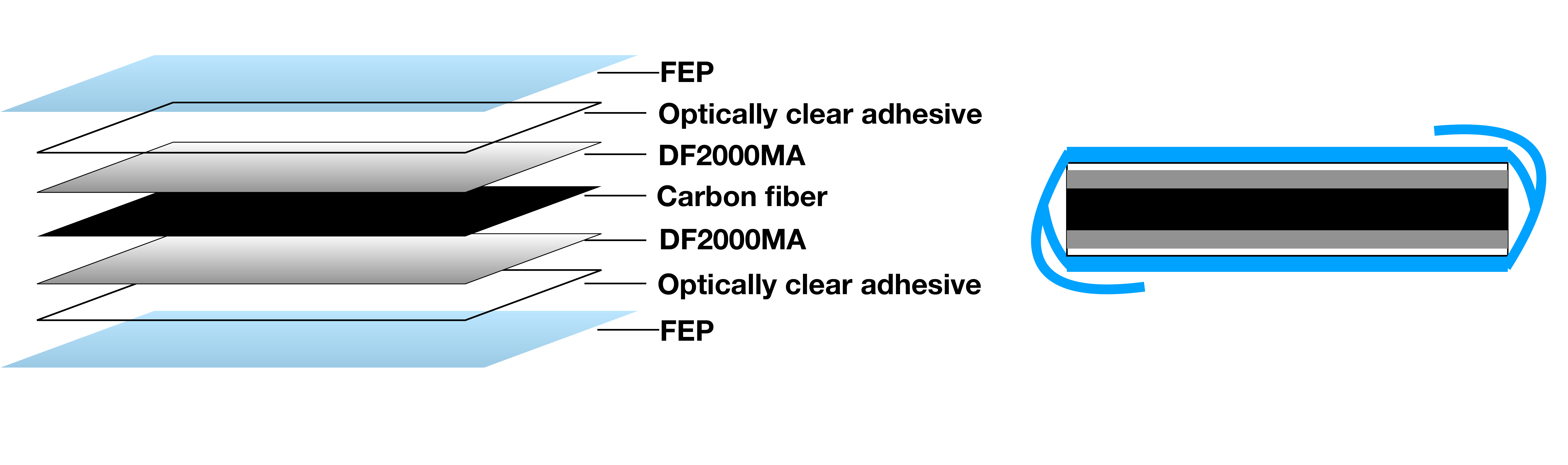} \\
\includegraphics[trim = 0.0cm 2.0cm 0.0cm 8.0cm, clip=true, width=0.35\textwidth]{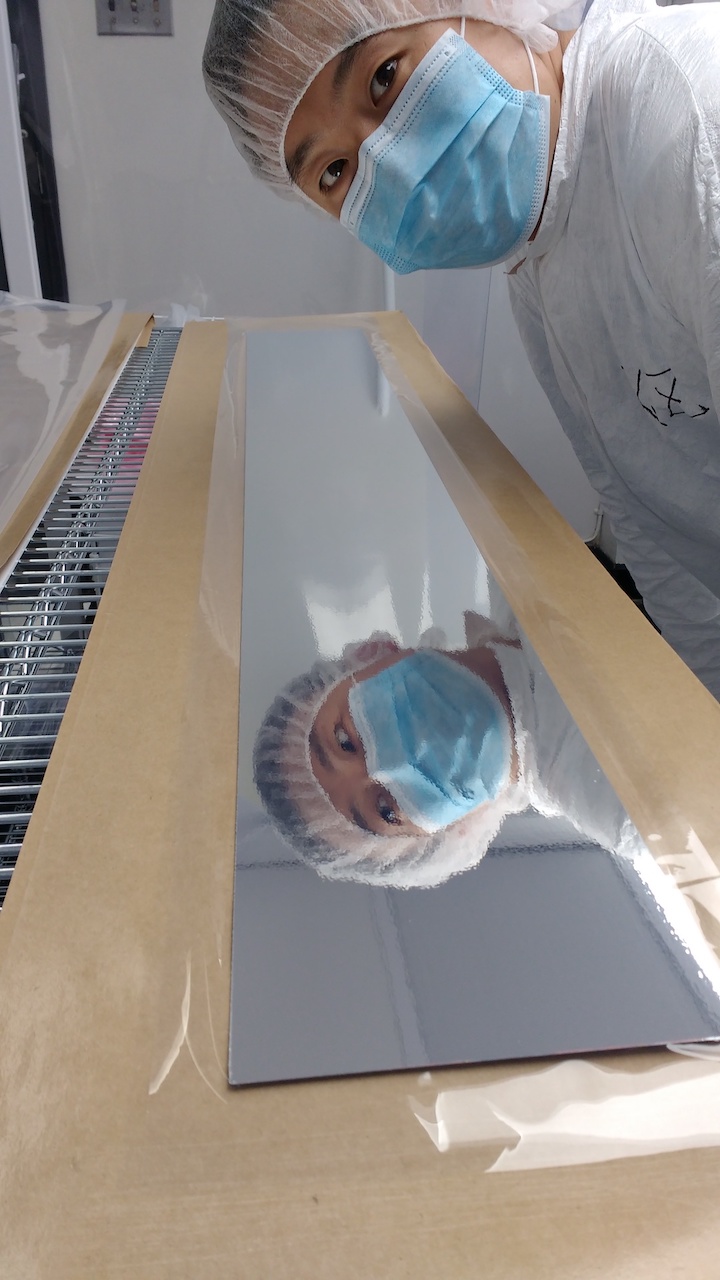}
\includegraphics[clip=true, width=0.6\textwidth]{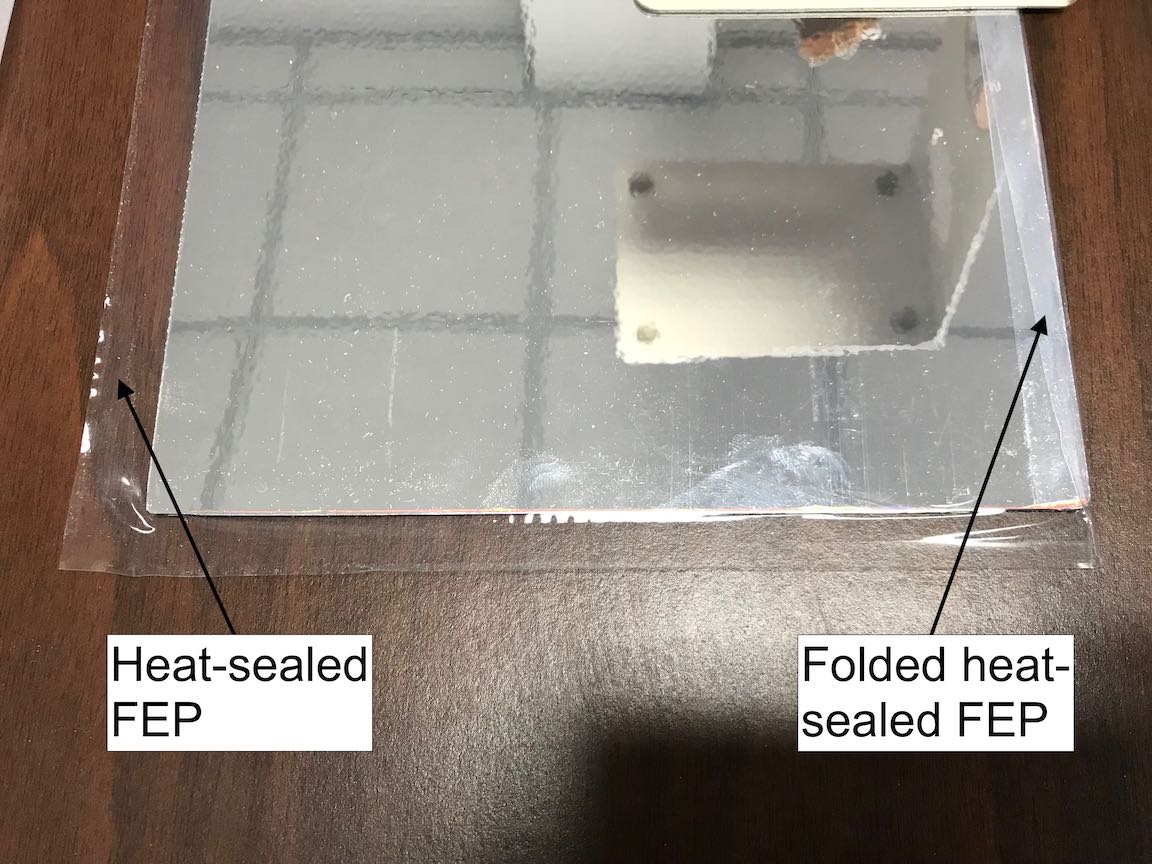}
\caption{(Top left) The illustration of the sandwich structure of a separator. 
(Top right) The illustration of the separator with  the overhung FEP folded. 
(Bottom left) A laminated separator. 
(Bottom right) Two edges of separator showing the folded edge of heal-sealed FEP.}
\label{fig:panel_scheme}
\end{figure}

The separators are composed of a sandwich of carbon fiber backbone, reflector layers, adhesive layers, and protective fluorinated ethylene propylene (FEP) layers, as shown in Figure~\ref{fig:panel_scheme}.  
The structural constituent (backbone) used for the separators is carbon fiber sheeting.  
In addition to acting as the structural component, the backbone material dictates the surface texture.
It was found that thin 0.6~mm (0.023~in) sheets of carbon fiber coated with epoxy resin are sufficiently structurally stiff and can be procured with a glossy finish on both sides.  
Carbon fiber sheets with a pre-impregnated resin system (prepregs) were identified to be mechanically uniform over large sizes and quantities.

\begin{figure}[h!]
\centering
\includegraphics[trim = 0mm 0cm 0cm 0cm, clip, width=0.49\textwidth]{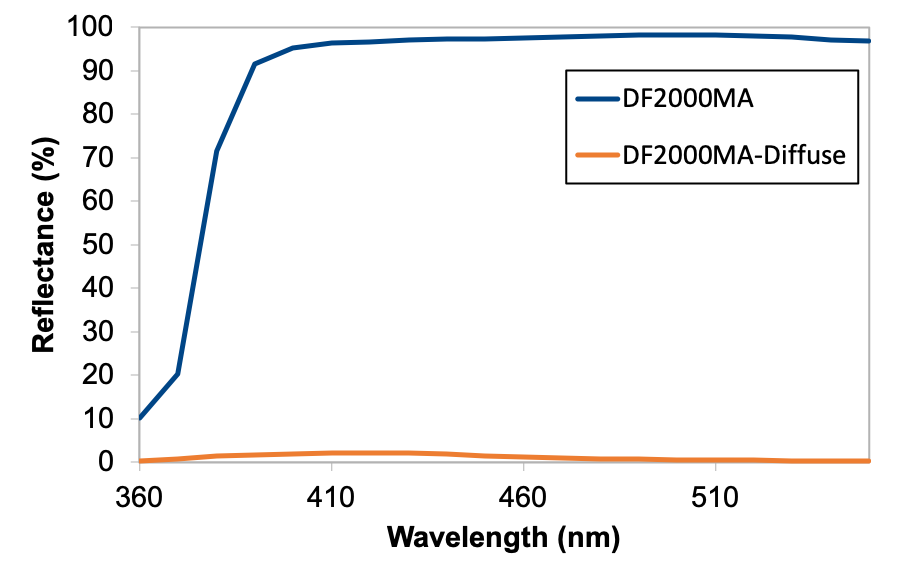}
\caption{Total reflectance and diffuse reflectance of DF2000MA.}
\label{fig:refcomp}
\end{figure}

The reflective material used is DF2000MA, an adhesive-backed organic reflecting film from 3M\footnote{Certain trade names and company products are mentioned in the text or identified in illustrations in order to adequately specify the experimental procedure and equipment used. In no case does such identification imply recommendation or endorsement, nor does it imply that the products are necessarily the best available for the purpose.}.  
Unlike metallic reflective coatings, DF2000MA is made of multiple polymer layers with varying refractive indices, which produce multiple total internal reflections and increased overall reflectivity~\cite{bib:ESR_science}.  
Figure \ref{fig:refcomp} shows the specular and diffuse reflectance spectra of the reflector materials measured with an integrating sphere spectrometer. 
With the pressure-sensitive adhesive as its backing material, DF2000MA film can be laminated on both sides of PROSPECT's carbon fiber backbone sheets to produce rigid sheets that are highly-reflective on both sides.  

On top of the reflector, a film of optically clear double-sided adhesive, General Formulations CON106, is laminated.  
This pressure sensitive adhesive has high transmission in the 400~nm to 700~nm wavelengths.
This adhesive layer ensures stable positioning between protective and reflective layers and maintains uniformity in the overall separator thickness by reducing wrinkles.  
The adhesive also ensures a uniform optical coupling between the reflective and FEP protective layers.  

Finally, a layer of transparent FEP is laminated on both side to guarantee the chemical compatibility of the separator with the $^{6}$LiLS. 
PROSPECT R\&D has shown that the $^{6}$LiLS is compatible with fluoropolymers in general and FEP in particular.
Thus, to ensure compatibility of the full separator with the scintillator, all separators are heat-sealed with an excess of FEP extending beyond the 4 edges of the laminated separator, as shown in Figure \ref{fig:panel_scheme}.
This seal prevents $^{6}$LiLS contact with the carbon fiber, adhesive and reflective layers of the separator.
The heat-sealed FEP films and seals are highly flexible and are folded back during optical grid assembly, as shown in Figure \ref{fig:panel_scheme}.  
The FEP film, having a substantially lower index of refraction than the $^{6}$LiLS ($\approx$1.3 versus $\approx$1.55, respectively), also ensures total internal reflection of grazing incident scintillation light back into the $^{6}$LiLS bulk.  

The designed dimensions of the separators are dictated by the designed dimensions of the segments, which is also governed by the PMT size.  
As the diameter of the PMTs used in PROSPECT AD is 12.7~cm, the PMT housings are built accordingly.  
Based on the PMT housing dimensions and extra width needed for securely coupling to the PLA rods, the nominal width of separators was designed to be 15.35~cm $\pm$ 0.04~cm (6.045~in $\pm$ 0.015~in).  
Therefore, the width of carbon fiber sheets were constrained to 15.3~cm (6.030~in), accounting for the expected additional width added by the extra layers of folded FEP coatings.
Procured rolls of DF2000MA, CON106 adhesive and FEP raw material to be used in separator fabrication are left wider to leave room for natural variabilities in alignment during the laminating procedure.  
Each layer is then trimmed to the right dimensions either by hand or using a custom-built slitter.  
After heat-sealing and trimming extra FEP, the total width of the separator including the overhanging FEP is slightly under 17.78~cm (7~in).
During the detector assembly, the overhanging FEP was wrapped inward as shown in Figure~\ref{fig:panel_scheme}, and the separator width is then constrained by the PLA rods and PMT housing structure determined by the designed segment width.
In practice, allowance for the separator thickness is wider than the sum of the all laminated layers' thicknesses, because of the flexibility of FEP and adhesive as well as the flexibility of the tabs on the PLA rods, as described in Section~\ref{sec:PLAdesign}.
Therefore, the thickness tolerances shown in Table~\ref{tab:material} are dictated by the gap between tab and body of PLA rods. 

\begin{table}[htpb]
\centering
\begin{tabular}{|c|c|}
\hline 
Material & Thickness and tolerance (mm) \\ \hline  \hline
Carbon Fiber & 0.58 $\pm$ 0.1 \\ \hline
DF2000MA & 0.1 \\ \hline
Adhesive & 0.05 \\ \hline
FEP & 0.05 \\ \hline
Thickness tolerance  & 1.176 to 1.244 \\ \hline
Measured thickness & 1.18 $\pm$ 0.05 \\
\hline
\end{tabular}
\caption{Specifications of the separator materials.}
\label{tab:material}
\end{table}

The length of the segment, $\approx$1.22~m (48~in), is dictated by the length of commercial glossy carbon fiber sheets available.  
When combined with the dimensions of the housing, the detector mechanical components, and the shielding material, this length matches the available experimental space at HFIR.
Accounting for the removal of the rough raw carbon fiber sheet edges, the total length of the separators, excluding the overhung FEP, was designed to be 120.65~cm $\pm$ 0.25~cm (47.5~in $\pm$ 0.1~in) long.  
The designed distance between the front surfaces of the two PMT housing is 117.4~cm (46.25~in), with the reflecting separator surface extending beyond the front windows of the PMT housings and out of the active optical volume of the segment.  
Since the separators extend past the faces of the PMT housings, the tolerances in the separator length are not stringent.  
The separators are secured in place by PLA rods.  

\subsection{PLA rods}
\label{sec:PLAdesign}
The main purpose of the PLA rods is to support the separators, as well  as to provide the interface between the optical grid and the other structural elements of the inner detector, such as the PMT housings and acrylic support plates.  
Meanwhile, PLA rod elements that physically constrain the separators must also have a small overall impact on the segment optics.
In addition, these PLA rods need to be low mass and designed to allow calibration and optical sources to freely pass through them.
The PLA rod material should be compatible with the $^{6}$LiLS and should withstand the stresses from the structure and $^{6}$LiLS over the experiment period.

After research and testing, Fused Deposition Modeling (FDM) 3D printing was deemed to be the best choice for the production of the PLA rods. 
This method of 3D printing has advantages of its ability to produce complicated geometries, wide choice of materials, ease of prototyping and minimal setup cost.

After investigating the compatibility of various thermoplastics with $^{6}$LiLS, the white-dyed PLA was selected as the material of choice for the production of 3D printed PLA rods. 

The PLA rods are rigid, longitudinal tubes of rectangular cross-section with short tabs forming a pinwheel shape at one to three positions along each PLA rod's profile.  
The PLA rods are made with white-dyed PLA to maximize diffuse reflection of the scintillation light and to minimize the light cross-talk between segments and light absorption.
Considering the part failure rate and limited size of existing filament-based 3D printers, all PLA rods are no more than 15.24~cm (6~in) in length.
To produce a total PLA rod axis of >1~m in length required to cover the span between PMT housings, these short printed rods were strung onto thin $\approx$1.8~m long acrylic rods or Polytetrafluoroethylene (PTFE) tubes prior to the detector assembly.  

\begin{figure}[h!]
\centering
\includegraphics[width=0.8\textwidth]{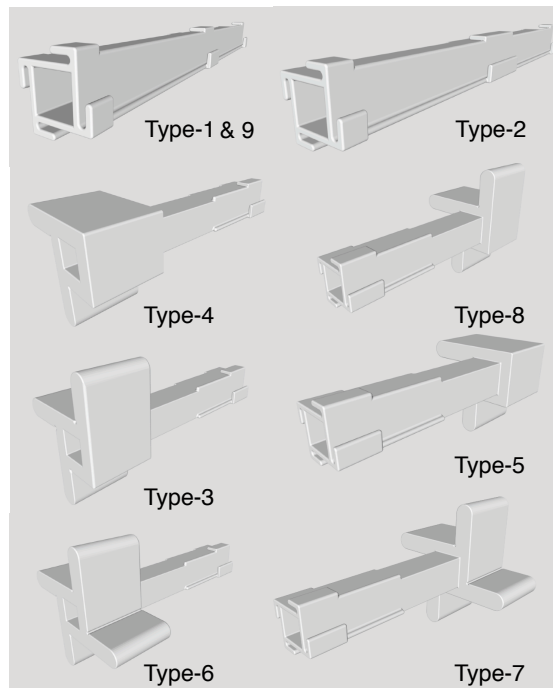}
\caption{Schematic of PLA rods labeled by type.}
\label{fig:pinwheel_types}
\end{figure}

\begin{figure}[h!]
\centering
\includegraphics[trim = 0cm 3cm 0cm 3cm, clip,width=0.8\textwidth]{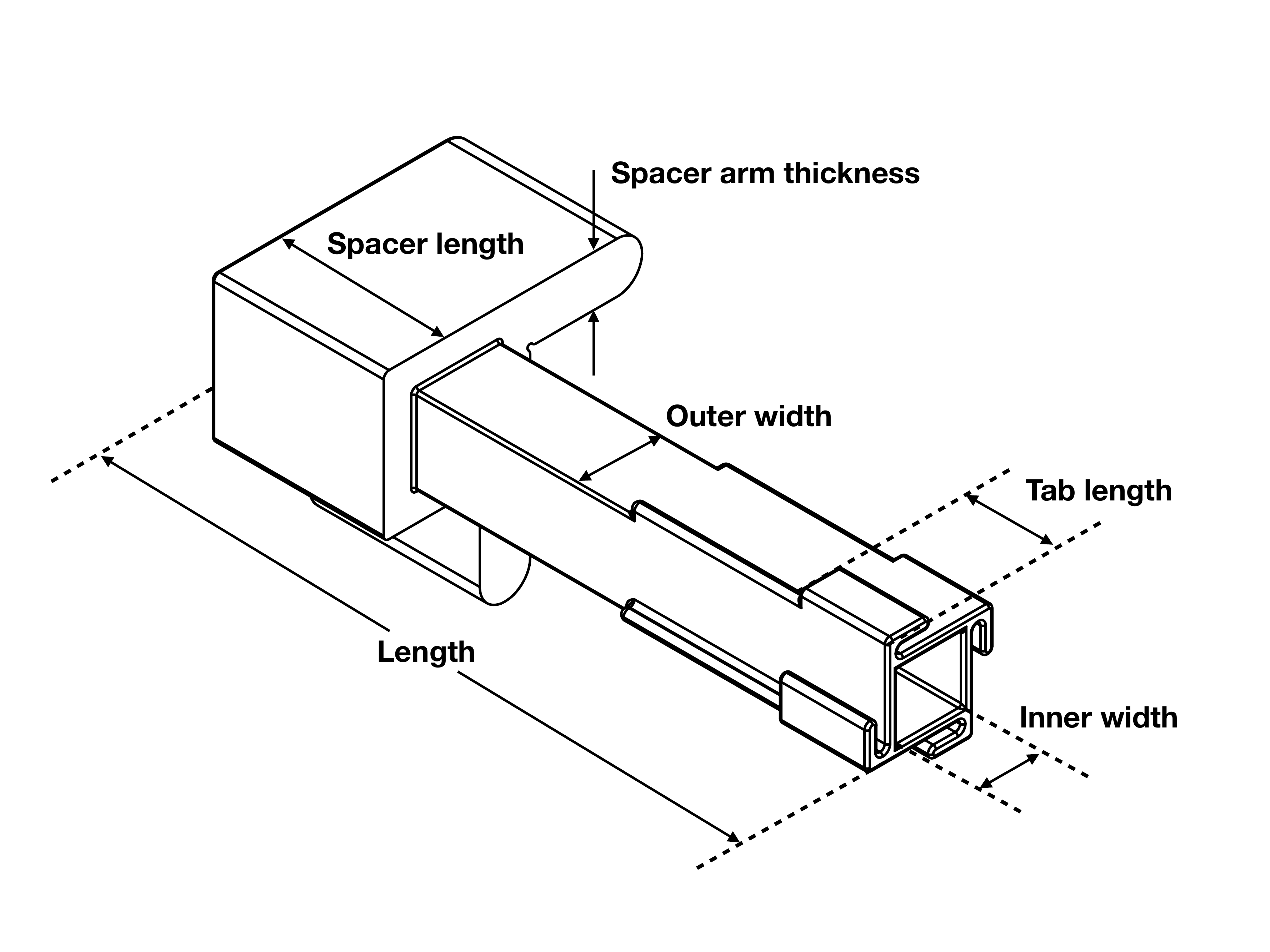}
\caption{The illustration of the key PLA rod dimensions.}
\label{fig:roddesign}
\end{figure}

There were nine types of PLA rods designed according to their location in the assembled detector, as shown in Figure \ref{fig:pinwheel_types}. These nine types can be categorized into three main categories listed below, whose precise designed dimensions are shown in Figure~\ref{fig:roddesign} and Table~\ref{tab:roddesign}.
\begin{itemize}
\item Standard PLA rods: A 15.69~cm (6.233~in) long rod with pinwheel-shaped tabs at its center and each end to allow the insertion of separators. 
The tabs on the ends are $\approx$6~mm (0.25~in) long and the tab at the center is $\approx$13~mm (0.5~in) long to balance increased structural stability and increased reflector exposure.
Among the standard PLA rods, there are PLA rods slightly longer to accurately fit the length of each segment, and ensure light-tight closure between segments.
These standard PLA rods are labeled as type-1 and type-9 respectively.
\item Center PLA rods: Similar to the standard PLA rod but with a 2.54~cm (1~in) wide center tab that allows further machining for the insertion of the optical calibration system components.
The center PLA rods are labeled as type-2.
\item End PLA rods: A 9.53~cm (3.75~in) long rod whose one end is a standard tab for the separator to insert and whose other end is a pinwheel-shaped, thick, rigid spacer to maintain set spacings between PMT housings and strung PLA rods. The number of arms on the spacers depends on the location of rods in the detector.
The end PLA rods are labeled as type-3 to type-8.
\end{itemize}

\begin{table}[htpb]
\centering
\begin{tabular}{|c|c|c|c|}
\hline 
Category  & Standard (longer) & Center & End \\ \hline  \hline
Designed quantity  & 720 (360) & 180   & 360 \\ \hline  \hline
Length  & 15.69~cm (15.83~cm) & 15.69~cm & 9.53~cm\\ \hline
Inner width & 0.98~cm & 0.98~cm & 0.98~cm \\ \hline
Outer width & 1.27~cm & 1.27~cm  & 1.27~cm   \\ \hline
Side tab length & 0.66~cm  & 0.66~cm & 1.31~cm  \\ \hline
Center tab length & 1.31~cm & 2.62~cm  & NA  \\ \hline
Spacer length   & NA & NA  & 2.54~cm \\ \hline
Spacer arm thickness & NA & NA  & 0.74~cm  \\ \hline
\end{tabular}
\caption{The required amount and designed dimensions of PLA rods for each category. 
These dimensions are illustrated in Figure~\ref{fig:roddesign}.}
\label{tab:roddesign}
\end{table}

\begin{figure}[h!]
\centering
\includegraphics[trim = 0cm 5cm 0cm 5cm, clip,width=0.8\textwidth]{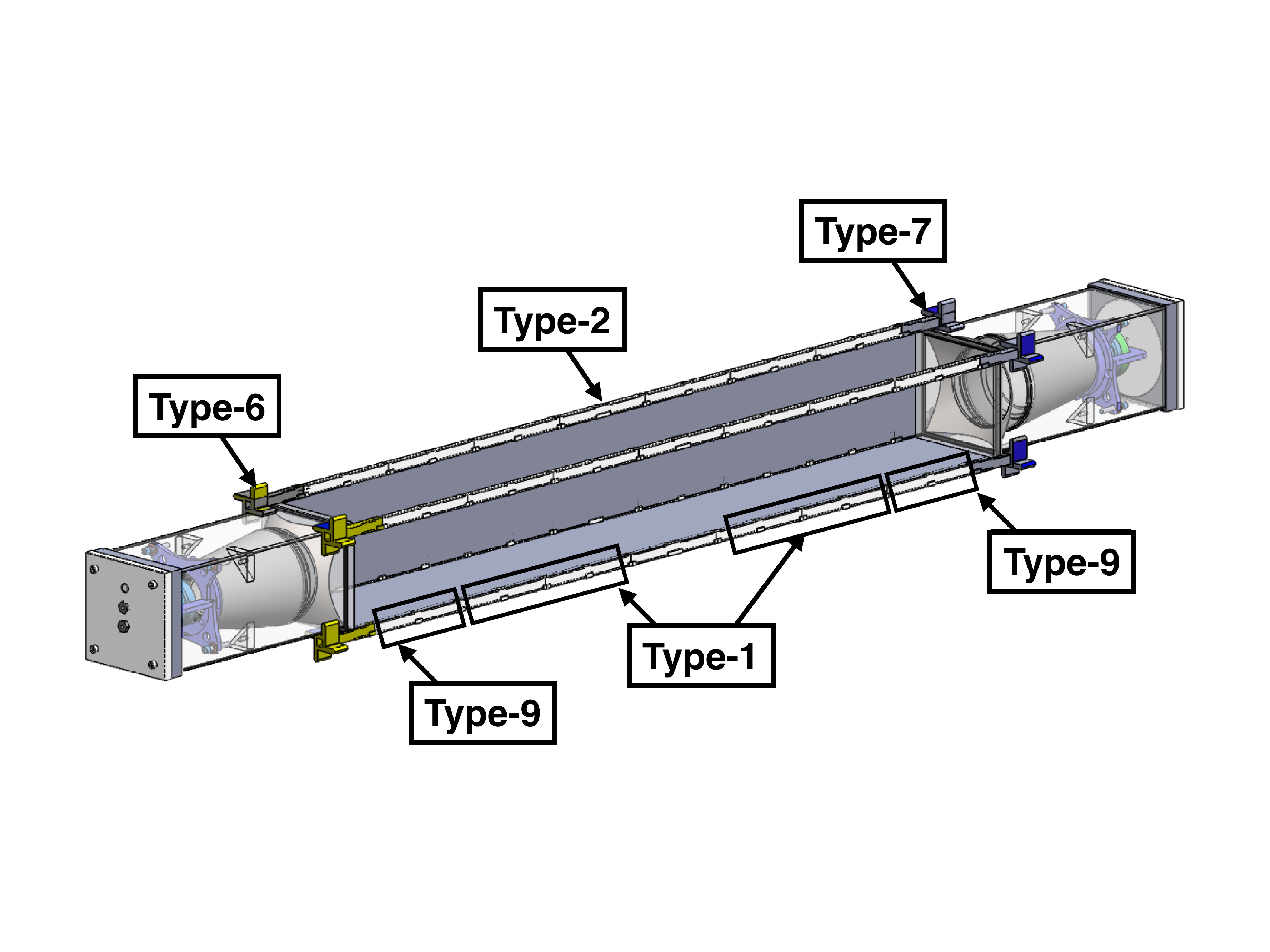}
\caption{The assembled locations of different types of PLA rods. In this figure, the end PLA rods are type-6 and -7. If a segment is at a corner of the detector, the end PLA rods at the specific corner of the segment would be type-4 and -5. Similarly, if the segment is on an edge of the detector, the end PLA rods on one edge would be type-3 and -8.}
\label{fig:pinwheelloc}
\end{figure}

Each completely strung PLA rod contains nine rods of different types. 
The scheme of the assembled locations is shown in Figure \ref{fig:pinwheelloc}.

To avoid scraping and puncturing of the protective FEP surface by the PMT housing faces and loss of compatibility between separators and $^{6}$LiLS, the optical grid was designed to allow no direct mechanical contact between PMT housings and separators.  
The dimensions of the spacers of the end PLA rods ensure $\approx$0.5~mm thick gaps between housing and separator produce marginal losses in an overall light collection, while also allowing the $^{6}$LiLS to more easily flow into each segment volume during AD filling.

\section{Fabrication}
\label{sec:Fab}
To realize the PROSPECT AD, 372 separators were fabricated, including 10\% spares and 5 rejects.
More than 2000 PLA rods were printed, of which 1620 rods were required to build the detector.  
The following section will describe the steps taken in fabricating these optical grid sub-components.  

\subsection{Separator Fabrication}
A total of 388 carbon fiber sheets were ordered from ACP Composites, Inc and delivered to Illinois Institute of Technology (IIT) for processing. 
Each sheet was CNC-cut by the manufacturer to meet the nominal dimensions. 
Following delivery to IIT, the sheet thicknesses were measured to ensure they met the specifications given in Table~\ref{tab:material}. 
To avoid potential damage to the FEP coating, the edges of the carbon fiber were filed to remove splinters, and the corners were rounded with steel hand files. 
In order to limit any changes in dimensions due to the filing, the sides were filed at an oblique angle, while the corners were filed perpendicular to the sheet. 
The sheets were then cleaned first with ethyl alcohol and then twice with water; afterwards, they were dried first with lint free wipes and then with  class 100 cleanroom polyester wipes. 
After filing and cleaning, the sheets were moved into a cleanroom awaiting lamination.

Separator fabrication was conducted in a class 10000 (ISO class 7) soft-wall cleanroom to reduce the incidence of dust and other particulate matter getting laminated into the reflectors, which could mark the reflective surface or lead to punctures of the thin protective FEP coating.
A 46~cm (18~in) wide silicone roll laminator was used to laminate all layers of separators at room temperature.
The structure of the laminated layers are shown in Figure~\ref{fig:panel_scheme}. 
The time sequence consisted of first performing DF2000MA and then adhesive laminations on one side of the carbon fiber, then identical laminations on the other side, and ending with FEP laminations on either side.  
This order was chosen to reduce dust pickup and scratching of the FEP, which, in contrast to the DF2000MA and adhesive, did not include removable protective coatings.  

\begin{figure}[h!]
\centering
\includegraphics[width=0.44\textwidth]{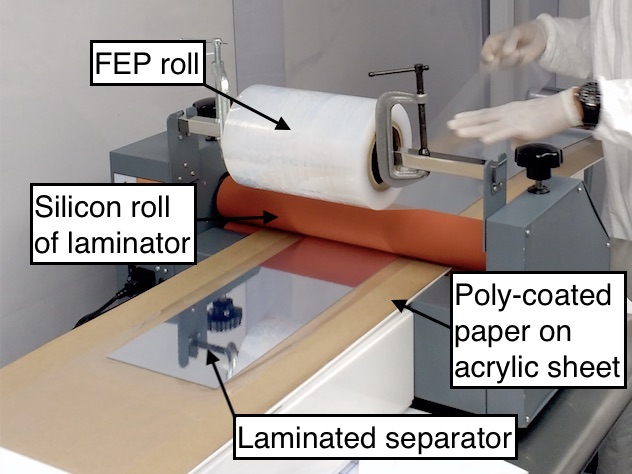}\quad
\includegraphics[width=0.5\textwidth]{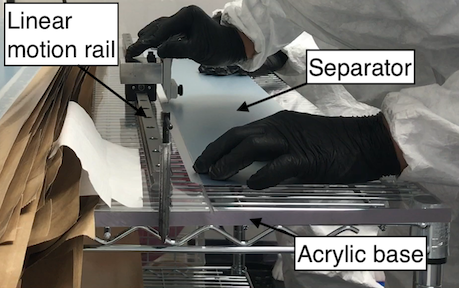}
\includegraphics[width=0.5\textwidth]{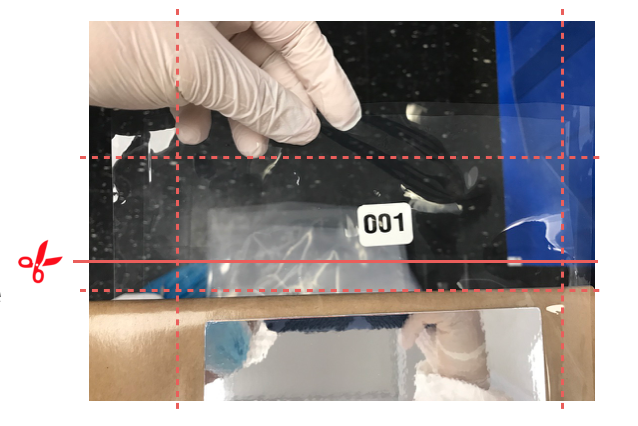}
\caption{(Left top) Photograph of lamination setup, when FEP film were being laminated on one side of separator. (Right top) Photograph of the slitting setup used to cut the excess DF2000MA. (Bottom) Photograph of a labeled separator, where the dashed lines represent the scheme of heat-sealing, and the solid line is the cutting line to remove the separator label.}
\label{fig:laminationPic}
\end{figure}

Lamination was performed in two-person shifts, with one person aligning the to-be-laminated separator and operating the laminator, and the other person keeping the roll of lamination straight and checking for defects. 
During lamination, two acrylic sheets (6.35~mm thick $\times$ 46~cm wide $\times$ 150~cm long) were used in rotation as rigid bases for the ease of aligning the separators. 
At the beginning of every shift, a test lamination was necessary to set up the correct pressure and alignment; incorrectly set pressure can cause non-uniform compression on the separator resulting in delamination, trapped air, and bubble formation.  
This pressure correction test was also necessary when a roll of DF2000MA, adhesive, and FEP was replaced.
During the lamination of each separator, a sheet of poly-coated paper was placed between the separator and the acrylic base sheet to provide a dust free surface and prevent the adhesive from gluing on the base. 
Once a separator was laminated with one layer, it was cut from the laminator roller along with the poly-coated paper, and the next separator was prepared for lamination. 
This process was then repeated until near the end of the roll of laminating material, where occasionally there were defects in the laminate roll, such as wrinkling and collected dust. 
To perform quality control and assurance checks, each separator was labeled with a unique ID between the excess of FEP films once the lamination of the FEP layer was finished as shown in Figure~\ref{fig:laminationPic}.

Between each lamination, the excess material around the separators was removed using a custom-built slitter, as pictured in Figure~\ref{fig:laminationPic}, with finer cuts being made by hand with an artist's knife.
The slitter is composed of a ball bearing linear guide rail that is fixed on an acrylic base, whose sockets for rail attachment and separator placement were CNC machined precisely.
The razor blade of the slitter is attached on the carriage of the rail with a precisely machined aluminum brick.
Cutting one edge with this slitter took $\sim$10 seconds per separator. 
The width variation caused by the cutting with the slitter was below the millimeter level.
Following a visual check and a series of quality assurance (QA) procedures (further described in Section \ref{sec:QA}) after the FEP lamination, separators were packed into non-scratching class-100 cleanroom polyethylene bags and shipped out for heat sealing.
The shipping preparation is detailed in Section~\ref{sec:Trans}.

Heat sealing was performed at Ingeniven (Hapmton, New Hampshire) with two impulse sealers, each of which employed custom elements were used to allow for sealing close to the edges of the separators.  
One of the two sealers that was capable of producing seals on the long sides while the second sealer capable of producing seals on the short side.

Five seals were performed on each separator. 
A $\sim$0.3~mm wide seal was placed at a $\sim$0.6~mm closest distance along each of the four edges of the carbon fiber - reflector - adhesive sandwich.  
To prevent the label from coming off and contaminating the separator surface during cleaning procedures (described in Section~\ref{sec:Assembly}), a fifth seal $\sim$5~cm away from the sandwich was also added to seal a paper label into each of the separators, as shown in Figure \ref{fig:laminationPic}.
The heat sealer was set up at the lowest temperature to achieve the seal, preventing the degradation of the sealer bar which would, in turn, affect the quality of seals.
After the sealing and subsequent QA (described in Section~\ref{sec:QA}) were completed, excess FEP beyond the seal was trimmed using a scalpel and a straight edge.  
Separators were then placed back into cleanroom bags for shipment to Yale University for PROSPECT detector assembly.  

\begin{figure}[h!]
\centering
\includegraphics[width=60mm]{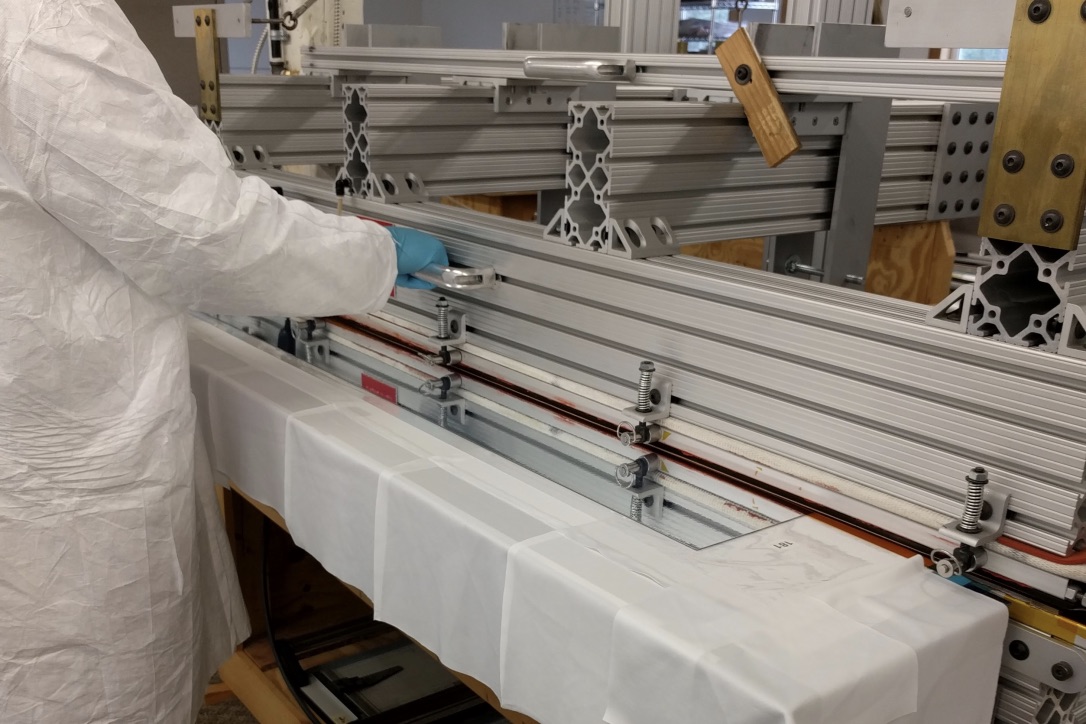}\quad
\includegraphics[width=60mm]{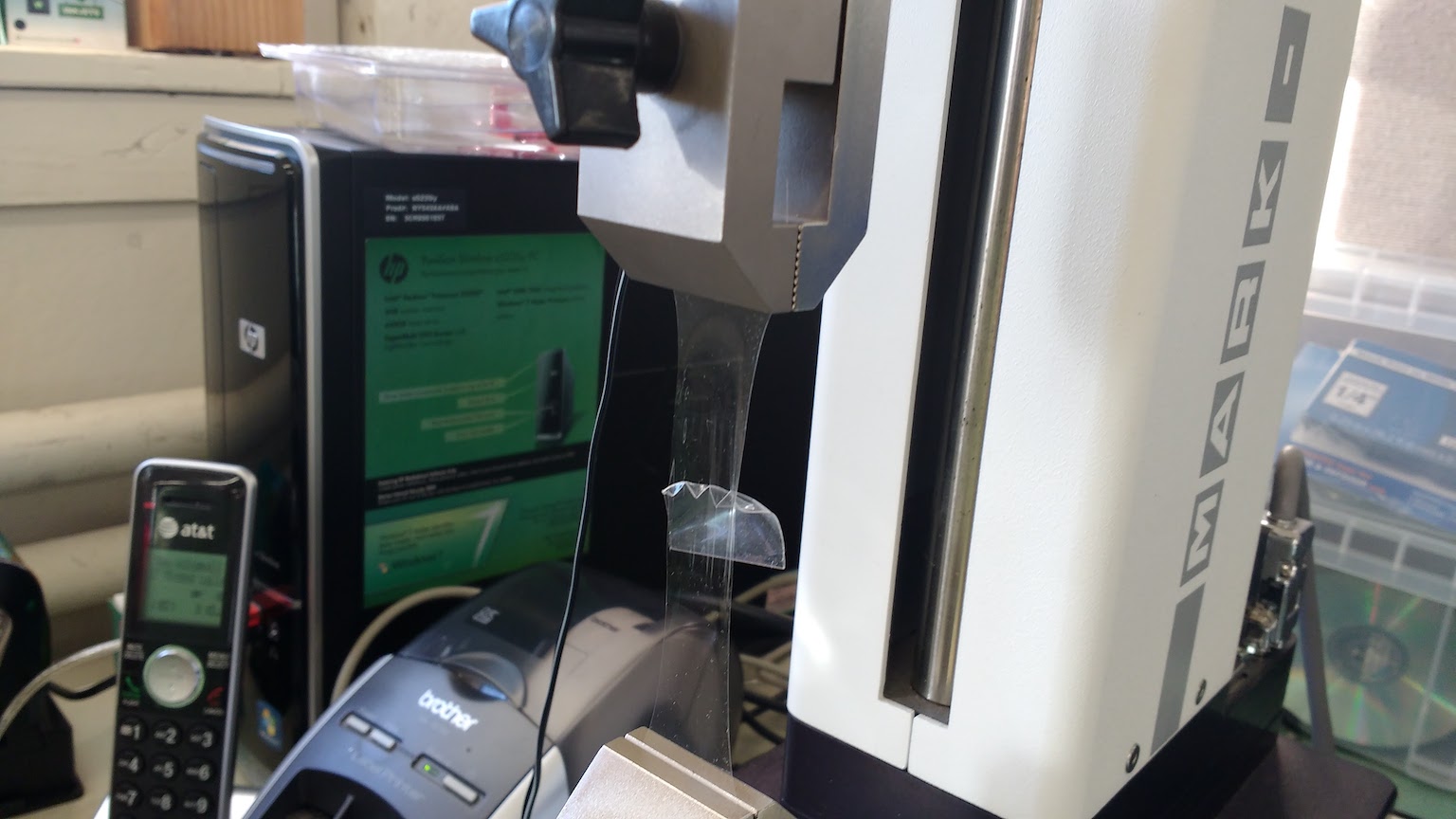}
\caption{(Left) The setup of heat sealing of separator. (Right) The pull test on a heat-sealed sample.}
\label{fig:HeatSealFabrication}
\end{figure}

To ensure the seal reliability, peel test using ASTM F88 was perfomed using a Mark-10 motorized pull testing machine before and after all sealing shifts.
Employing the same setup used for the sealing the separators, two 2.5~cm wide sealed FEP samples, one each using the sealing machines for separators' long sides and short sides were produced.
The samples were then subjected to an increasing amount of tension as shown in Figure~\ref{fig:HeatSealFabrication}.
Samples were deemed to pass the test if the seals did not peel off before the load reached 1.36~kg (3~lbs).
Three pull tests were performed each day, with additional tests before and after all sealing shifts.
All samples successfully passed the seal tests indicating consistent seal quality throughout the separator sealing process.  
    
\subsection{PLA Rod Fabrication}
\begin{figure}[h!]
\centering
\includegraphics[width=60mm]{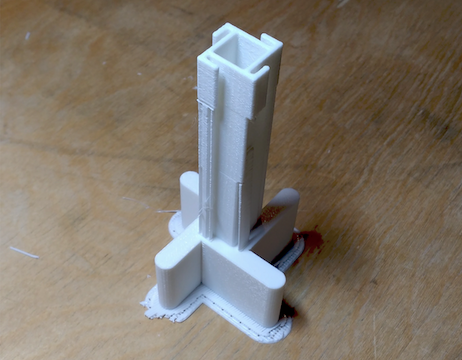}\quad
\includegraphics[width=60mm]{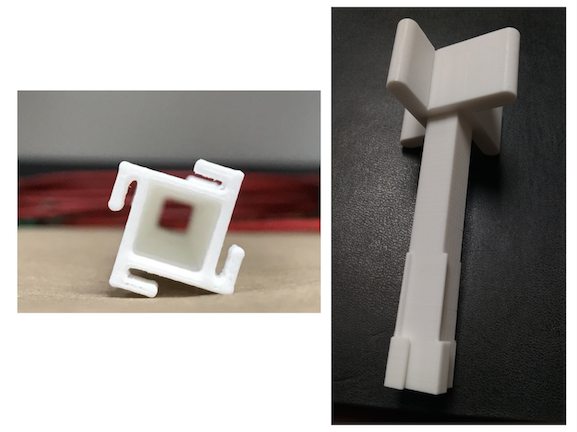}
\caption{(Left) The PLA rods before removal of the support material. (Center, right) The PLA rods after removal of the support material.}
\label{fig:deliveredPinwheel}
\end{figure}

All nine types of suppport rods were 3D printed with PLA as the raw material by a commercial 3D printing services company. 
According to the manufacturer, the PLA rods were printed with the 100~\textmu m PLA filament.
PLA was also used as the support material, which is necessary to support suspended parts of the rods, during 3D printing. 
Shown in Figure \ref{fig:deliveredPinwheel} is an example of a PLA rod before and after the support material was removed. 

\section{Quality Control/Quality Assurance}
\label{sec:QA}
The volume uniformity, high specular reflectance, and chemical stability of the optical segments are essential requirements of the experiment. 
Therefore, extensive dimensional, optical, and compatibility QA measurements were performed during the fabrication, pre-assembly, and detector assembly.  
    
\subsection{Separator QA/QC}
\label{subsec:PanelQA}

The first set of QA procedure was performed in the lamination cleanroom at IIT during the separator fabrication.  
After the DF2000MA and adhesive were laminated on both sides of a separator's carbon fiber backbone, the width of each separator was measured with a 25.4~\textmu m (0.001~in) precision caliper.  
After the FEP lamination on both sides of the separator, a 25.4~\textmu m (0.001~in) precision thickness gauge was used to measure the thickness of the labeled separators.
In addition, any visible imperfections (diameter > 1 mm) on the separator were photographed and recorded. 
The separators were rejected if their measured dimensions did not satisfy the design tolerances or they contained defects that might significantly impact the reflectivity or compatibility of the separators, such as wrinkles, dust inclusions, or punctures.
10\% of the separators were chosen randomly for optical QA via UV-Vis reflectance measurements; the results of optical tests are summarized in Section~\ref{sec:optical}.   
After the QA procedures in the cleanroom was completed, the separators were packed for shipping to Ingeniven for heat sealing.

\begin{figure}[h!]
\centering
\includegraphics[trim = 0.0cm 0.0cm 0.0cm 0.0cm, clip=true, width=0.44\textwidth]{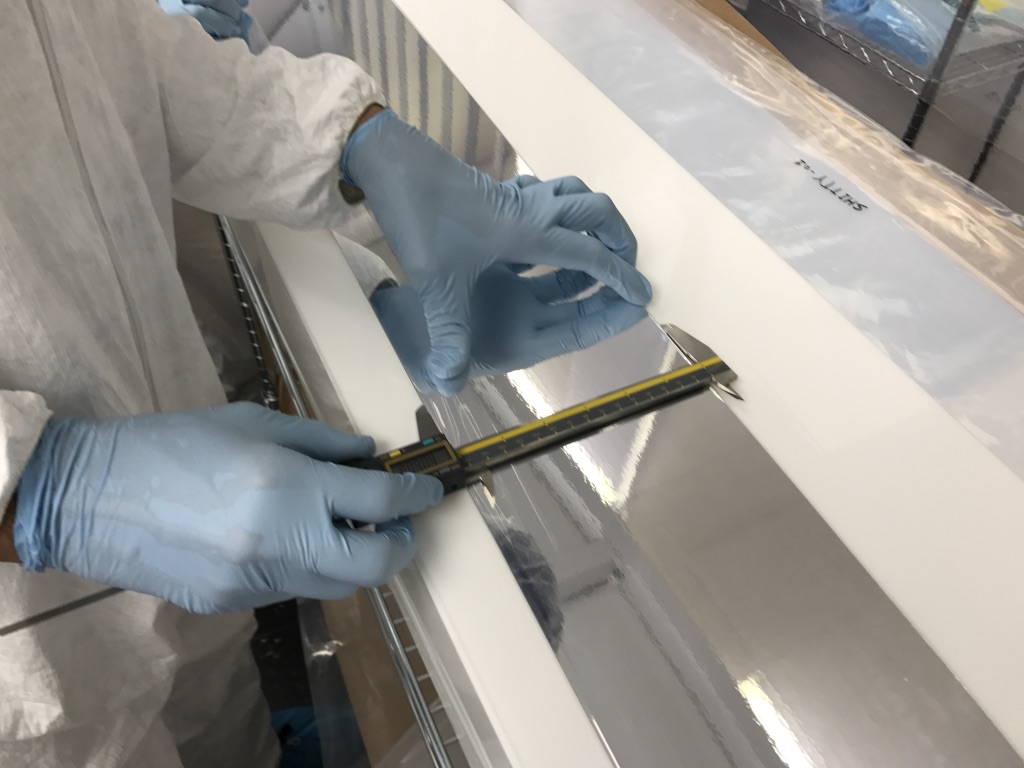}\quad
\includegraphics[trim = 0.0cm 0.0cm 5.0cm 19.0cm, clip=true, width=0.44\textwidth]{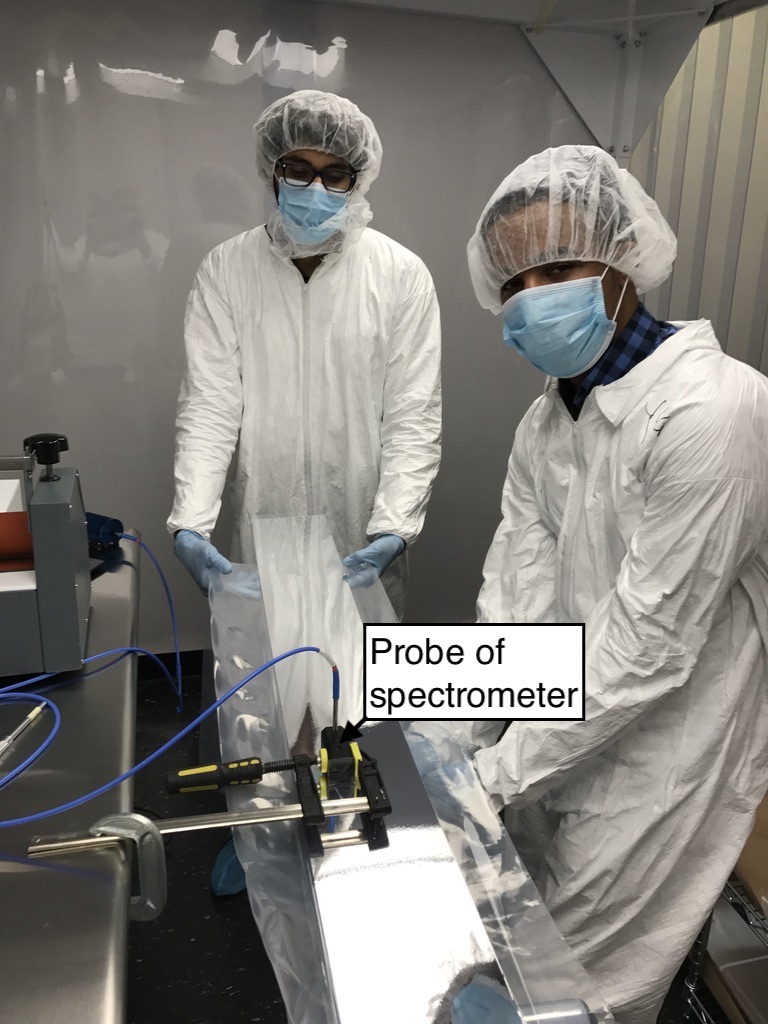}
\caption{Photographs of separator QA in the cleanroom. (Left) Width measurement with caliper before the lamination of FEP. (Right) Optical measurement by visible light reflectance spectrometry setup.}
\label{fig:PanelQA}
\end{figure}

\begin{figure}[h!]
\centering
\includegraphics[width=.55\textwidth]{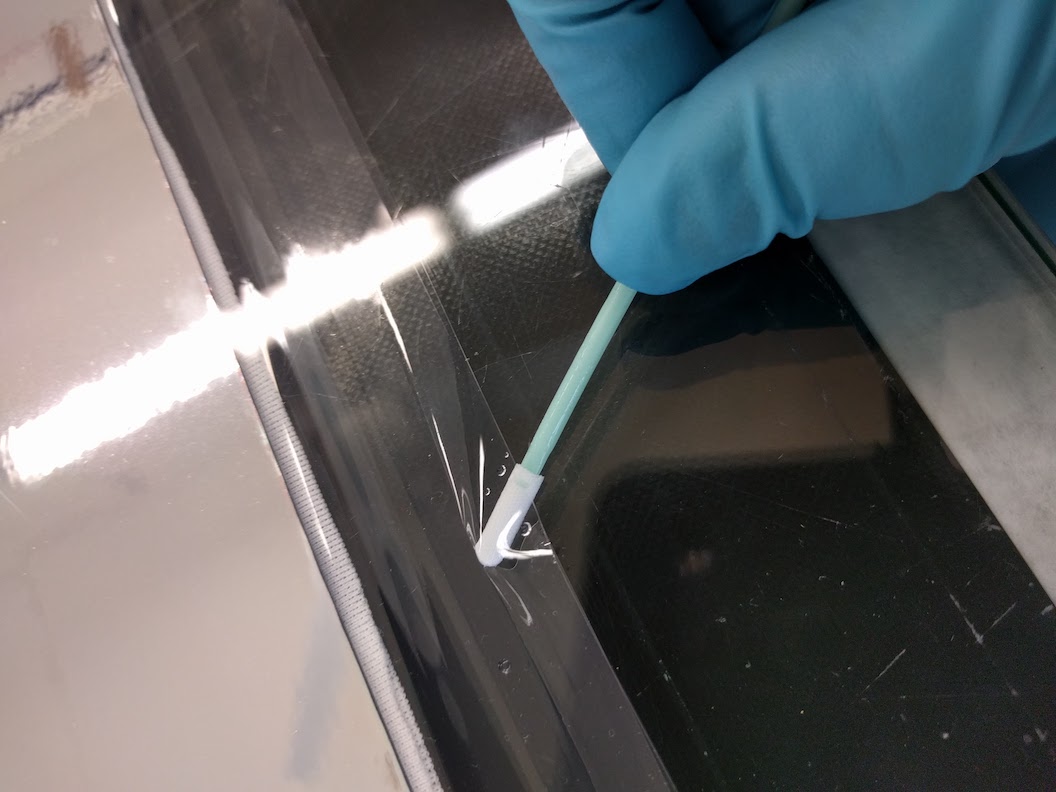}
\caption{Photographs of the heat sealing QA.}
\label{fig:HeatSealingQA}
\end{figure}

The second QA check was performed when the separators were delivered to Ingeniven for possible new flaws that appeared during shipping.
Seals were then chemically tested by running a low-lint polyester swab soaked in ethyl alcohol between the two sheets of FEP along each seal.
An incomplete seal allows alcohol to seep in and react with the adhesive in the separator sample producing a visible whitening of the adhesive.  
Any observation of whitening of the separator sample within a few minutes of application would thus indicate a puncture in the FEP or an incomplete seal and the separator was rejected.  
By gently pushing at the seal with the swab, as shown in the Figure \ref{fig:HeatSealingQA}, this procedure also provided a test of mechanical integrity of the seal. 
    
The quality control (QC) of separators was based on the thresholds set on QA measurements. 
The total number of separators that passed each level of QC is shown in Table~\ref{tab:separatorQA}.

\begin{table}[htpb]
\centering
\begin{tabular}{|c|c|}
\hline 
QC level & Count of separators\\
\hline
Laminated & 372 \\
\hline
Surface quality & 371 \\
\hline
Optical QA & 370 \\
\hline
Dimensional QA & 369 \\
\hline
Heat sealing quality & 367 \\
\hline
Used in detector assembly & 333\\
\hline
\end{tabular}
\caption{The count of laminated separators that passed each level of QC. 
98.6\% (367 out of 372) of laminated separators passed QC, in which 333 separators assembled into the PROSPECT AD.}
\label{tab:separatorQA}
\end{table}

\subsection{PLA Rod QA/QC}

Once 3D printed by Autotiv, PLA rods of same types were bagged together and shipped to IIT.
Each bag was measured through the full QA/QC procedure individually; since the PLA rods are printed with FDM, there is a possibility that the 3D printing may burn the layer below the current printing layer, leaving a visible dark spot.

Given that $^{6}$LiLS chemical compatibility tests were not performed on burnt PLA material, if burns could not be filed off, the PLA rod was rejected. 
There are 10-20\% of PLA rods rejected by the burns, contributing the majority of the rejects. 
PLA rods with sharp irregularities that might damage the FEP layer of the separators were also rejected. 
    
To ensure that PLA rod tabs could properly hold a separator without causing abrasion to the separator's FEP exterior due to slight movements, a filing with metal hand file was performed on the interior of each tab to smooth out the surfaces in direct contact with the separator.  
Special care was taken to remove bulges naturally produced during the printing process at corners and edges.  
Every PLA rod, regardless of the type, thus needed to be filed to reduce the risk of separator surface abrasion. 
Following the filing, a 5~cm $\times$ 5~cm separator sample was swiped through each tab on every PLA rod to check for excess abrasion as a separator-fitting test. 
If a PLA rod punctured the separator sample's FEP layer, it was either rejected or filed again until no punctures were made in the separator-fitting test. 

\begin{figure}[h!]
\centering
\includegraphics[width=0.4\textwidth]{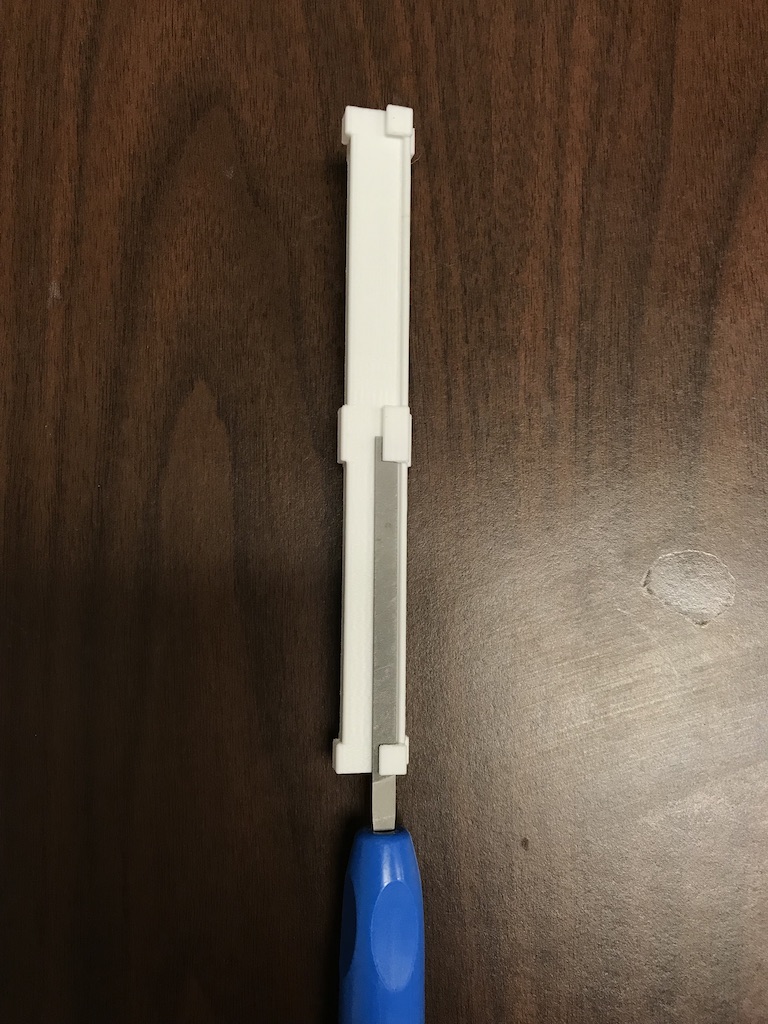}\quad
\includegraphics[width=0.4\textwidth]{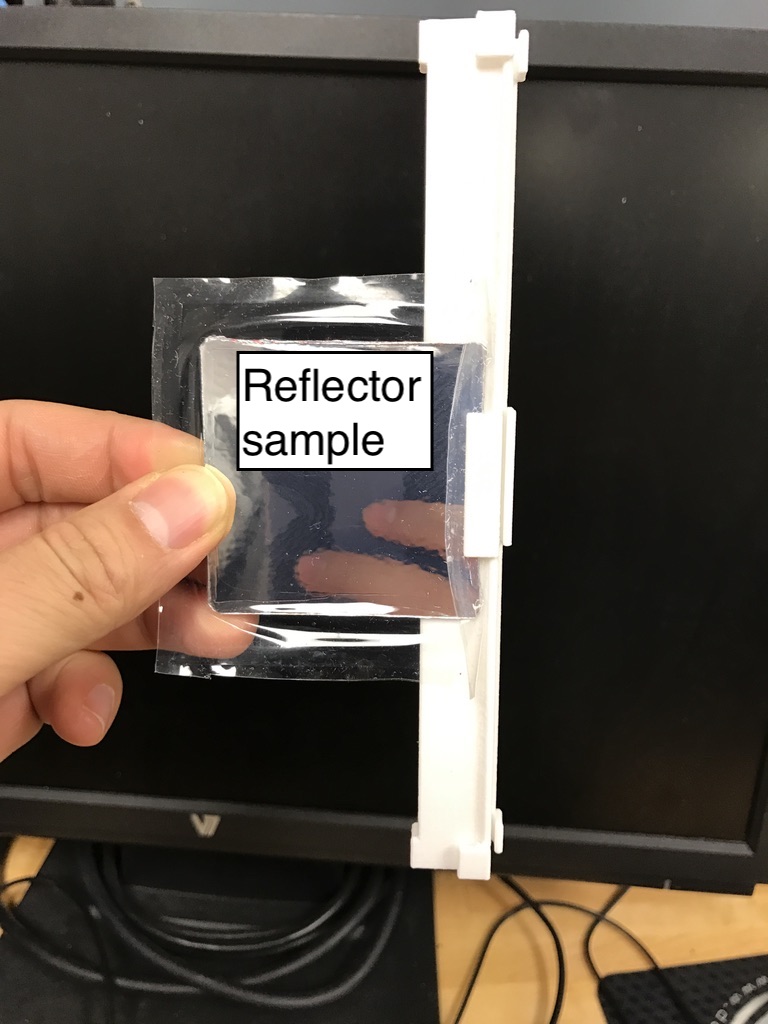}
\caption{Photographs of QA of PLA rods in action. (Left) An example shows the filing of the rods. (Right) An illustration of test fitting with separator sample.}
\label{fig:PinwheelQA}
\end{figure}

After filing, a $\sim$30~mm long sample of 3/8~in outer-diameter PTFE tube was run through the central shaft of each PLA rod to ensure easy integration of PLA rods with PROSPECT calibration system components during the PROSPECT pre-assembly process. 
If remaining PLA filament flashing inside the shaft obstructed the PTFE tube's path, an additional filing of the shaft interior was performed using a steel pipe brush.
If filing could not remove the obstruction, that PLA rod was rejected.  

Following the fitting tests, the PLA rod dimensions were measured.  
The tab dimensions, total lengths and outer cross-sections were measured for 5\% of all standard PLA rods with digital calipers.
As the spacers of the end PLA rods play an important role in the proper seating and alignment of the PMT housings, the thickness of the spacers were measured with digital calipers for all end PLA rods.  
These measurements are further detailed in Section~\ref{sec:Char}.  

\section{Shipping}
    \label{sec:Trans}

The separators and PLA rods were transported to the heat sealing company and detector assembly site, respectively after the QA/QC processes at IIT were completed. 
The packaging scheme was designed to reduce mechanical damage to the PLA rods and separators during shipping.

\subsection{Separator Packing and Shipping}

\begin{figure}[h!]
\centering
\includegraphics[width=0.32\textwidth]{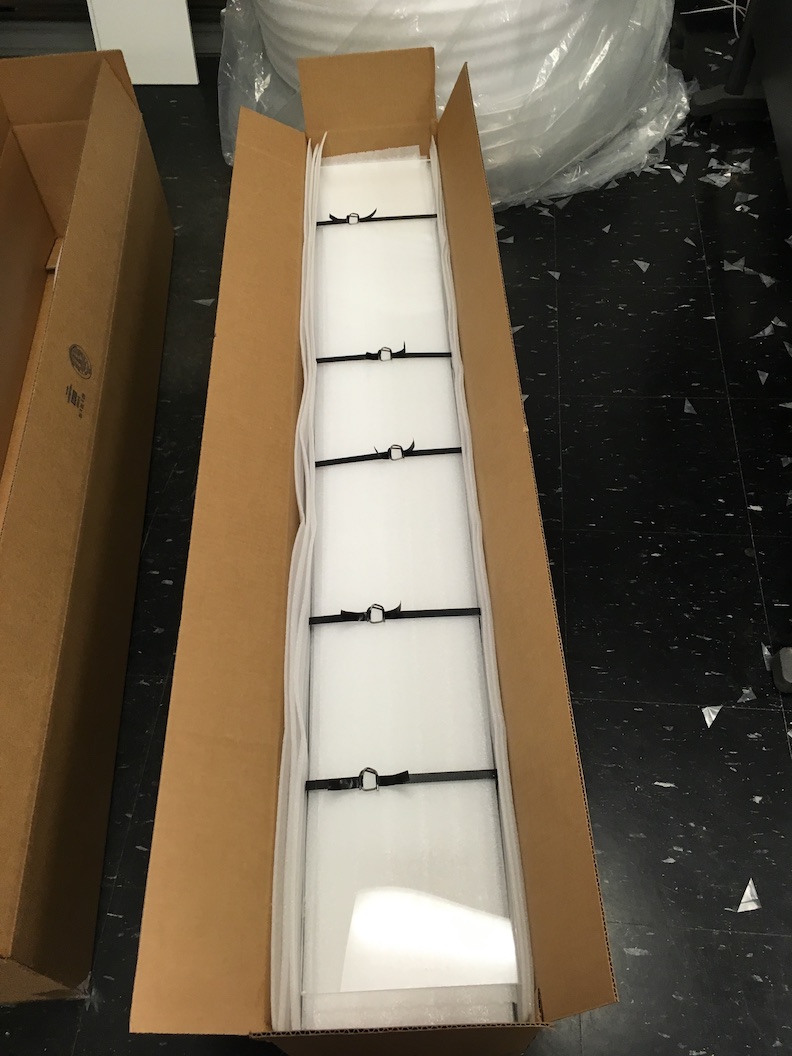}\quad
\includegraphics[width=0.55\textwidth]{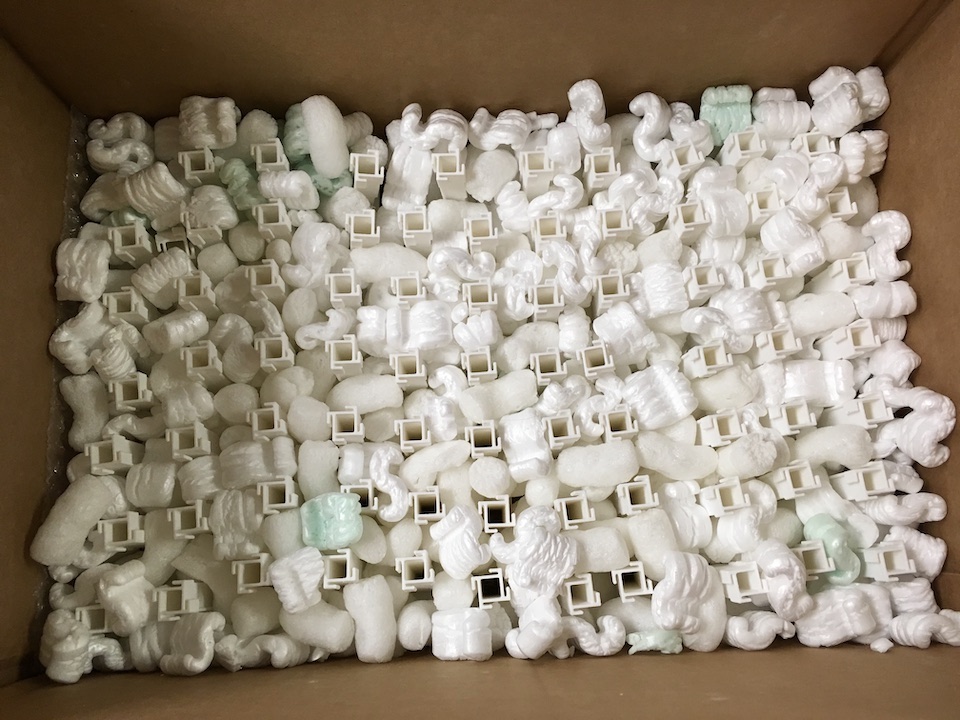}
\caption{(Left) One group of separators packed for shipping. (Right) PLA rods aligned in box for shipping.}
\label{fig:PanelShipping}
\end{figure}

The separators were packed into class-100 cleanroom bags that were heat sealed on both ends to avoid direct exposure of the bag interior to dust during packing and shipping.  
The separators were stacked in batches of 30, with a layer of foam placed between each bagged separator to avoid scratching of FEP surfaces against dust grains present on the exterior of the cleanroom bags.  
With the separators stacked and secured, the corners of the bags containing the separators were then cut to allow the release of excess trapped air.  
The stack of separators was then placed between two acrylic panels and additional foam layers and secured using poly straps and metal clasps.  
The separators were then placed inside a shipping box lined with multiple layers of foam wrap to protect from mechanical damage and also to create a tight fit in the box. 
After placing edge protectors on the box exterior, the separator batch was considered fully prepared for shipping.  
A photograph of a separator batch for transportation is in Figure~\ref{fig:PanelShipping}.  

\subsection{PLA rod Packing and Shipping}

The primary objective in developing a PLA rod packing and shipping method was to minimize any relative motion between rods, thus reducing the risk of breakage.  
Standard PLA rods of common type without spacers were packed in zip-lock bags in groups of 40.  
These bags were then packed in shipping boxes with foam and paper cushioning.  
For end PLA rods, the packing box was lined with multiple layers of foam wrap on bottom and sides.  
The space between the PLA rods was then filled with packing peanuts to minimize torque on the lighter portion of the rod profile, as shown in Figure~\ref{fig:PanelShipping}. 

\section{Assembly}
	\label{sec:Assembly}

\subsection{Separator Cleaning and Pre-Assembly}

Heat sealed separators were cleaned before the beginning of assembly. 
Before every cleaning shift, all working areas were covered by class-100 non-scratching cleanroom wipes. 
Then, groups of 20 separators were taken out of their shipping boxes and cleanroom bags and placed on racks in a class 1000 cleanroom.  
Each separator was then cleaned individually by scrubbing it with a 1\% solution of Alconox, transferring it to a separate clean work area, and rinsing it multiple times with 10 $\mathrm{M\Omega}$-cm deionized water (DI water).  
Once the conductivity of the rinse water coming off from the separator surface was measured to be 0.1~\textmu S/cm, the separator was considered clean.  
The separators were then air dried and placed on a rack. 
After drying, the separators were placed in individual cleanroom bags and shelved until assembly.  

\subsection{Rod Cleaning and Pre-Assembly}
The PLA rods were cleaned in a class-1000 cleanroom with $1\%$ Alconox solution in an ultrasonic cleaner.
Before cleaning, the PLA rods were transferred into large bags grouped according to their type.  
The bags were then filled with the Alconox solution and placed into the ultrasonic cleaner.  
After about 20 minutes cleaning with Alconox at room temperature, the PLA rods were ultrasonic cleaned with 10~$\mathrm{M\Omega}\cdot$cm DI water, and then rinsed with DI water until the conductivity of the rinse water was measured to be 0.1~\textmu S/cm.
Cleaned rods were then air-dried at room temperature. 

\begin{figure}[h!]
\centering
\includegraphics[width=0.7\textwidth]{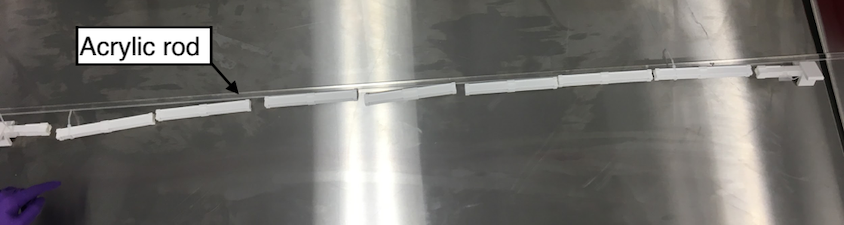}\\

\includegraphics[width=0.7\textwidth]{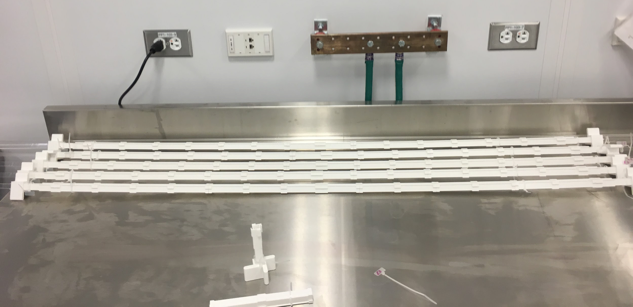}
\caption{(Top) The PLA rods and acrylic rod before stringing. (Bottom) The pre-assembled long PLA rods.}
\label{fig:pinwheelPreclean}
\end{figure}

Once dried, the PLA rods were strung onto clear, longitudinal acrylic rods, as shown in Figure \ref{fig:pinwheelPreclean}. 
These acrylic rods were used to position the PLA rods from their center during the assembly and simplify the detector assembly. 
With the exception of radioactive calibration source axes, these acrylic rods remain in place in the assembled PROSPECT detector.  
For calibration source axes, the acrylic rods were removed from the assembled detector and were replaced with the source-guiding PTFE tubes.
The procedure of the PLA rods pre-assembly differs with respect to setups for different calibrations. 
The radioactive source calibration rods and the standard PLA rods that do not carry calibration purpose were strung from one end of the acrylic rod in order. 
The assembled PLA rods were then held in position with a Telfon pin stopper.
The center PLA rod for optical calibration was machined to keep the PTFE guide tube and optical fiber at its center. 
Thus the PLA rods were strung through two shorter acrylic rods with the sequence of one type-6/-7 rod followed by one type-9 and two type-1 rods.
The two conjunct half rods were then assembled to the center PLA rod.
Finally, the strung PLA rods were packed into class-100 cleanroom bags and ready for detector assembly.

\subsection{Optical Grid Assembly}
The optical grid was assembled as part of the assembly of the PROSPECT inner detector, which consists of the optical grid, PMT housings, and acrylic support plates, as described in Section~\ref{sec:Design}.  
Before the detector assembly, the pre-assembled components (pre-strung PLA rods, separators, PMT housings, and acrylic support plates and connection hardware) necessary for building a single horizontal row of 14 PROSPECT segments were transported into the detector assembly cleanroom.  
The separators were unpacked and staged in a pre-defined order corresponding to their placement in the assembled row of segments so that each separator's position in the detector could be properly recorded.  
A final visual check was performed on the separators prior to staging for any flaws appeared/disappeared between cleaning and staging.   
Following staging, the label of the separator was then removed.

\begin{figure}[h!]
\centering
\includegraphics[width=0.5\textwidth]{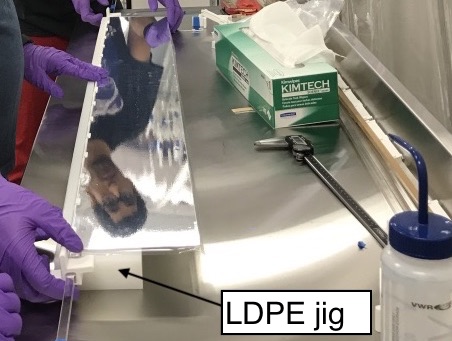}
\caption{An example of the assembly of strung PLA rods and separator using the LDPE jig to fix the PLA rods.}
\label{fig:preassemble}
\end{figure}

Next, combinations of one separator and one strung PLA rod were pre-assembled into 'segment side' geometries.
To do this, a machined Low-Density PolyEthylene (LDPE) jig was attached on a working desk that firmly held one pre-assembled string of PLA rods in place, as shown in Figure~\ref{fig:preassemble}. 
With the string secured in the jig, the FEP film at the edge of one separator was then folded facing the body of the string of rods, and the separator was then inserted into the tabs of the secured string.  

\begin{figure}[h!]
\centering
\includegraphics[width=0.48\textwidth]{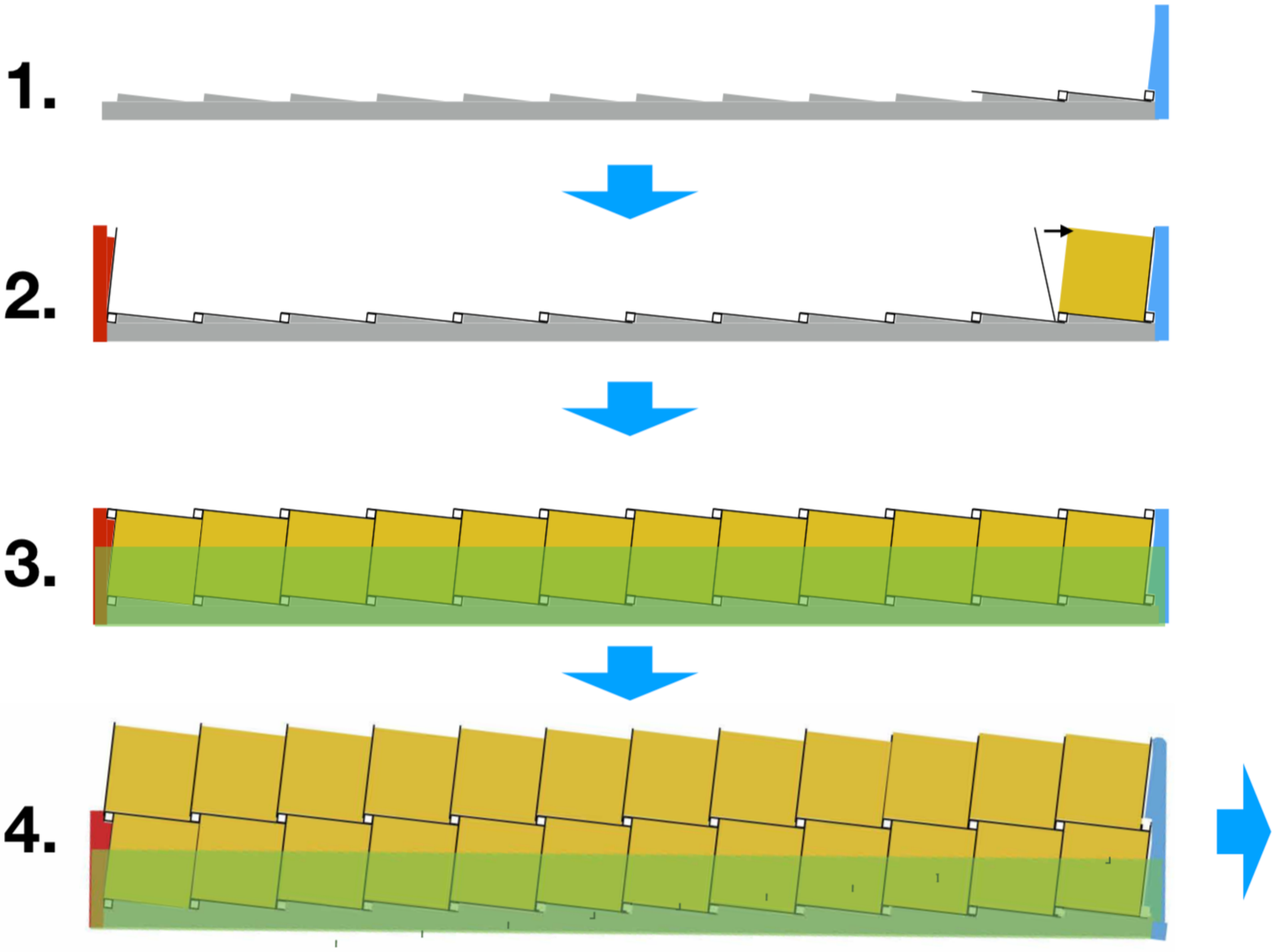}
\includegraphics[width=0.48\textwidth]{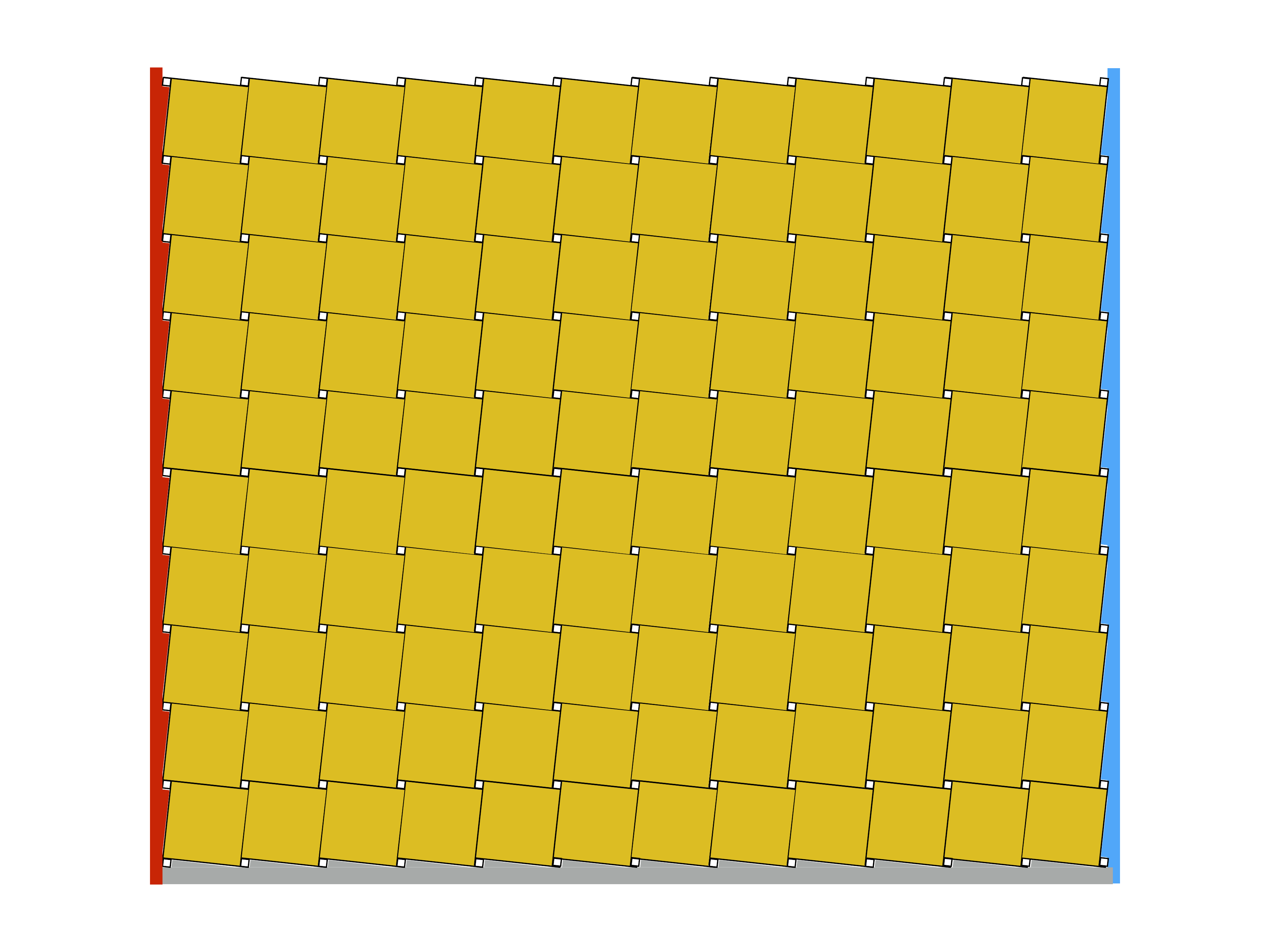}
\caption{The illustration of the optical grid assembly procedure, where the black lines represent separators, the white squares represent PLA rods, the yellow squares represent PMT housings, and other colored parts are acrylic support plates.
(1) The first-layer horizontal separators and PLA rods were being assembled upon the base acrylic support plates.
(2) The first-layer vertical separators assembled together with PMT housings.
(3) The second-layer horizontal separators and PLA rods assembled upon the first layer, closing the bottom row segments, while green shade representing end acrylic support plates that constrained the PMT housing positions.
(4) The following rows of segments were assembled similarly until the detector optical grid was assembled.
}
\label{fig:assemble_procedure}
\end{figure}

\begin{figure}[h!]
\centering
\includegraphics[width=0.45\textwidth]{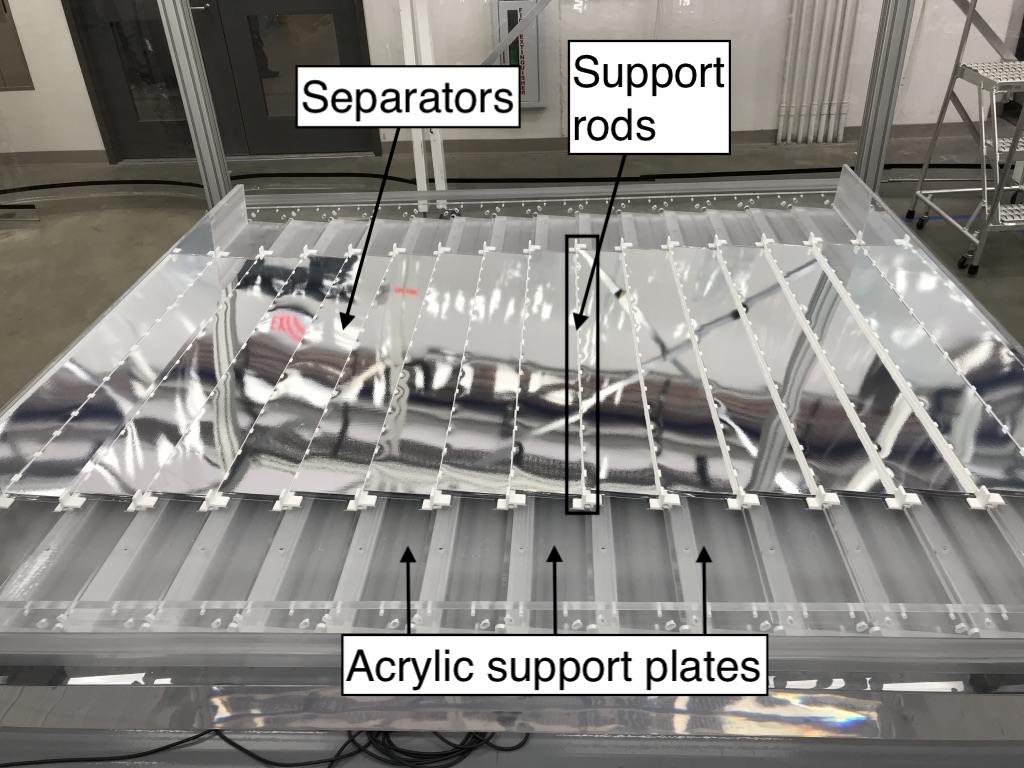}\quad
\includegraphics[width=0.45\textwidth]{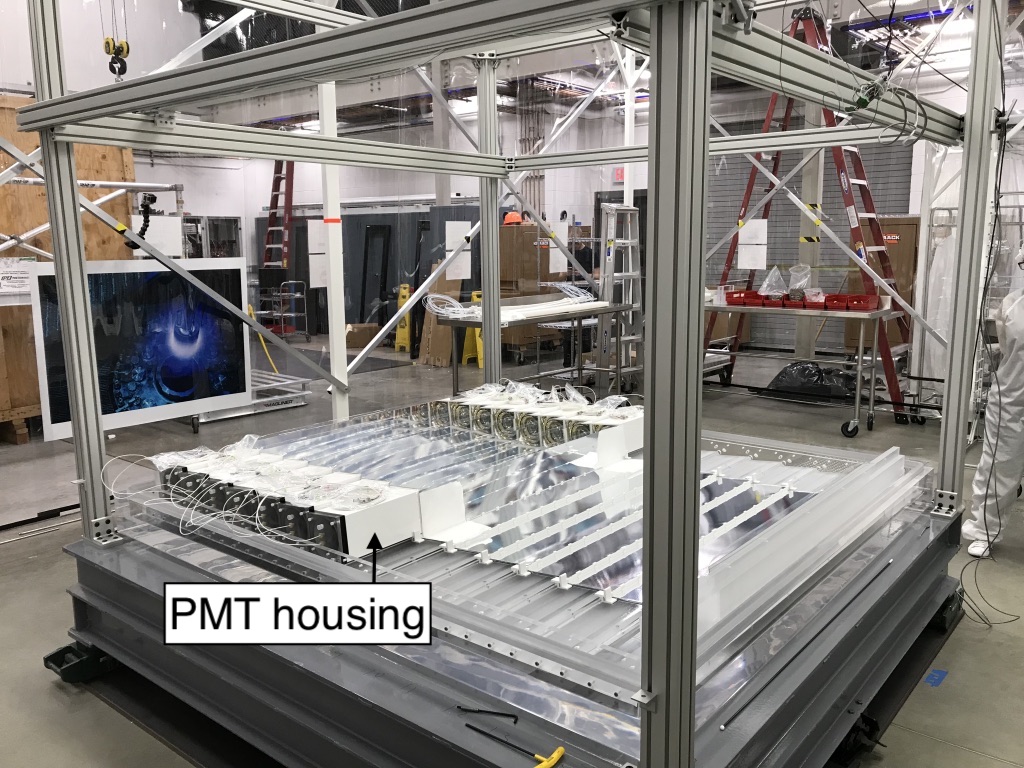}\\
\includegraphics[width=0.45\textwidth]{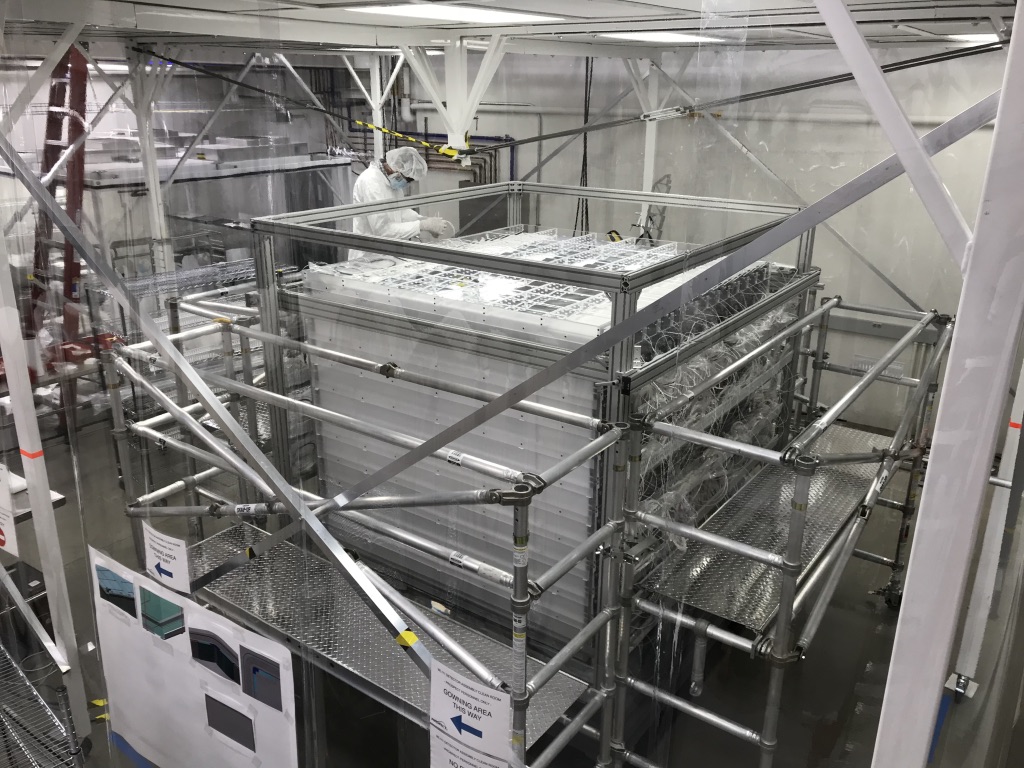}
\caption{
(Top left) The bottom layer of the assembled PLA rods and separators on the acrylic support plates.
(Top right) The assembly of the bottom layer with PMT housings stacking on the acrylic support plates. 
(Bottom) Photograph of the assembled inner detector.}
\label{fig:ADAssemble}
\end{figure}

The assembly of the optical grid includes assembly shifts with groups of separator and PLA rods for the horizontal layers and vertical layers.
In Figure~\ref{fig:assemble_procedure}, the procedure of the optical grid assembly is illustrated.
The assembly of the bottom horizontal layer of separators began with a vertical pre-assembled separator-rod combination at one corner of the AD, with the strung PLA rod string sitting at the bottom resting against a groove in one of the bottom acrylic support plates.  
Another horizontal separator-rod combination was then combined with the first to form the side and bottom of the first (corner) segment.  
Each of two assemblers held one end of the separator-rod combination while folding and inserting the separator into the tabs of the previously assembled separator-rod combination.
The assembly of separator-rod combinations was repeated until the layer was fully assembled, as shown in Figure~\ref{fig:ADAssemble} (top left).
Once all separator-rods of one horizontal layer were assembled and all spacers were fit securely into the grooves below them, the installed layer was visually inspected.  

The vertical separators were installed into the AD together with PMT housings based on the assembled horizontal layer.
With 0.13~mm (0.005~in) thick FEP sheets protecting the assembled separator below, two PMT housings were inserted securely between the installed spacers on both ends one segment.
A separator was then vertically inserted into the tabs of the assembled PLA rods, as shown in Figure~\ref{fig:assemble_procedure} (step 2). 
This procedure was repeated horizontally from one side of the AD to the other.
Figure~\ref{fig:ADAssemble} (top right) shows an on-going assembly of the first vertical layer.
Once the assembly of a vertical layer was finished, a visual inspection was also conducted. 
A horizontal layer of separator-rod combinations was then assembled on top of the vertical separators and PMT housings to fully close one row of segments, the procedure of which was similar to the horizontal layer assembly described in the paragraph above.

Following the assembly of each segment row, acrylic support plates were installed to fully secure the entire newly assembled row. 
A full metrological survey of the installed row was then performed to measure segment height difference in-situ.  
To maintain dimensional uniformity in the presence of tolerance stack-up effects and occasional out-of-tolerance components, additional $\sim$0.25~mm FEP shims were installed between PLA rod end spacers and PMT housing bodies.  
Shim placements and thicknesses were dictated by the results of the metrological survey of the previous row.  
Subsequent rows were then installed in a manner consistent with that described above.  
Assembly of the entire 14$\times$11 optical grid was completed on November 17, 2017.
Photograph of the fully-assembled PROSPECT inner detector can be seen in Figure~\ref{fig:ADAssemble} (bottom).  

The components of the PROSPECT optical grid were designed in 2015 and 2016, and fabricated and assembled in 2017. 
The fully assembled inner PROSPECT AD was shipped to ORNL in January 2018.
The timelines and primary locations in the life cycles of each component are outlined in Table~\ref{tab:timeline}.  

\begin{table}[htpb]
\centering
\begin{tabular}{|c|c|c|c|c|}
\hline 
\multirow{2}{*}{Milestone} & \multicolumn{2}{|c|}{Separators} & \multicolumn{2}{|c|}{PLA rods} \\ \cline{2-5}
& Date & Location & Date & Location \\ \hline  \hline
Begin Fabrication & 11/22/2016 & IIT & 10/20/2016 & Autotiv \\ \hline
End Fabrication/QA & 10/24/2017 & Ingeniven  & 07/14/2017 & IIT \\ \hline
Delivered to Yale & 10/26/2017 & Yale & 07/26/2017  & Yale\\ \hline
Assembled at Yale & 11/17/2017 & Yale & 11/17/2017& Yale\\ \hline
Shipped to Oak Ridge & 1/31/2018 & ORNL & 1/31/2018 & ORNL\\ \hline
\end{tabular}
\caption{The fabrication and assembly timelines, as well as the primary locations of the optical grid separators and support.}
\label{tab:timeline}
\end{table}

\section{Characterization}
    \label{sec:Char}
The measurements of the mass, dimension and optical properties of the detector components, as well as the tests of stability and compatibility of materials were made.
These characterizations were made during the R\&D phase to select compatible materials, and after the fabrication for QA/QC to quantify the quality and accuracy of our fabrication methods for potential future reference.
The key parameters of the detector components were also recorded for detector simulation and physics measurement.
    
\subsection{Mass Measurements}
Because of the limited space and the limitations in the dynamic range of available cleanroom scales, instead of weighing separators directly, a 15-separator batch, including bags and labels, were weighed prior to assembly.  
Once the separator preparation in the cleanroom was finished, the empty bags and disconnected labels for that batch were then measured, enabling determination of the separator batch by itself.  
The precision of the scale used for this measurement was 0.2~kg.  
Measured average separator masses are given in Table~\ref{tab:mass}; the average separator mass among all batches is 326~g, with a standard deviation of 10~g in the average separator mass per batches.  
The total mass of all separators in the PROSPECT target is 108.7~kg $\pm$ 1.7~kg.  

After PLA rod QA procedures were completed at IIT, they were weighed by a 0.5~g precision scale in groups of 20 to 50 for the standard and end PLA types, respectively.
Average calculated rod masses for the different rod types are shown in Table~\ref{tab:mass}. 
The average masses of the various type of PLA rods range from 12.3 to 29.7 g.  
Based on these average masses, and the designed amount of PLA rods from Table~\ref{tab:roddesign}, the total mass of all rods in the PROSPECT target is estimated to be 26.0~kg $\pm$ 0.2~kg.  

\begin{table}[htpb]
\centering
\begin{tabular}{|c|c|c|c|}
\hline
Category & Average mass(g) &  & Total mass(kg)\\
\hline
Separator & $326\pm10$ & 333 & $108.7\pm1.7$\\
\hline
Standard PLA rod (type-1) & $12.3\pm0.1$ & 720 & $8.86\pm0.07$\\
\hline
Center PLA rod (type-2) & $12.8\pm0.2$ & 180 & $2.30\pm0.04$\\
\hline
Standard PLA rod (type-9) & $12.5\pm0.03$ & 360 & $4.50\pm0.01$\\
\hline
Four arms end PLA rod (type-6\&7)& $29.7\pm0.2$ & 260 & $7.72\pm0.05$ \\
\hline
Three arms end PLA rod (type-3\&8)& $26.8\pm0.1$ & 92 & $2.47\pm0.01$\\
\hline
Two arms end PLA rod (type-4\&5) & $20.8\pm0.1$ & 8 & 0.2\\
\hline
\end{tabular}
\caption{The results of mass measurements on the separators and the PLA rods. 
All separators and PLA rods were weighed in batches and the quoted uncertainties reflect variation in the average component mass per batch. $1\sigma$ uncertainties are shown.}
\label{tab:mass}
\end{table}

In addition, the total mass of FEP tubes is approximately 3.7~kg, and the total mass of the acrylic rods is approximately 19.6~kg. 
The total mass of the optical grid in the detector active volume is 134.8~kg $\pm$ 1.9~kg (158.1~kg with PTFE tube and acrylic rod), contributing $\approx3\%$ (3.5\% with PTFE tube and acrylic rod) of dead mass to the PROSPECT active target region.

\subsection{Dimensional Measurements}
\begin{figure}[h!]
\centering
\includegraphics[width=0.48\textwidth]{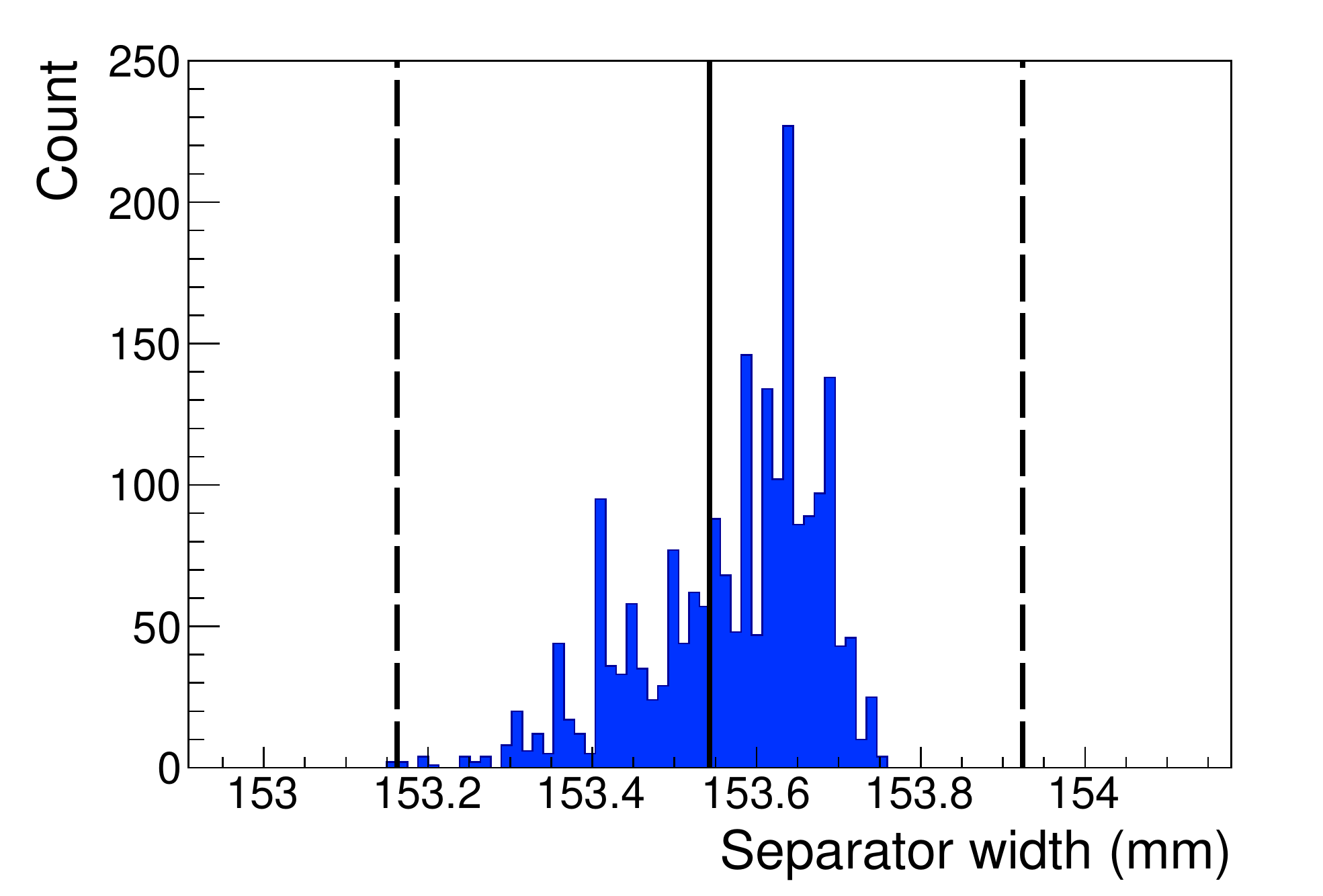}\quad
\includegraphics[width=0.48\textwidth]{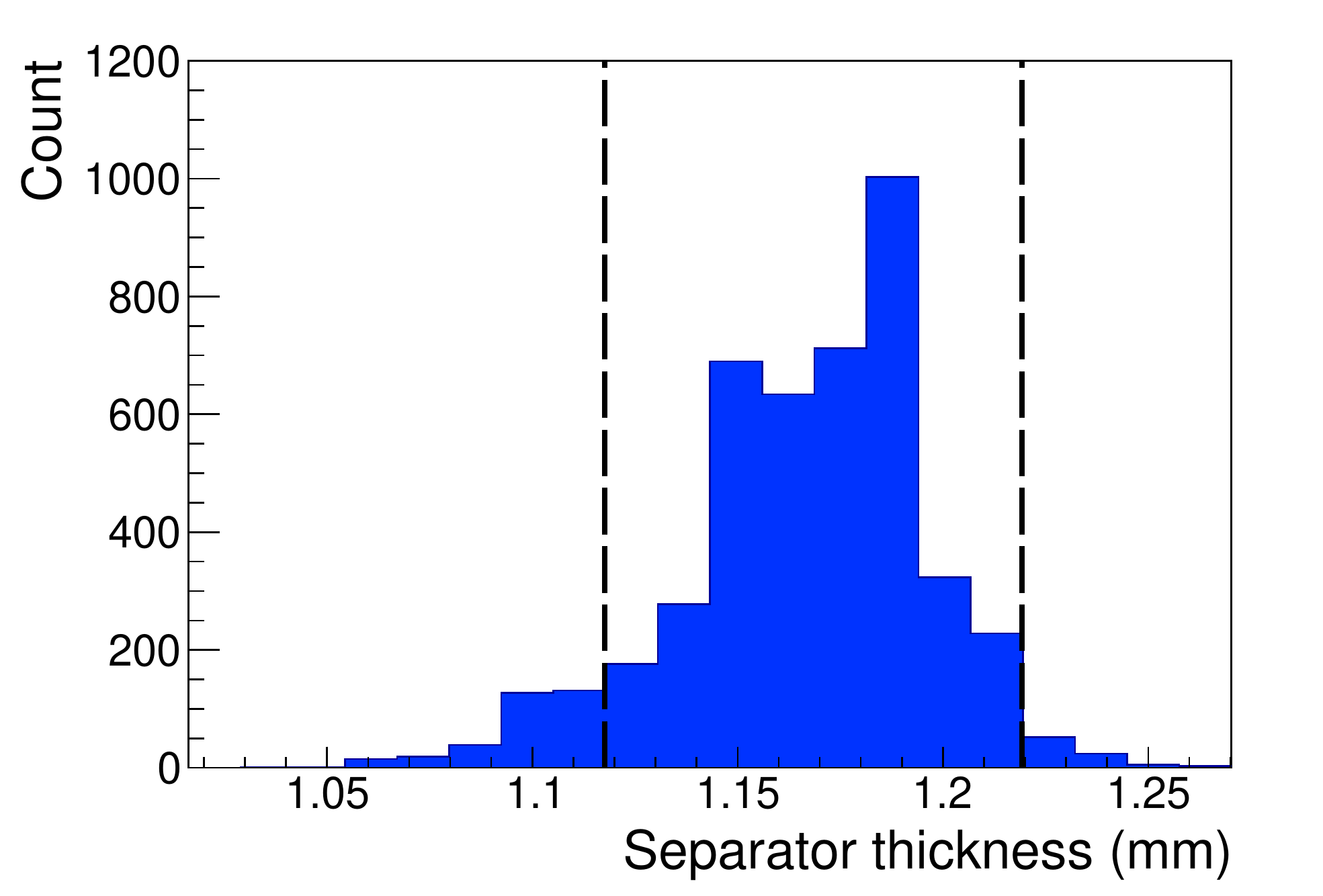}
\caption{The dimensional measurements made on 351 PROSPECT separators, where the solid line represents the nominal dimension and the dashed line represents tolerances. 
(Left) The distribution of widths measured on all separators.  
Separators outside the tolerance were rejected.
(Right) The thickness measurements from all separators. 
Separators outside the tolerance were rejected.}
\label{fig:panelMeasure}
\end{figure}

The width and thickness of the separators were measured. 
To prevent damaging the FEP film, the width was measured with a 25.4~\textmu m (0.001~in) precision caliper prior to the lamination of the FEP coating. 
After the FEP lamination, a 25.4~\textmu m (0.001~in) precision thickness gauge was used to measure the thickness at twelve different positions on each separator's surface.  
Thickness and width measurements are given in Figure~\ref{fig:panelMeasure}.  
The average separator thickness was measured to be 1.18~mm $\pm$ 0.05~mm (0.0463~in $\pm$ 0.0021~in), including an uncorrelated variation of 0.01~mm (0.0006~in) among all measurements. 
The main contributor of the 0.0021~in uncertainty is a 0.002~in systematic variation caused by differing pressure applied by the human testers.  
The average separator width was measured to be 15.36~cm $\pm$ 0.06~cm (6.048~in $\pm$ 0.024~in).  
The largest uncertainty associated with this value is the statistical variation.  
As noted in Section~\ref{sec:Design}, the length of the separators did not have any bearing on mechanical integration of the optical grid with other detector components; for this reason, precision measurements of this dimension were not documented.  

\begin{figure}[h!]
\centering
\includegraphics[width=.48\textwidth]{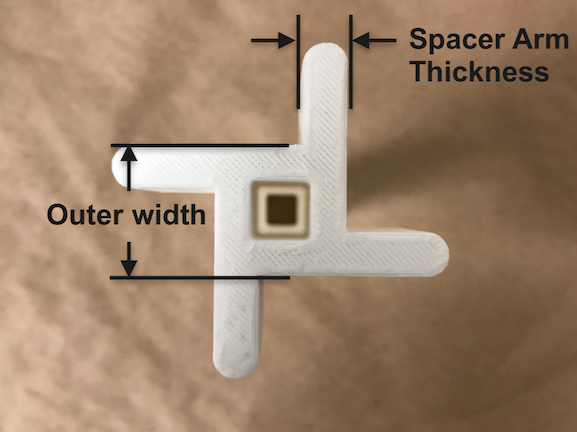}\quad
\includegraphics[width=.48\textwidth]{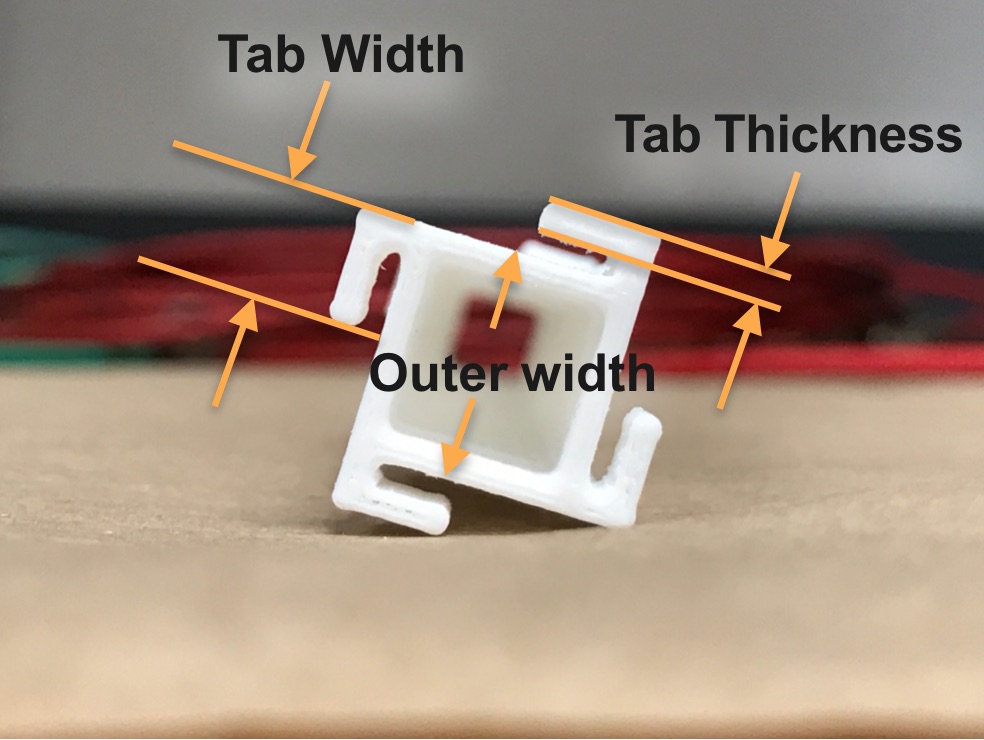}
\caption{(Left) Schematic of the measurements made on end PLA rods. (Right) Schematic of the measurements made on standard PLA rods. }
\label{fig:pinwheelscheme}
\end{figure}

After the filing and cleaning of the PLA rods at IIT, measurements of key rod dimensions were made.
The schematics of the dimensional measurements on the PLA rod are shown in Figure~\ref{fig:pinwheelscheme}.
The thicknesses of all end PLA rod arms were measured, because they are important in determining the orientation of PMT housings with respect to one another.
All measurements were performed with a 25.4 \textmu m (0.001~in) precision caliper.  
The recorded thickness for each arm was the result of averaging measurements at three different locations along the arm profile.  
Both outer cross-section dimensions of all end PLA rod shafts were also measured.  

\begin{figure}[h!]
\centering
\includegraphics[width=.45\textwidth]{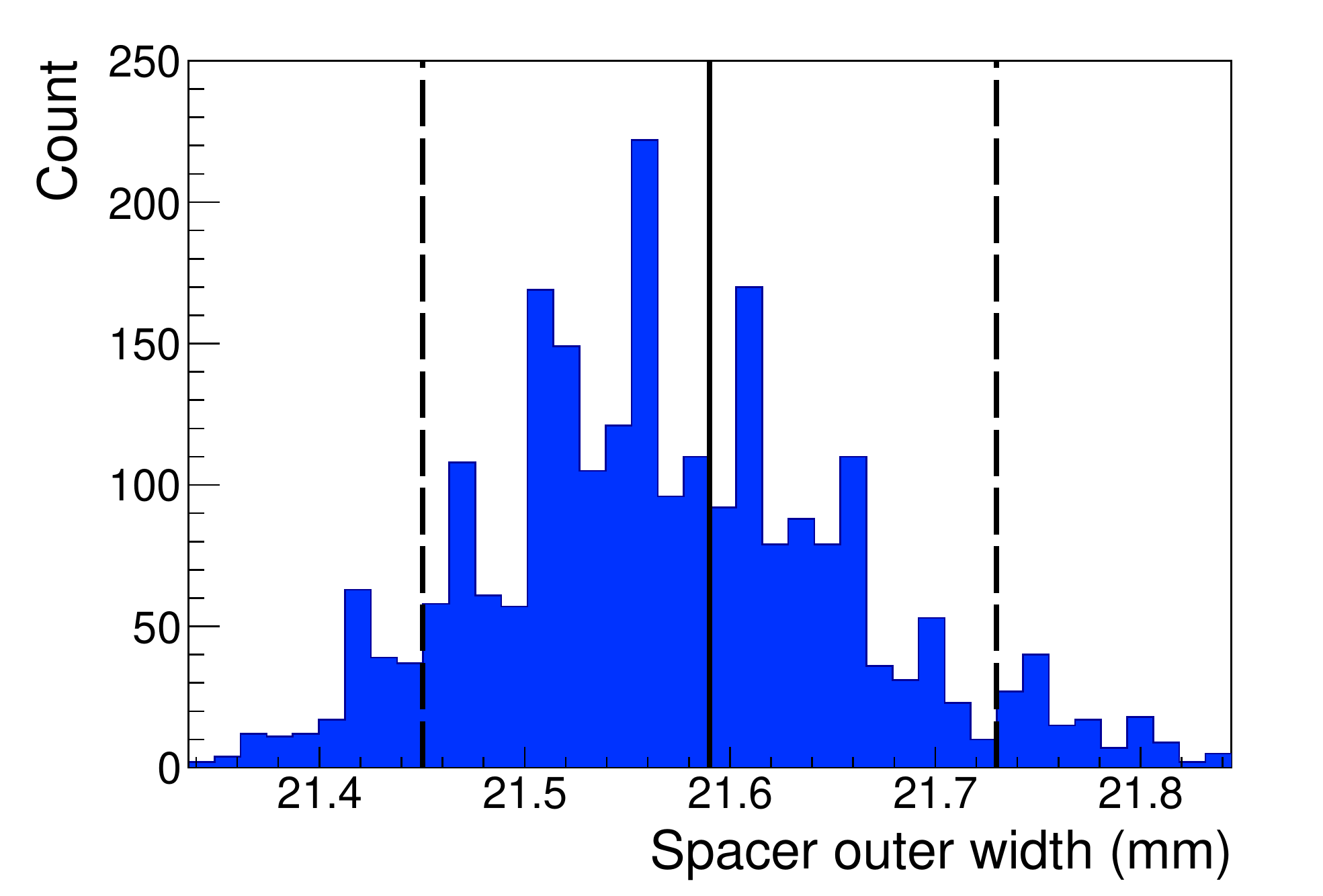}
\includegraphics[width=.45\textwidth]{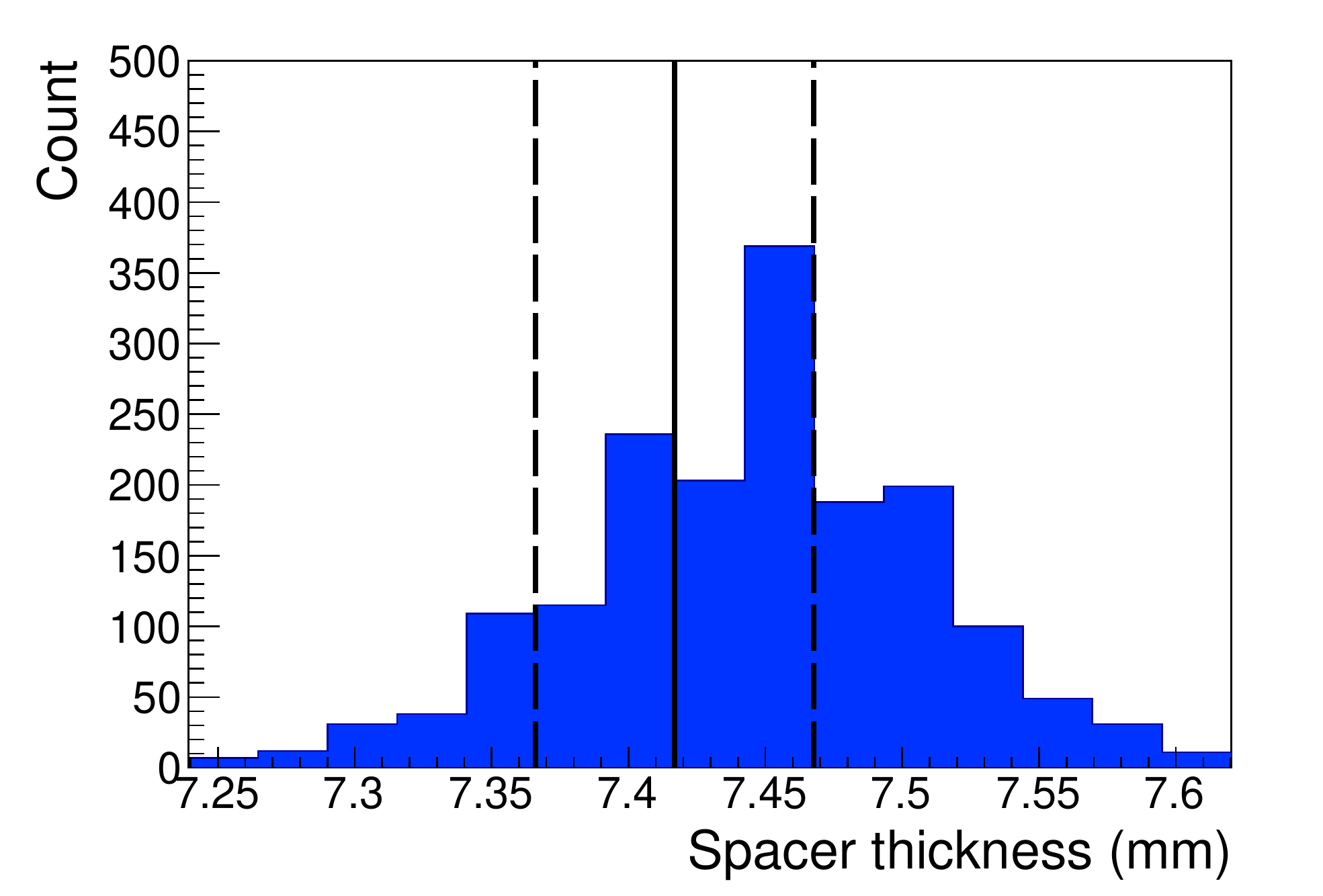} \\
\includegraphics[width=.45\textwidth]{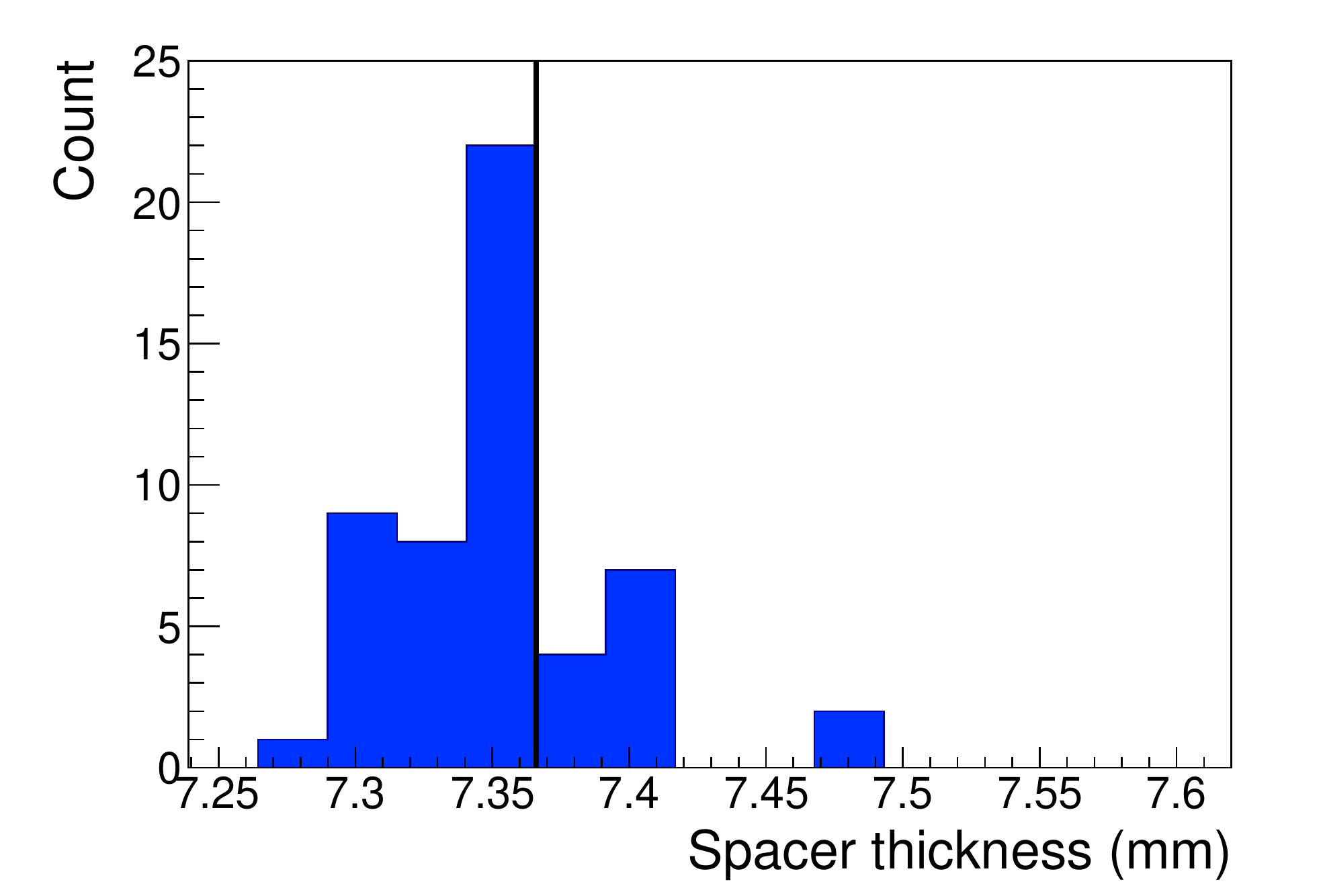}
\caption{The dimensional measurements on spacers of end PLA rods, where the solid line represents the nominal dimension. The dashed lines represent tolerance.
Spacer dimensions outside the tolerance were rejected
(Left) The outer width of the spacers. (Right) The thickness of spacers when delivered. (Bottom) The thickness of spacers after CNC machining. Smaller amount of samples were measured.}
\label{fig:SpacerDimensionSchem}
\end{figure}

Arm thickness and shaft cross-sectional dimensions are pictured in Figure~\ref{fig:SpacerDimensionSchem}, along with the tolerances given for each dimension. 
While most measured spacer outer widths were within their tolerances, a large number of spacer arm thicknesses were not.  
To correct for this, as well as slightly larger-than-expected PMT housing cross-sections, spacer arms were milled with a CNC machine to 7.37~mm (0.290~in) thickness prior to assembly.  
Spacer arm thickness before and after milling operations can also be seen in Figure~\ref{fig:SpacerDimensionSchem}. 
To quantify the centrality of each spacer shaft on each PLA rod, the length from each spacer arm end to the shaft was also measured; no substantial deviations of shaft centering were observed.  
Taken together, these measurements indicate that the dimensional repeatability of filament-based 3D printing can reasonably be expected at the $\pm$0.13~mm ($\pm$0.005~in) level quoted by the manufacturer.  

\begin{figure}[h!]
\centering
\includegraphics[width=.48\textwidth]{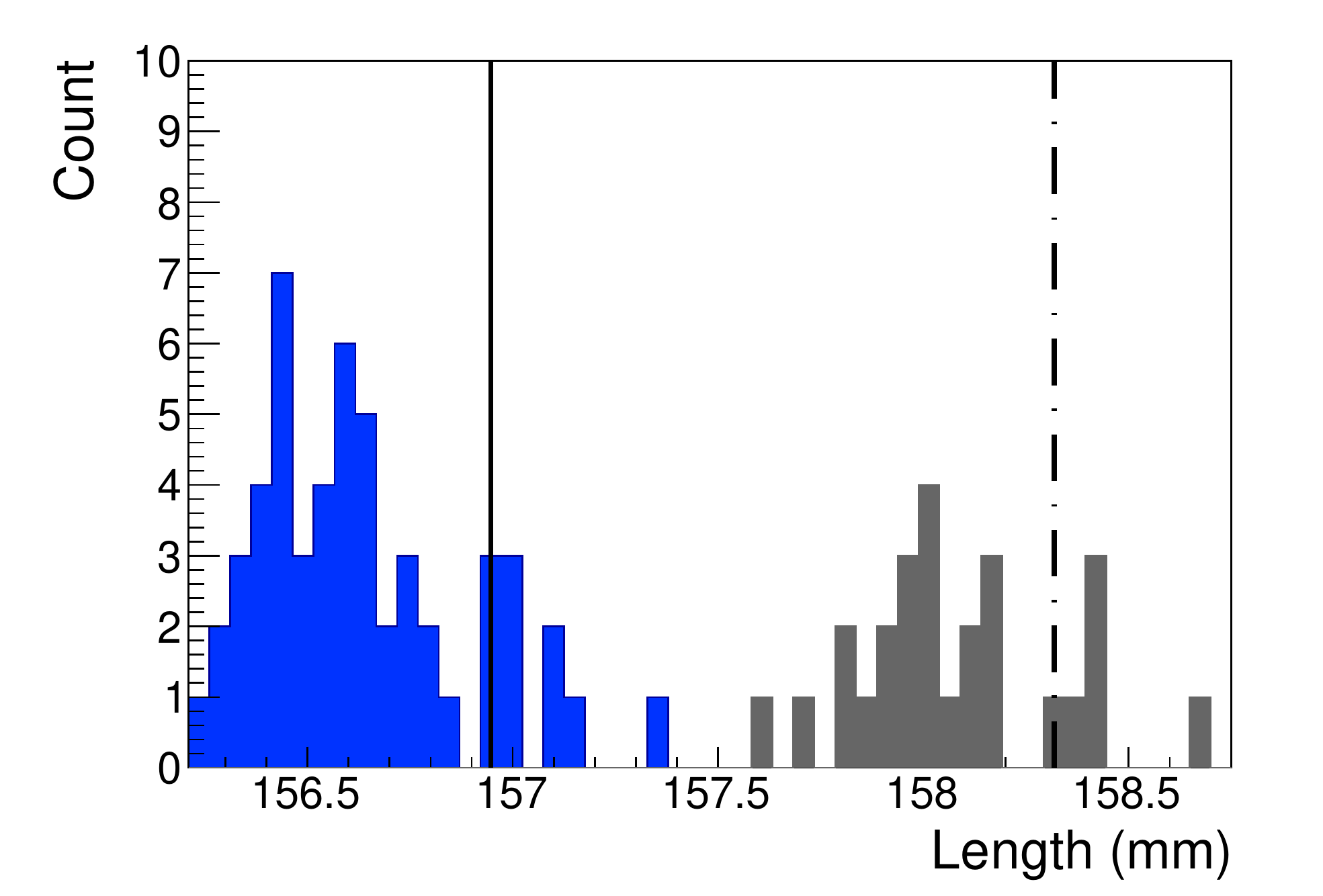}\quad
\includegraphics[width=.48\textwidth]{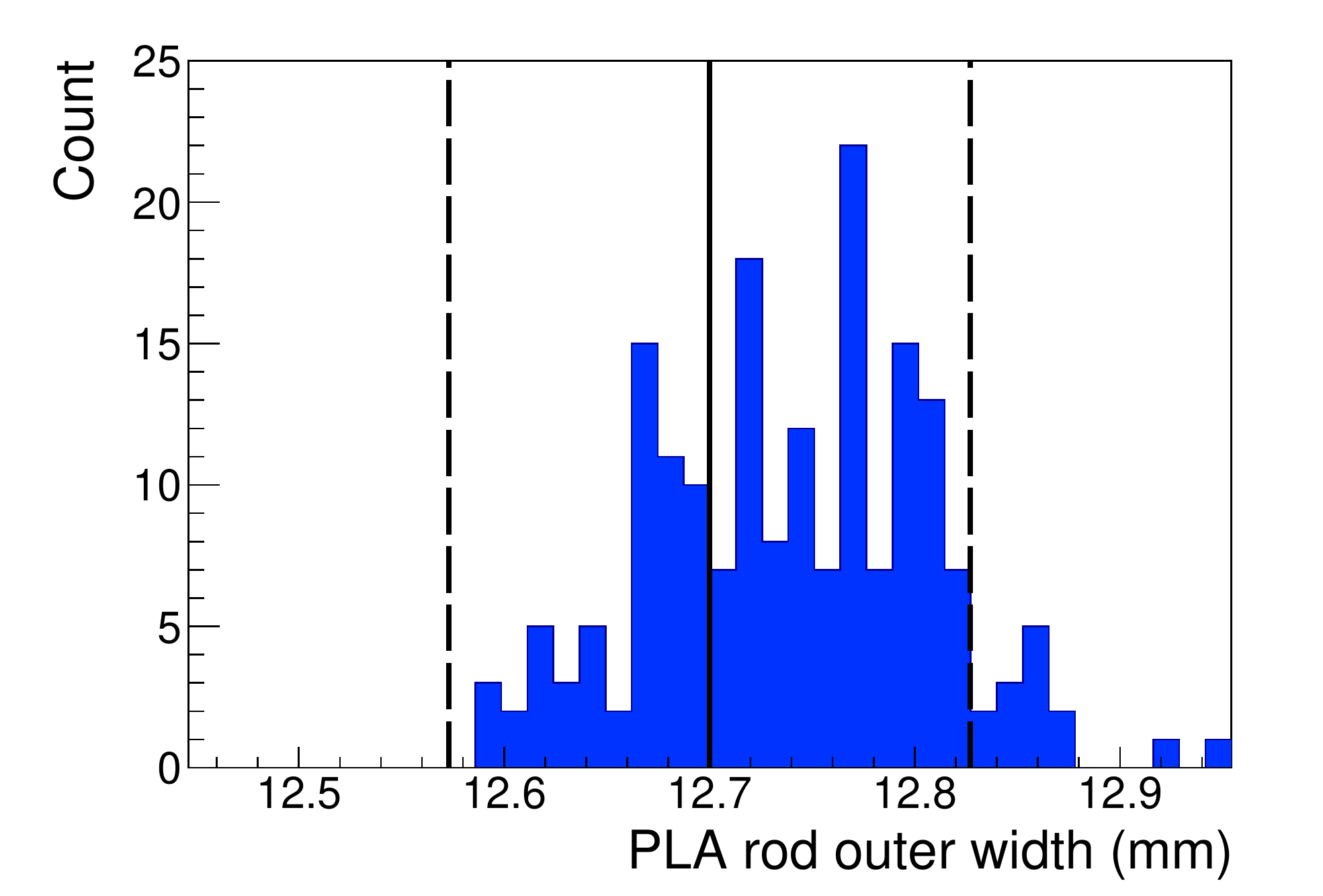}\\
\includegraphics[width=.48\textwidth]{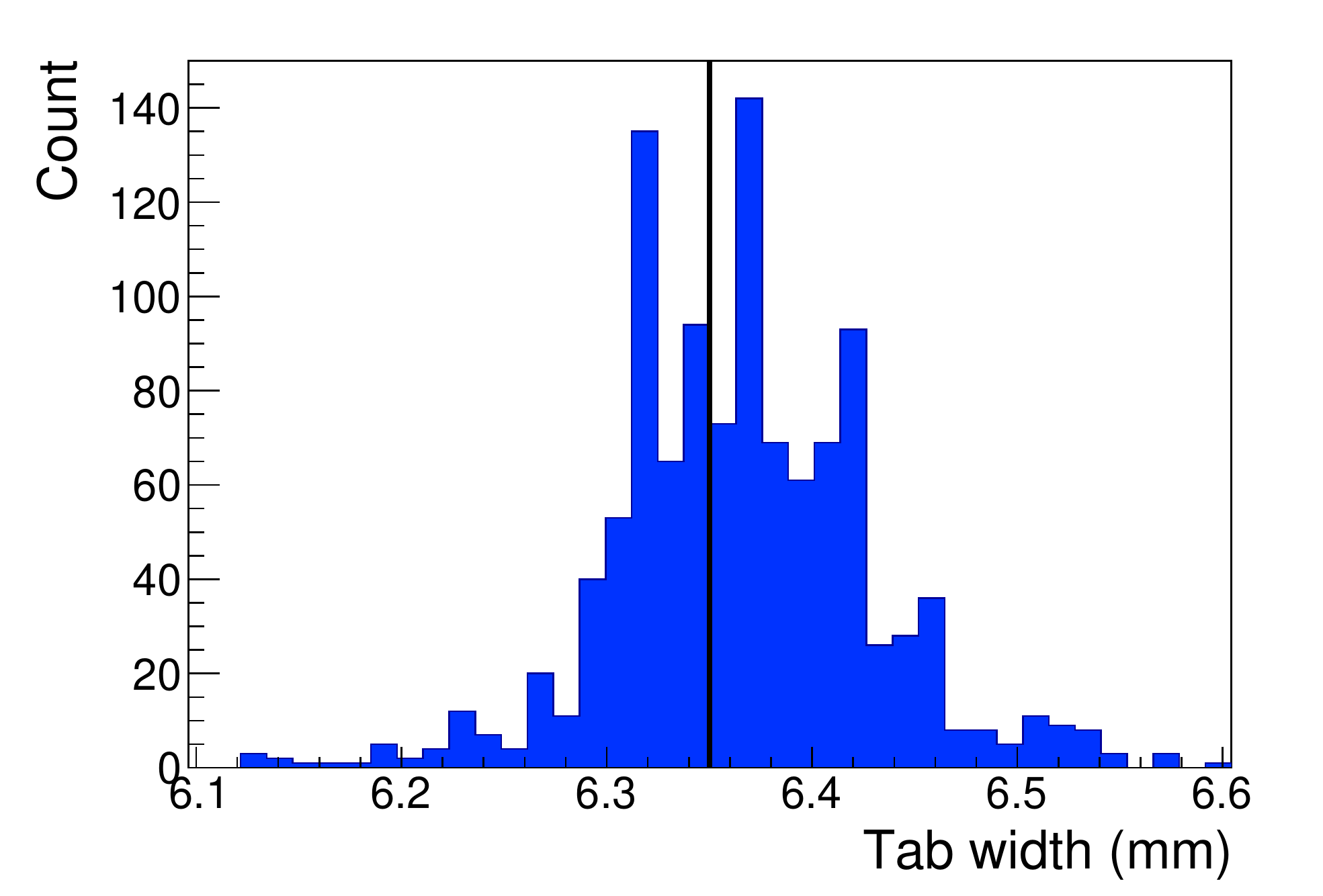}\quad
\includegraphics[width=.48\textwidth]{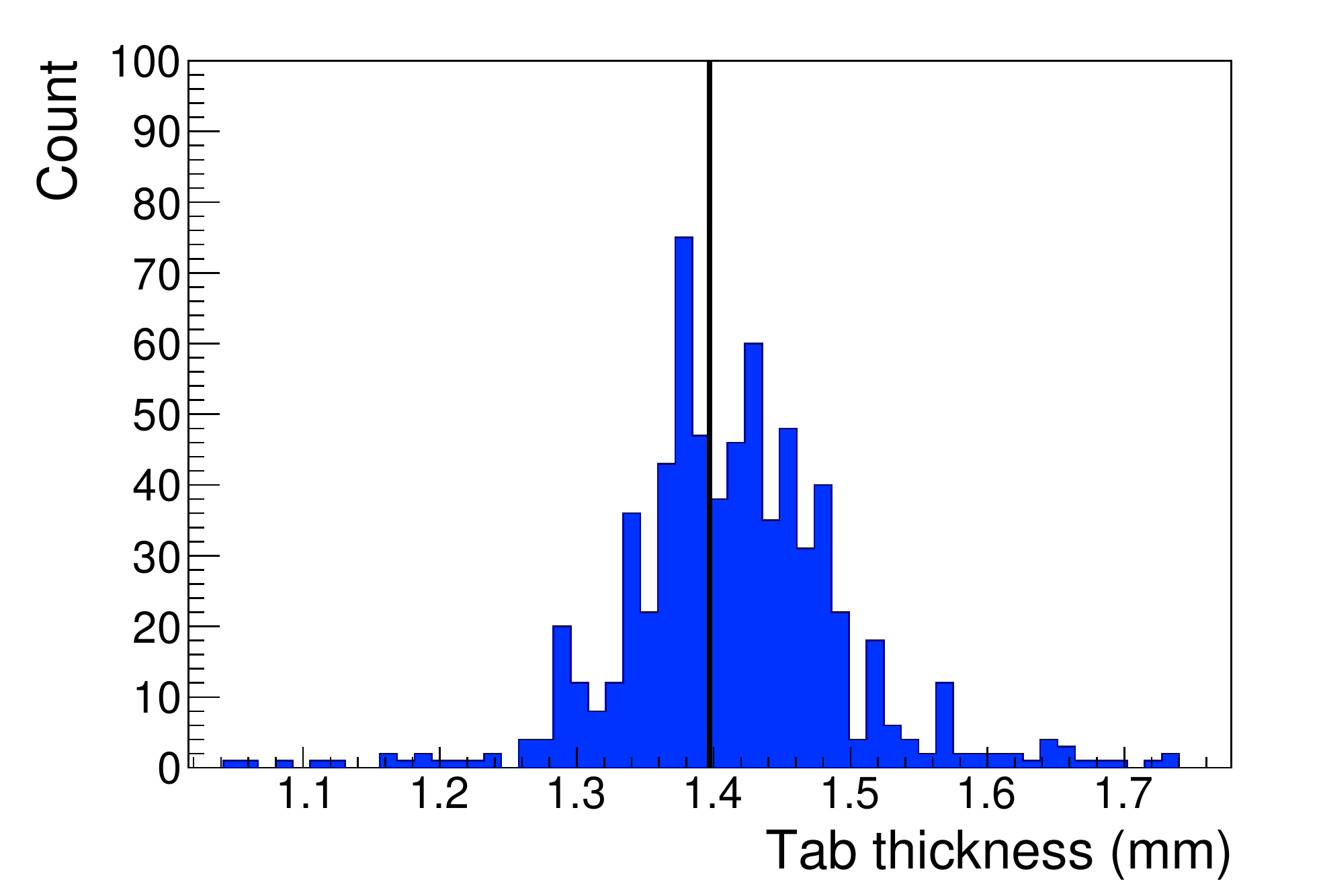}
\caption{Dimensional measurements on the standard PLA rods, solid lines are nominal values. The dashed lines are design tolerance. (Top left) The length of the standard (type-1, blue) and longer standard PLA rods (type-9, gray), where the dashdotted line represents the longer's nominal value. (Top right) The outer width of the PLA rods. (Bottom left) The width of the tabs. (no design tolerance). (Bottom right) The thickness of tabs (no design tolerance).}
\label{fig:PinwheelMeasure}
\end{figure}

Dimensional measurements of shaft cross-sections, rod tab lengths, and total rod length were made on 5\% of all standard PLA rod types.  
Measured dimensions are summarized in Figure~\ref{fig:PinwheelMeasure}, along with the tolerances provided to the manufacturer for each dimension.  
Once again, dimensional repeatability of the printing is demonstrated at the $\pm$0.13~mm ($\pm$0.005~in) level.  
However, the overall rod lengths were measured to be systematically shorter than the specified value; this feature was attributed by the manufacturer to either mis-calibration of their printers or to a slight unexpected shrinking of the filament diameter during printing.  
This small deviation from the requested dimensions was remedied by printing a final batch of slightly longer standard PLA rods to make up the overall shortness in length (the type-9 rods), as shown in Figure~\ref{fig:PinwheelMeasure} (top left).  

\newpage
\subsection{Optical Properties}
\label{sec:optical}
\begin{figure}[h!]
\centering
\includegraphics[width=.45\textwidth]{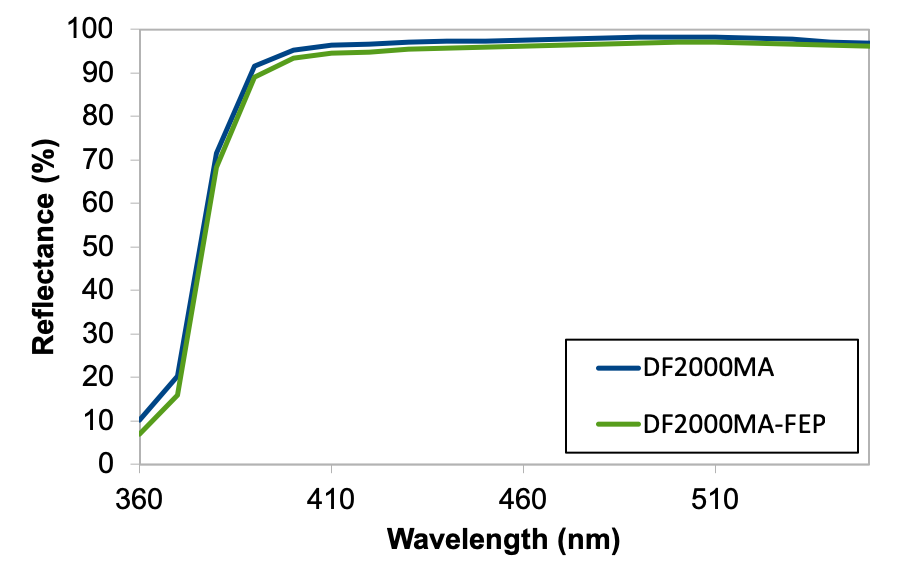} \\
\includegraphics[width=.45\textwidth]{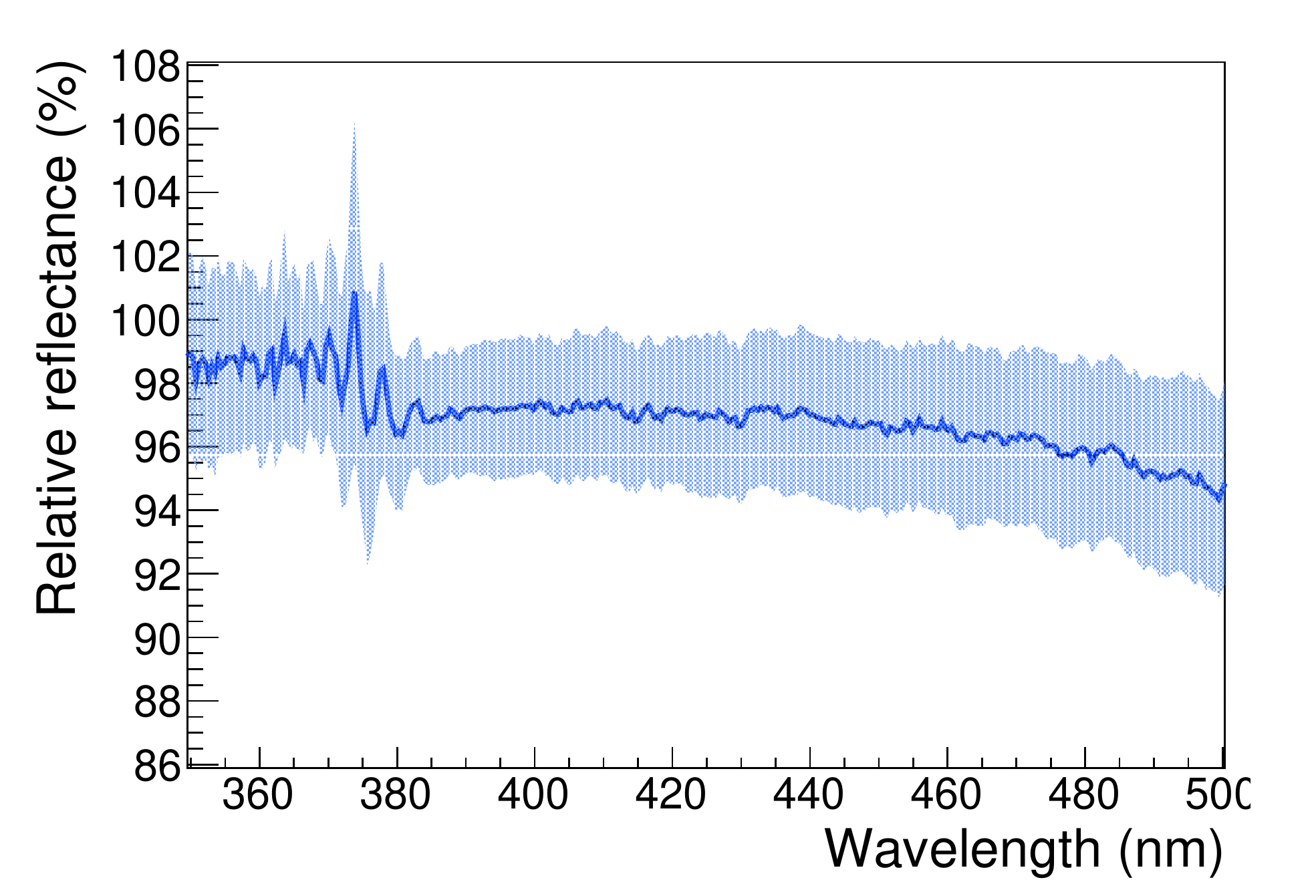}\quad
\includegraphics[width=.45\textwidth]{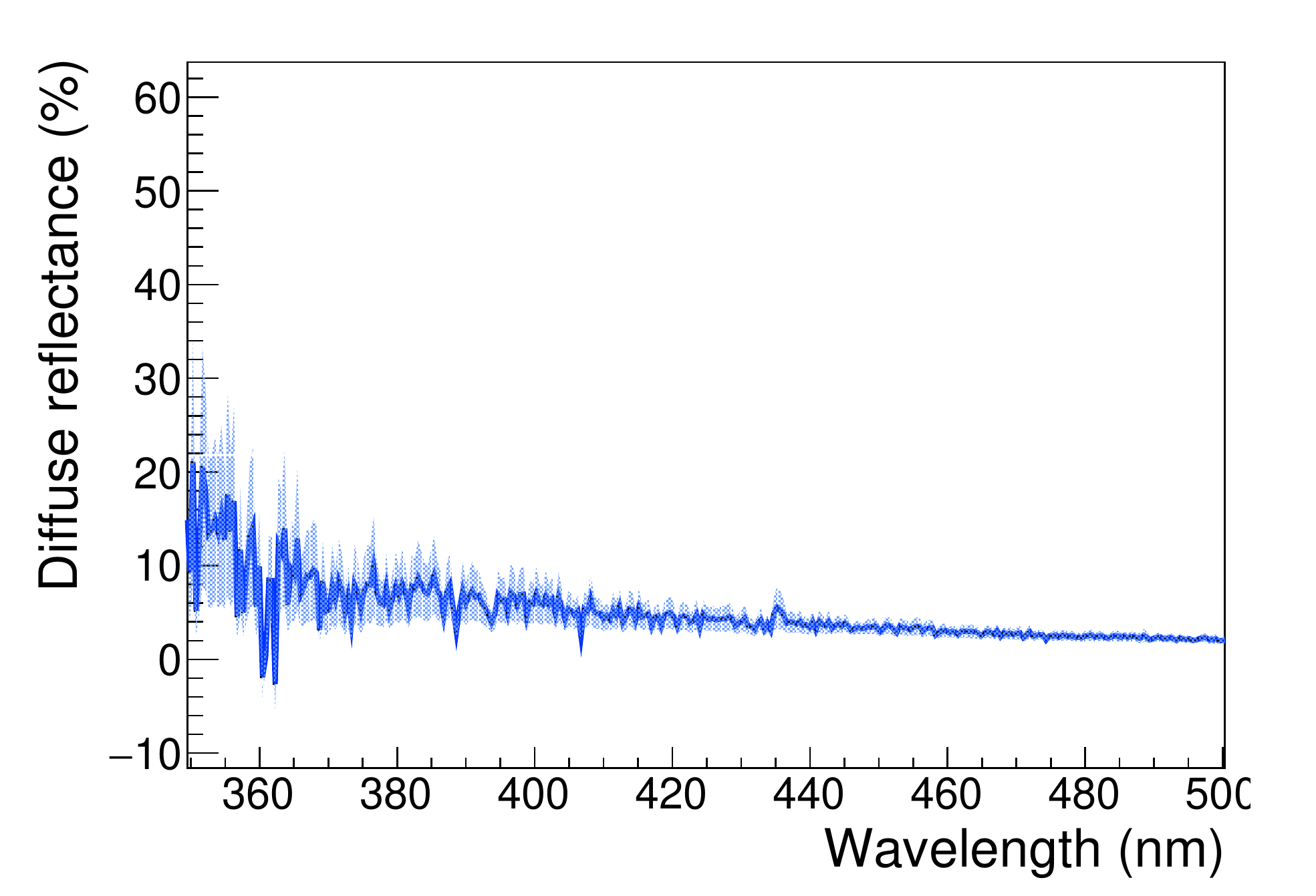}
\caption{The reflectance measurements on the separators. (Top) The absolute total reflectance of FEP coated separator (green) compared with bare DF2000MA (blue). 
(Bottom left) The relative reflectance of the mass production separators compared aganst a 5~cm $\times$ 5~cm separator sample with 1$\sigma$ error, showing 5\% $\pm$ 2\% of difference from the sample. 
(Bottom right) The diffuse reflectance of the mass production separators with 1$\sigma$ error, showing <10\% absolute diffuse reflectance.}
\label{fig:Reflector}
\end{figure}

The specular and diffuse reflectance of separators were measured to ensure high optical quality and for use in PROSPECT optical simulations.
During the R\&D phase, an UltraScan PRO Spectrophotometer was utilized to measure the total (specular+diffuse) and diffuse-only reflectance of 5~cm $\times$ 5~cm samples of the separator materials. 
These results were compared with the other separator candidates during the R\&D phase and led to the decision of utilizing DF2000MA and FEP films.
This spectrometer provided absolute reflectance and transmission of material in the 400~nm to 550~nm wavelength region of interest.
A comparison of total reflectance between bare DF2000MA and a sample of the laminated separator is shown in Figure~\ref{fig:Reflector}, indicating <3\% reduction of reflectance from the adhesive and 50~\textmu m (0.002~in) thick FEP coating.

During the separator QA, 10\% of separators were also randomly chosen to compare their relative variation of reflectance to a small laminated sample of the separator sandwich and a diffuse standard. 
The Ocean Optics STS-VIS spectrometer connected to an optical fiber probe was used for these measurements.
These measurements are also provided in Figure~\ref{fig:Reflector}, along with the 1$\sigma$ band provided by all QA measurements.  
Because of the flexibility of the optical fiber and high light background in the lamination cleanroom, the variations between measurements are significant.
The fabricated separators were found to have total reflectance roughly 5\% lower than the clear laminated sample with $\approx$2\% variation.
The diffuse reflectance measurements showed the <10\% absolute diffuse reflectance among all measured separators.

\begin{figure}[h!]
\centering
\includegraphics[trim = 5cm 5cm 0cm 5cm, clip, width=0.66\textwidth]{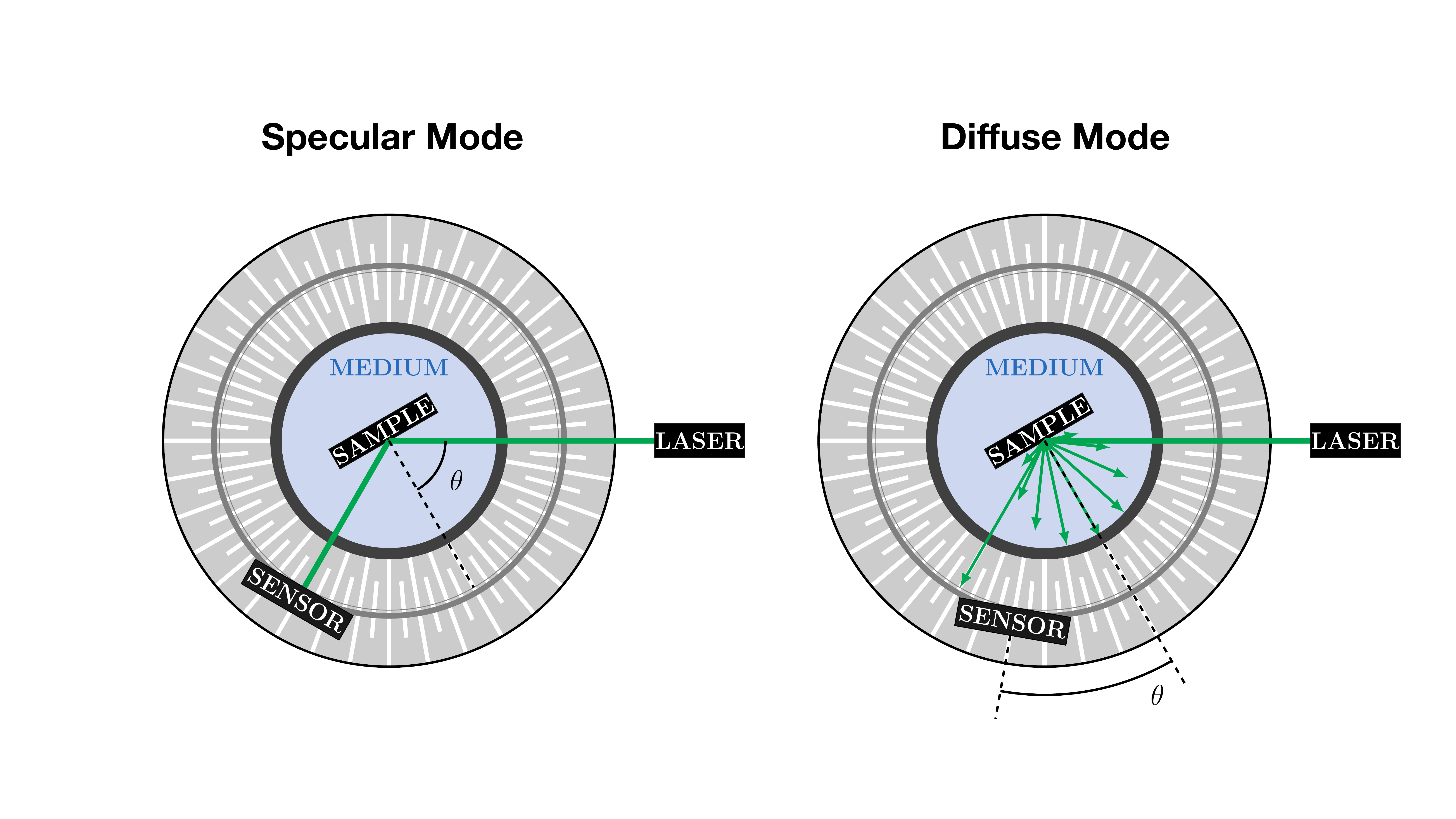}
\includegraphics[width=0.3\textwidth]{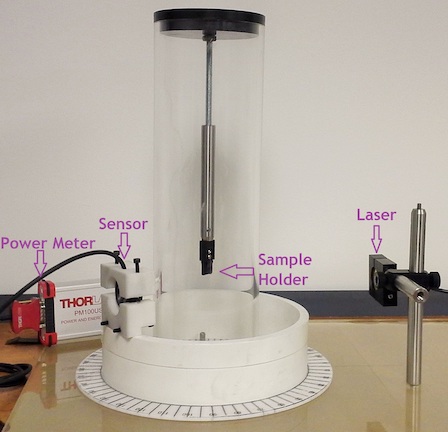}
\caption{The setup of the laser goniometer reflectance measurement setup. (Left) The setup for specular reflectance measurements. The power sensor moves with respect to the track of incident and reflected laser. (Middle) The setup for diffuse reflectance measurements. The incident angle is fixed, while the power sensor is moved from 0$^{\circ}$ to 180$^{\circ}$. 
(Right) Photograph of the custom-built goniometer, where the acrylic tank enables measurement of samples in liquid.}
\label{fig:reflectorgonio}
\end{figure}

In addition to the spectrometer measurement, a laser goniometer was designed and built to measure the sample reflectance vs. incident light angle for samples in different environmental media. 
The goniometer is made up with a 532~nm wavelength laser diode as the light source, a ThorLab S120C light power sensor and PM100USB power meter, and a cylindrical acrylic goniometer with 5$^{\circ}$ precision. 
The goniometer is also a container of liquid media.
In addition to characterizing reflectance vs. incident light angle, this setup allows direct measurements of total internal reflection and the angular dependence of diffusely reflected light. 
Schematics and a photograph of the goniometer system are shown in Figure~\ref{fig:reflectorgonio}, along with descriptions of specular and diffuse measurement procedures.

\begin{figure}[h!]
\centering
\includegraphics[width=0.49\textwidth]{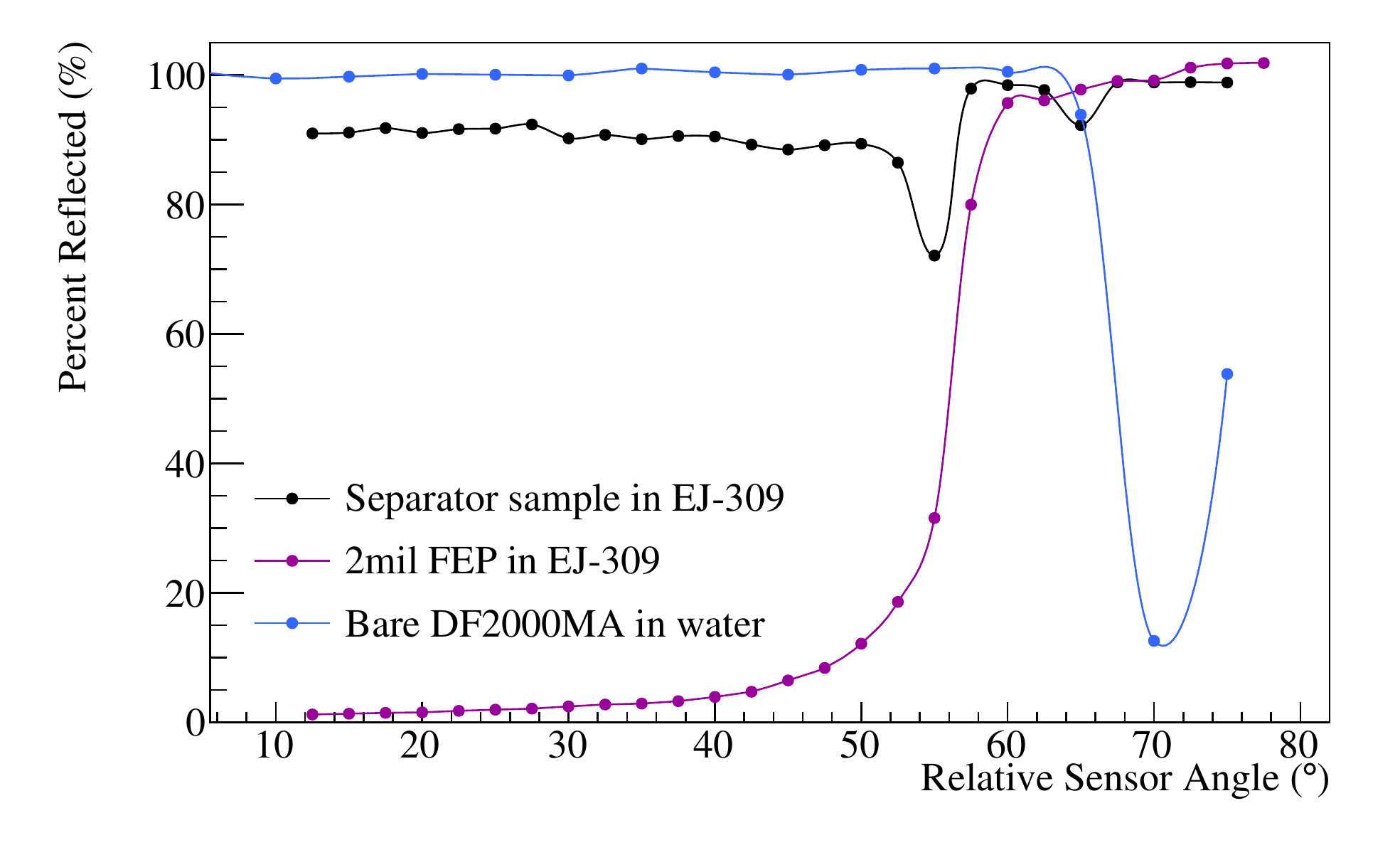}
\includegraphics[width=0.49\textwidth]{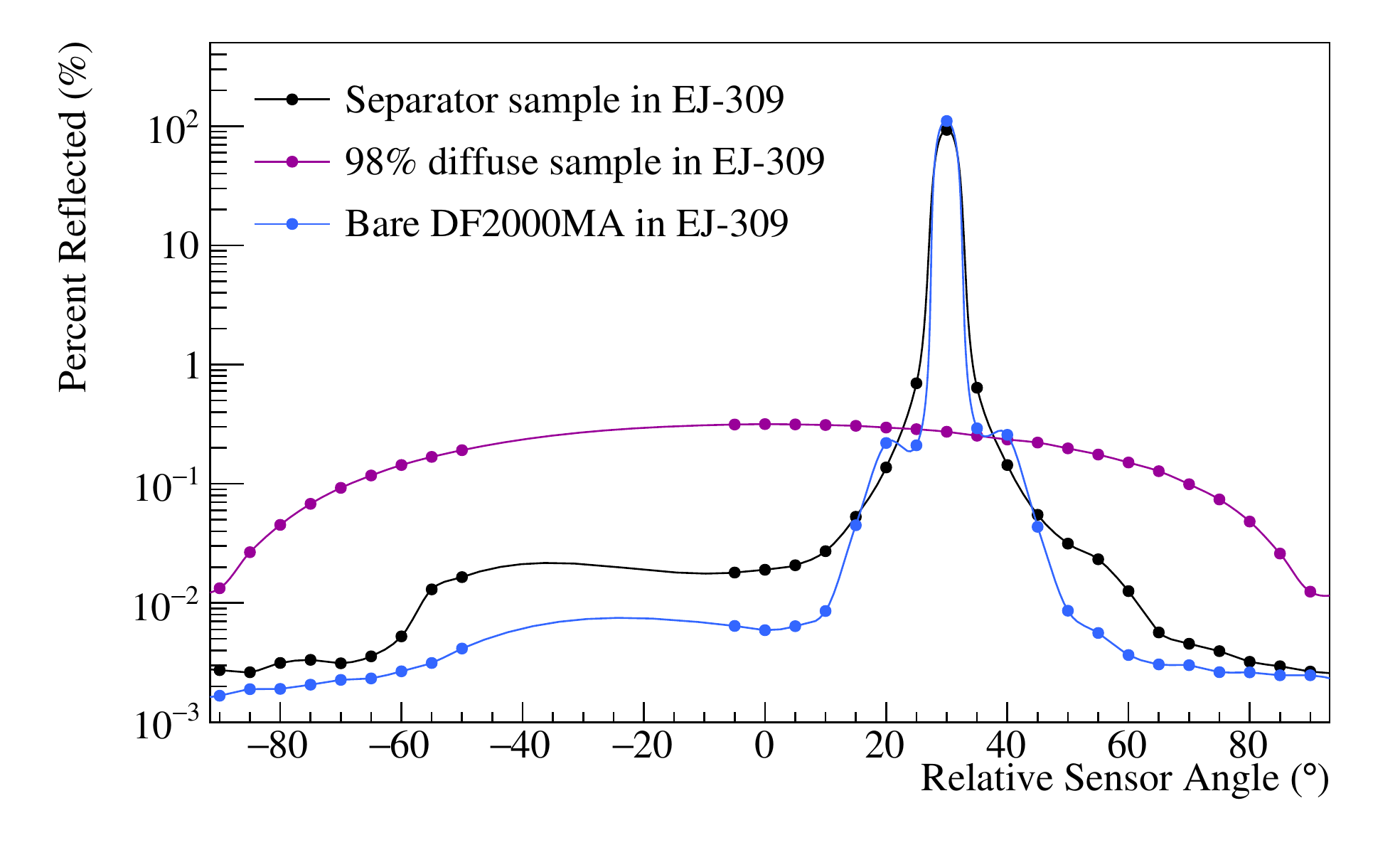}
\caption{The results of goniometer measurements on PROSPECT's separator sample. (Left) The specular reflectance of the laminated separator sample, FEP and bare DF2000MA in clear liquid, with respect to incident angle, shows the total internal reflection effect at large angle. (Right) The diffuse reflectance measured by the goniometer. The bare and laminated reflector compared against a 98\% diffuse reflective sample.}
\label{fig:gonioreflect}
\end{figure}

The results of the goniometer measurement are shown in Figure~\ref{fig:gonioreflect} (left).
The specular reflectance vs. light incident angle indicates good specular reflectance in liquid of bare DF2000MA until its reflectance drops at high incident angles; this non-reflective behavior at high incident angles while in contact with materials of high refractive index has been previously observed with similar type of material~\cite{bib:stereo}. 
These specular measurements also showed the turn-on of total internal reflection in bare FEP film in EJ-309 and a modest reduction in specular reflection of the full separator panel with increasing incident angle prior to the critical angle for total internal reflection in the FEP.
These results underscore the key role played by the external FEP coating in achieving good optical performance, as well as good chemical compatibility, in the optical grid sub-system. 
The diffuse reflectance measurement (Figure \ref{fig:gonioreflect} right) shows diffuse reflection at 30$^{\circ}$ light incidence of the bare DF2000MA, separator sample and 98\% diffuse reflective sample (White98).
A diffuse reflection component was visible in the bare reflector but was clearly sub-dominant to the specular component. 
Diffuse light appeared again to be greater when FEP is introduced; diffuseness appeared to deviate from a perfectly Lambertian reflection. 

\begin{figure}[h!]
\centering
\includegraphics[width=0.48\textwidth]{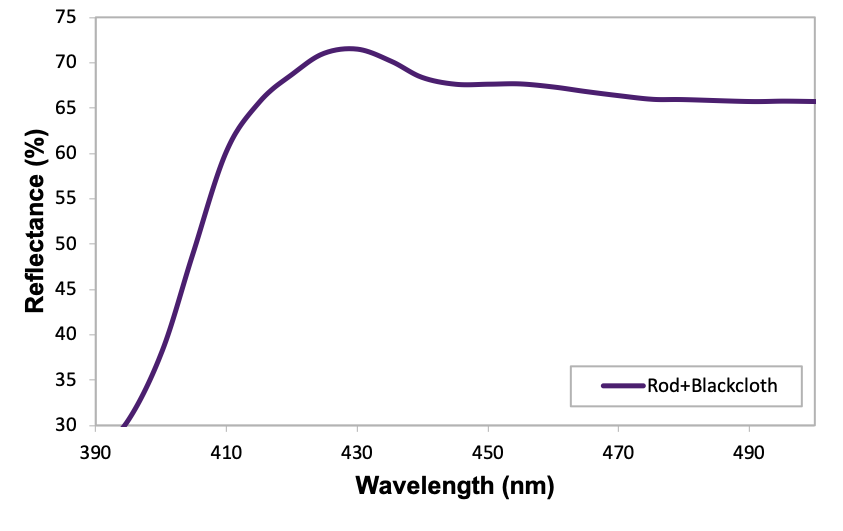} \quad
\includegraphics[width=0.48\textwidth]{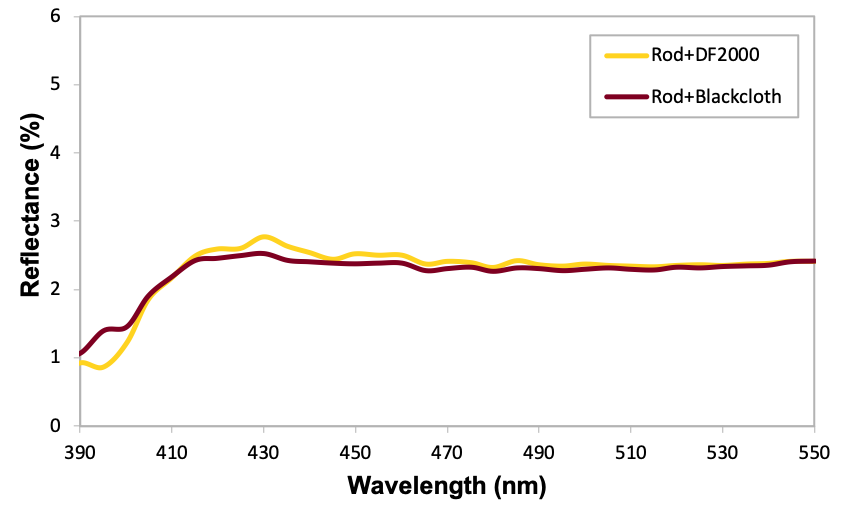}

\caption{(Left) The absolute diffuse reflectance of PLA rods. (Right) Total reflectance of PLA rods relative to bare DF2000MA. When PLA was backed by DF2000MA reflector, the reflectance can be compared against the measurement of PLA backed by black cloth to indicate the transmission of light. }
\label{fig:PinwheelOptic}
\end{figure}

The optical specification of the 3D printed PLA rods was also measured with the Ocean Optics STS-VIS spectrometer.
The diffuse and specular reflectance measurements are shown in Figure~\ref{fig:PinwheelOptic}.
The diffuse reflectance measurement showed 65\% to 75\% of total reflection compared with a standard diffuse reflector sample.
In addition, because the white 3D PLA wall of the PLA rods is not perfectly opaque to visible light, the specular reflectance was measured to characterize both its reflectivity and its light cross-talk.
The specular reflectance was measured when the PLA wall was backed by a piece of black cloth to serve as light trap, indicating 2\% to 3\% of specular reflectance relative to bare DF2000MA.
The cross-talk was evaluated when the PLA wall was backed by a bare DF2000MA, so the reflected light could transmit through the PLA material, showing <1\% increase of reflection light intensity, demonstrating the light transmission through the PLA is negligible.

\subsection{Mechanical test on rods}
\begin{figure}[h!]
\centering
\includegraphics[width=0.5\textwidth]{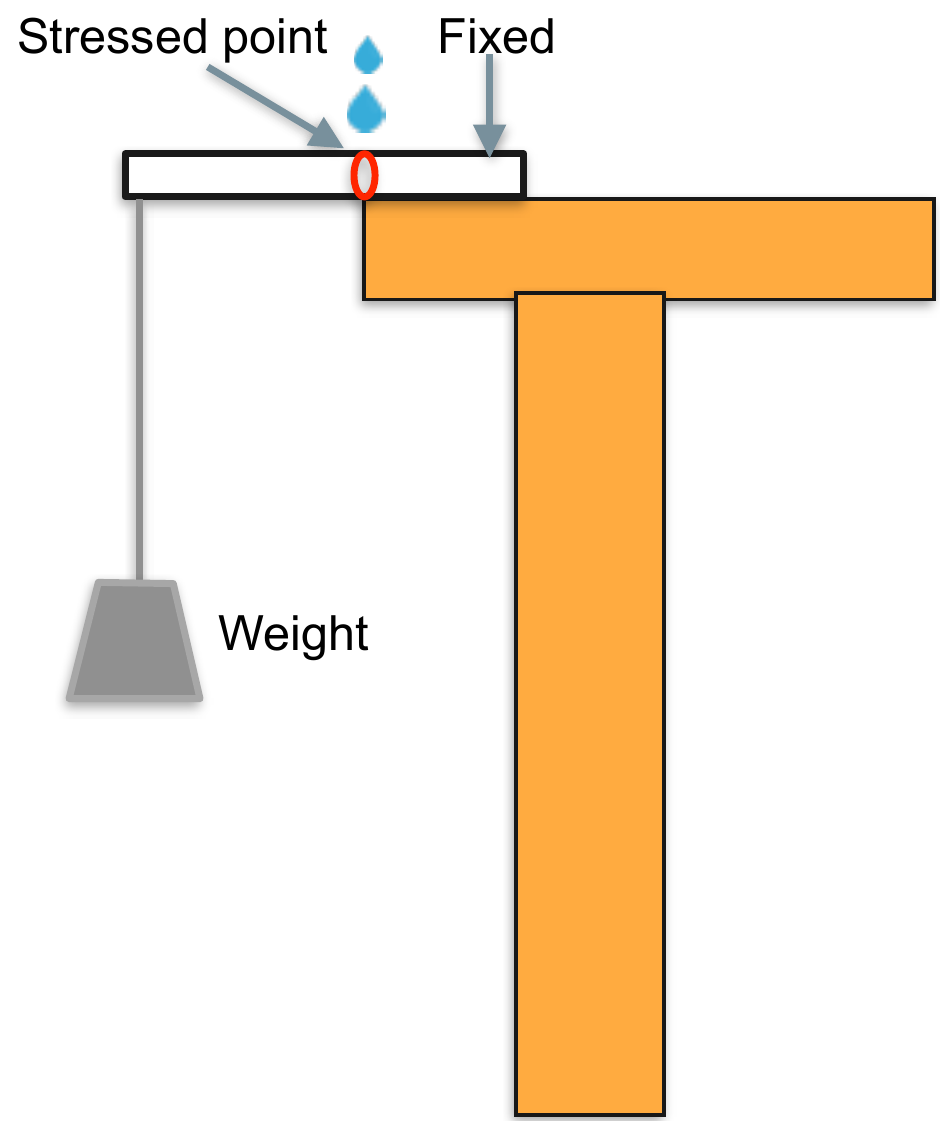}
\caption{The scheme of the lever test, where the liquid drops represent $^{6}$LiLS. }
\label{fig:leverPic}
\end{figure}

Due to the nature of the 3D printing procedure, it is easier to break the PLA rods along the filament.
Thus, a long-term stress test on the PLA rods was made to ensure the PLA rods are rigid and stable under stress when in contact with PROSPECT $^{6}$LiLS.
Stress was applied to individual standard PLA rods by cantilevering 2/3 of the length of each PLA rod from a bench and then attaching water bottles with a known mass on the other end.
In this lever system, the lengths of lever arms and the mass of the water-bottle weight were precisely measured to ensure the desired stress at the fulcrum, as shown in Figure~\ref{fig:leverPic}.
By approximating the cross-sectional shape of the PLA rod as square, the stress at the fulcrum can be calculated based on the length of lever arm and mass of the bottle. 
This test system applied 100 kPa (substantially larger stress than that expected on the optical grid in the deployed PROSPECT detector) to four rods during the test: two were used as references, two were subjected to 1~mL EJ-309 applied each day.
There was no difference between the test samples and reference samples over 3 months. 
After 3 months, the same liquid was applied once a week for another 3 months. 
Then the liquid application was stopped while all test and reference PLA rods were kept in the same situation for another 8 months. 
During the 14-month stress testing, no tested rod bent or broke, and all tested rods exhibited similar results as the references.
During the R\&D phase, other candidates, including the 3D printed ABS rods and extruded acrylic rods, were found unstable under stress while contacting EJ-309.

\subsection{Compatibility with $^{6}$LiLS}
The chemical compatibility of the candidate separator materials with $^{6}$LiLS was characterized by observing the changes of the liquid and materials' quality while in contact with each other.  
The components exposed to $^{6}$LiLS including the PMT housing, PLA rods and separators
There for materials tested included separator samples with the FEP coating, 3D printed white-dyed PLA.
If a material is incompatible with the $^{6}$LiLS and produces degradation in the chemical composition of the liquid, the light absorbance of the liquid is expected to change. 
Thus, compatibility was tested by performing UV-Vis absorbance spectrum measurements of $^{6}$LiLS with an Agilent Technologies Cary 5000 UV-Vis-NIR spectrometer after long liquid-material exposure times.   

\begin{figure}[h!]
\centering
\includegraphics[width=.7\textwidth]{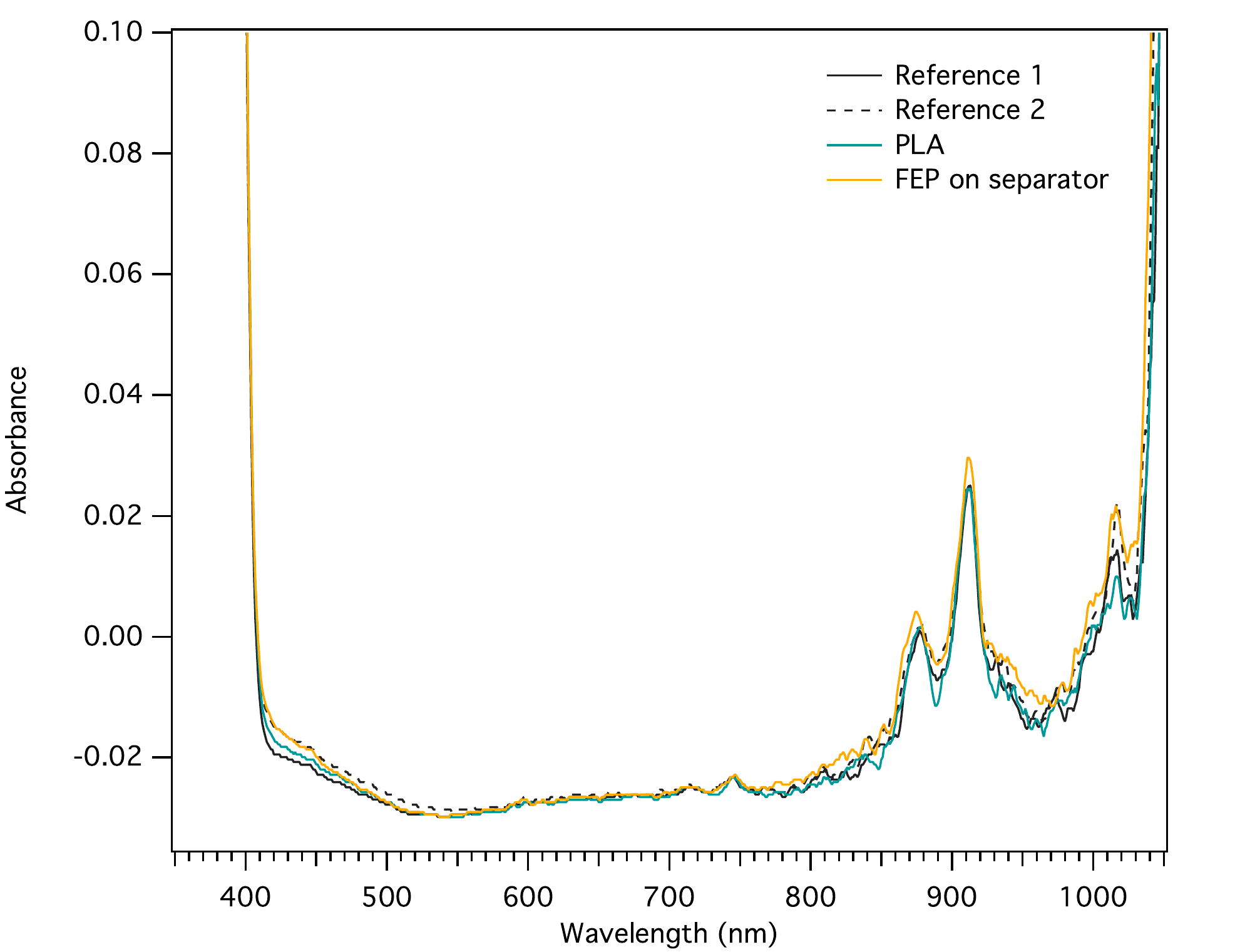}
\caption{The absorbance spectra of $^{6}$LiLS in contact with selected samples, relative to air.
The negative value is caused by the different refraction index between LiLS and air.
After 6 months of contact with different detector components, all test liquid samples showed similar absorbance in 400~nm to 500~nm within the differences caused by sample handling, as shown by the difference between two references.}
\label{fig:compat}
\end{figure}

To perform these tests, the separator samples of 5~cm $\times$ 2~cm size and 5~cm long 3D printed sample of PLA rods were individually soaked in 100~mL of $^{6}$LiLS or EJ-309 inside sealed glass bottles.
The reference liquid samples were $^{6}$LiLS stored in the same type of bottles at the same time.
To perform a measurement, bottles were opened and 3~mL to 5~mL of liquid sample was moved into a 1~cm quartz cuvette, which was then measured in the UV-Vis spectrometer described above.  
Relative absorbance measurement results from $^{6}$LiLS in contact for a 6-month period with the tested samples are shown in Figure~\ref{fig:compat}.  
The samples produced negligible liquid absorbance shift over the 6-month period.
Other materials, including cast acrylic dyed with different colors, polyether ether ketone screws and ACRIFIX used in the PMT housing fabrication, have also been tested compatible to the $^{6}$LiLS during PROSPECT's prototype testing~\cite{bib:prospect_50}.

Incompatibility of various optical grid materials with the $^{6}$LiLS was also characterized by observing changes in the optical grid materials themselves while in contact with the liquid.  
To perform these tests, materials were soaked in $^{6}$LiLS or EJ-309 inside sealed glass bottles over weeks-long timescales.  
During the first 3-week observation, it was found that all optical adhesives failed, resulting in total delamination and physical separation of all separator sandwiches not encased in sealed FEP.  
In addition, the carbon fiber sheet itself showed signs of degradation over few-week timescales, sloughing off a small quantity of black particulate onto the bottom of the test bottle. 
The FEP-sealed separator sandwich samples containing a small puncture were also soaked in $^{6}$LiLS, and exhibited increased levels of sandwich delamination along the edges of the sample over time; however, due to the presence of the intact and unchanged FEP encapsulation, these punctured, delaminated samples still maintained their original geometry and approximate sizes.  
It should also be noted that for the fully-delaminated separator sandwich samples containing the acrylic-based optically clear adhesive chosen for the PROSPECT separators, no visible degradation in UV-Vis absorbance of the host liquid was observed over months-long timescales. 

The separator delamination will cause reflection non-uniformity, if its inner components were exposed to $^{6}$LiLS.
The QA and assembly procedures in Section \ref{sec:QA} and \ref{sec:Assembly} were developed to minimize the potential of punctures on FEP film caused by PMT housing and PLA rods.
The QA tests with relative movements between separators and PLA rods verified the minimal possibility of separator damage.

\section{Summary}
The optical grid sub-system of the PROSPECT AD is a low mass stable subsystem that transports scintillation light efficiently to the segment-end PMTs.
With the optical grid, the detector is able to precisely measure the baseline of incident $\overline{\nu}_e$ events.
This system was successfully fabricated, characterized, transported and assembled into the PROSPECT detector. 
The QA processes were organized to ensure that all of the fabricated components met the requirements of the experiment on the stability, reflectivity, and uniformity of the segments.
The realization of this system relied on high precision machining and 3D printing technology and demonstrated the potential of these technologies for future application in particle detection experiments.

\section*{Acknowledgement}
This material is based upon work supported by the U.S. Department of Energy Office of Science and the Heising-Simons Foundation. 
Additional support for this work is provided by Yale University, the Illinois Institute of Technology, the National Institute of Standards and Technology, and the Lawrence Livermore National Laboratory LDRD program. 
We gratefully acknowledge the support and hospitality of the High Flux Isotope Reactor at the Oak Ridge National Laboratory, managed by UT-Battelle for the U.S. Department of Energy.

\input{main.bbl}
\end{document}

%% file: AuthorList.tex
\author[p]{J.\,Ashenfelter,}
\author[n]{A.\,B.\,Balantekin,}
\author[p]{H.\,R.\,Band,}
\author[f]{C.\,D.\,Bass,}
\author[g]{D.\,E.\,Bergeron,}
\author[j]{D.\,Berish,}
\author[e]{N.\,S.\,Bowden,}
\author[e]{J.\,P.\,Brodsky,}
\author[h]{C.\,D.\,Bryan,}
\author[n]{J.\,J.\,Cherwinka,}
\author[e]{T.\,Classen,}
\author[c]{A.\,J.\,Conant,}
\author[o]{D.\,Davee,}
\author[i]{D.\,Dean,}
\author[h]{G.\,Deichert,}
\author[d]{A.\,E.\,Detweiler,}
\author[a]{M.\,V.\,Diwan,}
\author[b]{M.\,J.\,Dolinski,}
\author[c]{A.\,Erickson,}
\author[i]{M.\,Febbraro,}
\author[p]{B.\,T.\,Foust,}
\author[p]{J.\,K.\,Gaison,}
\author[i,k]{A.\,Galindo-Uribarri,}
\author[d]{Y.\,Gebre,}
\author[i,k]{C.\,E.\,Gilbert,}
\author[d]{K.\,E.\,Gilje,}
\author[d]{I.\,F.\,Gustafson,}
\author[i,k]{B.\,T.\,Hackett,}
\author[a,1]{S.\,Hans,%
\note{Also at: Department of Chemistry and Chemical Technology, Bronx Community College, Bronx, NY, USA.}}
\author[j]{A.\,B.\,Hansell,}
\author[p]{K.\,M.\,Heeger,}
\author[d]{K.\,H.\,Hermanek,}
\author[b]{J.\,Insler,}
\author[a]{D.\,E.\,Jaffe,}
\author[j]{D.\,C.\,Jones,}
\author[b]{O.\,Kyzylova,}
\author[b]{C.\,E.\,Lane,}
\author[p,2]{T.\,J.\,Langford,%
\note{Also at: Yale Center for Research Computing, Yale University, New Haven CT 06520}}
\author[g]{J.\,LaRosa,}
\author[d]{B.\,R.\,Littlejohn,}
\author[i,k]{X.\,Lu,}
\author[d]{D.\,A.\,Martinez\,Caicedo,}
\author[i]{J.\,T.\,Matta,}
\author[o]{R.\,D.\,McKeown,}
\author[e]{M.\,P.\,Mendenhall,}
\author[b]{J.\,M.\,Minock,}
\author[i]{P.\,E.\,Mueller,}
\author[g]{H.\,P.\,Mumm,}
\author[j]{J.\,Napolitano,}
\author[b]{R.\,Neilson,}
\author[p]{J.\,A.\,Nikkel,}
\author[p]{D.\,Norcini,}
\author[g]{S.\,Nour,}
\author[l]{D.\,A.\,Pushin,}
\author[a]{X.\,Qian,}
\author[i,k]{E.\,Romero-Romero,}
\author[a]{R.\,Rosero,}
\author[l]{D.\,Sarenac,}
\author[d,*]{P.\,T.\,Surukuchi,}
\author[g]{M.\,A.\,Tyra,}
\author[i]{R.\,L.\,Varner,}
\author[a]{B.\,Viren,}
\author[d]{C.\,White,}
\author[j]{J.\,Wilhelmi,}
\author[p]{T.\,Wise,}
\author[a]{M.\,Yeh,}
\author[b]{Y.-R.\,Yen,}
\author[a]{A.\,Zhang,}
\author[a]{C.\,Zhang,}
\author[d,**]{X.\,Zhang}

\collaboration{The PROSPECT Collaboration}
\emailAdd{$^*$psurukuc@hawk.iit.edu,~$^{**}$xzhan135@hawk.iit.edu}

\affiliation[a]{Brookhaven National Laboratory, Upton, NY, USA}
\affiliation[b]{Department of Physics, Drexel University, Philadelphia, PA, USA}
\affiliation[c]{George W.\,Woodruff School of Mechanical Engineering, Georgia Institute of Technology, Atlanta, GA USA}
\affiliation[d]{Department of Physics, Illinois Institute of Technology, Chicago, IL, USA}
\affiliation[e]{Nuclear and Chemical Sciences Division, Lawrence Livermore National Laboratory, Livermore, CA, USA}
\affiliation[f]{Department of Physics, Le Moyne College, Syracuse, NY, USA}
\affiliation[g]{National Institute of Standards and Technology, Gaithersburg, MD, USA}
\affiliation[h]{High Flux Isotope Reactor, Oak Ridge National Laboratory, Oak Ridge, TN, USA}
\affiliation[i]{Physics Division, Oak Ridge National Laboratory, Oak Ridge, TN, USA}
\affiliation[j]{Department of Physics, Temple University, Philadelphia, PA, USA}
\affiliation[k]{Department of Physics and Astronomy, University of Tennessee, Knoxville, TN, USA}
\affiliation[l]{Institute for Quantum Computing and Department of Physics and Astronomy, University of Waterloo, Waterloo, ON, Canada}
\affiliation[m]{Department of Physics, University of Wisconsin, Madison, Madison, WI, USA}
\affiliation[n]{Physical Sciences Laboratory, University of Wisconsin, Madison, Madison, WI, USA}
\affiliation[o]{Department of Physics, College of William and Mary, Williamsburg, VA, USA}
\affiliation[p]{Wright Laboratory, Department of Physics, Yale University, New Haven, CT, USA}

%% file: main.bbl
\providecommand{\href}[2]{#2}\begingroup\raggedright\endgroup

%% file: main.bbl
\begin{thebibliography}{10}

\bibitem{bib:prospect_physics}
J.~Ashenfelter et~al., {\scshape PROSPECT} collaboration, \emph{{PROSPECT
  Physics Program}}, {\emph{J. Phys. G} {\bfseries 43} (2016) 113001}.

\bibitem{bib:prospect_nim}
J.~Ashenfelter et~al., {\scshape PROSPECT} collaboration, \emph{{The PROSPECT
  Reactor Antineutrino Experiment}},
  \href{https://doi.org/https://doi.org/10.1016/j.nima.2018.12.079}{\emph{Nuclear
  Instruments and Methods in Physics Research Section A: Accelerators,
  Spectrometers, Detectors and Associated Equipment} {\bfseries 922} (2019) 287
  }.

\bibitem{bib:prospect_first}
J.~Ashenfelter et~al., {\scshape PROSPECT} collaboration, \emph{{First Search
  for Short-Baseline Neutrino Oscillations at HFIR with PROSPECT}},
  \href{https://doi.org/10.1103/PhysRevLett.121.251802}{\emph{Phys. Rev. Lett.}
  {\bfseries 121} (2018) 251802}.

\bibitem{bib:prospect_spectrum}
J.~Ashenfelter et~al., {\scshape PROSPECT} collaboration, \emph{{Measurement of
  the Antineutrino Spectrum from $^{235}$U Fission at HFIR with PROSPECT}},
  \href{https://arxiv.org/abs/1901.05569}{{\ttfamily 1901.05569}}.

\bibitem{bib:mention2011}
G.~Mention et~al., \emph{{The Reactor Antineutrino Anomaly}},
  \href{https://doi.org/10.1103/PhysRevD.83.073006}{\emph{Phys. Rev. D}
  {\bfseries 83} (2011) 073006}.

\bibitem{bib:huber}
P.~Huber, \emph{{On the determination of anti-neutrino spectra from nuclear
  reactors}}, \href{https://doi.org/10.1103/PhysRevC.85.029901,
  10.1103/PhysRevC.84.024617}{\emph{Phys. Rev. C} {\bfseries 84} (2011)
  024617}.

\bibitem{Dyb_Evol}
F.~P. An et~al., {\scshape Daya Bay} collaboration, \emph{{Evolution of the
  Reactor Antineutrino Flux and Spectrum at Daya Bay}}, {\emph{Phys. Rev.
  Lett.} {\bfseries 118} (2017) 251801}.

\bibitem{Giunti_235239}
C.~Giunti, \emph{Improved determination of the $^{235}\mathrm{U}$ and
  $^{239}\mathrm{Pu}$ reactor antineutrino cross sections per fission},
  \href{https://doi.org/10.1103/PhysRevD.96.033005}{\emph{Phys. Rev. D}
  {\bfseries 96} (2017) 033005}.

\bibitem{bib:Giunti2017yid}
C.~Giunti, X.~P. Ji, M.~Laveder, Y.~F. Li and B.~R. Littlejohn, \emph{{Reactor
  Fuel Fraction Information on the Antineutrino Anomaly}},
  \href{https://doi.org/10.1007/JHEP10(2017)143}{\emph{JHEP} {\bfseries 10}
  (2017) 143} [\href{https://arxiv.org/abs/1708.01133}{{\ttfamily
  1708.01133}}].

\bibitem{surukuchi_flux}
Y.~Gebre, B.~R. Littlejohn and P.~T. Surukuchi, \emph{Prospects for improved
  understanding of isotopic reactor antineutrino fluxes},
  \href{https://doi.org/10.1103/PhysRevD.97.013003}{\emph{Phys. Rev. D}
  {\bfseries 97} (2018) 013003}.

\bibitem{bib:DYB}
F.~P. An et~al., {\scshape Daya Bay} collaboration, \emph{{Measurement of the
  Reactor Antineutrino Flux and Spectrum at Daya Bay}},
  \href{https://doi.org/10.1103/PhysRevLett.116.061801}{\emph{Phys. Rev. Lett.}
  {\bfseries 116} (2016) 061801}.

\bibitem{bib:DC}
Y.~Abe et~al., \emph{{Improved measurements of the neutrino mixing angle
  $\theta$ 13 with the Double Chooz detector}},
  \href{https://doi.org/10.1007/JHEP10(2014)086}{\emph{Journal of High Energy
  Physics} {\bfseries 2014} (2014) 86}.

\bibitem{bib:RENO}
J.~H. Choi et~al., {\scshape RENO} collaboration, \emph{{Observation of Energy
  and Baseline Dependent Reactor Antineutrino Disappearance in the RENO
  Experiment}},
  \href{https://doi.org/10.1103/PhysRevLett.116.211801}{\emph{Phys. Rev. Lett.}
  {\bfseries 116} (2016) 211801}.

\bibitem{bib:haser}
C.~Buck, A.~Collin, J.~Haser and M.~Lindner, \emph{{Investigating the spectral
  anomaly with different reactor antineutrino experiments}},
  \href{https://doi.org/https://doi.org/10.1016/j.physletb.2016.11.062}{\emph{Physics
  Letters B} {\bfseries 765} (2017) 159 }.

\bibitem{bib:huber2016xis}
P.~Huber, \emph{{NEOS Data and the Origin of the 5 MeV Bump in the Reactor
  Antineutrino Spectrum}},
  \href{https://doi.org/10.1103/PhysRevLett.118.042502}{\emph{Phys. Rev. Lett.}
  {\bfseries 118} (2017) 042502}
  [\href{https://arxiv.org/abs/1609.03910}{{\ttfamily 1609.03910}}].

\bibitem{bib:lspaper}
J.~Ashenfelter et~al., {\scshape PROSPECT} collaboration, \emph{{Lithium-loaded
  Liquid Scintillator Production for the PROSPECT experiment}},
  \href{https://arxiv.org/abs/1901.05569}{{\ttfamily 1901.05569}}.

\bibitem{bib:prospect_50}
J.~Ashenfelter et~al., {\scshape PROSPECT} collaboration, \emph{{Performance of
  a segmented 6 Li-loaded liquid scintillator detector for the PROSPECT
  experiment}}, {\emph{Journal of Instrumentation} {\bfseries 13} (2018)
  P06023}.

\bibitem{bib:P20}
J.~Ashenfelter et~al., {\scshape PROSPECT} collaboration, \emph{{Light collection and pulse-shape discrimination
  in elongated scintillator cells for the PROSPECT reactor antineutrino
  experiment}}, {\emph{Journal of Instrumentation} {\bfseries 10} (2015)
  P11004}.


\bibitem{bib:ESR_science}
M.~F. Weber, C.~A. Stover, L.~R. Gilbert, T.~J. Nevitt and A.~J. Ouderkirk,
  \emph{{Giant Birefringent Optics in Multilayer Polymer Mirrors}},
  \href{https://doi.org/10.1126/science.287.5462.2451}{\emph{Science}
  {\bfseries 287} (2000) 2451}
  [\href{https://arxiv.org/abs/http://science.sciencemag.org/content/287/5462/2451.full.pdf}{{\ttfamily
  http://science.sciencemag.org/content/287/5462/2451.full.pdf}}].

\bibitem{bib:stereo}
N.~Allemandou et~al., \emph{{The STEREO experiment}}, {\emph{Journal of
  Instrumentation} {\bfseries 13} (2018) P07009}.

\end{thebibliography}
